\documentclass[ twocolumn]{aastex7}
\usepackage{graphics}
\usepackage{amsmath}
\usepackage{amssymb}
\usepackage{array}
\usepackage{placeins}
\usepackage{tabularx}
\usepackage{booktabs}
\usepackage{array}
\usepackage{multirow}
\usepackage{xcolor}
\usepackage{colortbl}
\usepackage{array}
\usepackage{hyperref}
\usepackage{rotating}
\usepackage{graphicx}
\usepackage{booktabs}

\received{\today}
\revised{\today}
\accepted{\today}
\submitjournal{ApJ}

\shorttitle{Waterloo 01}
\graphicspath{{./}{figures/}}

\begin{document}

\title{The Interplay of Stellar Feedback and Gravity in the Wat01–S208 Star-forming Complex}

        \author[0000-0002-0444-0439, sname='Kaur']{Harmeen Kaur}
        \affil{Aryabhatta Research Institute of Observational Sciences (ARIES),
        Manora Peak, Nainital 263 002, India}
        \affil{Department of Physics
        DSB Campus, Kumaun University, Nainital, 263002, India}
        \email[show]{harmeenkaur.kaur229@gmail.com}
        \email[show]{harmeen@aries.res.in}

        \author[0000-0001-5731-3057, sname='Sharma']{Saurabh Sharma}
        \affil{Aryabhatta Research Institute of Observational Sciences (ARIES),
        Manora Peak, Nainital 263 002, India}
        \email{saurabh@aries.res.in}
        
        \author[0000-0001-6725-0483, sname='K. Dewangan']{Lokesh K. Dewangan}
        \affil{Physical Research Laboratory, Navrangpura, Ahmedabad - 380 009, India}
        \email{lokeshd@prl.res.in}

        \author[0000-0001-9312-3816, sname='K. Ojha']{Devendra K. Ojha}
        \affil{Tata Institute of Fundamental Research (TIFR),
        Homi Bhabha Road, Colaba, Mumbai - 400 005, India}
        \email{ojha@tifr.res.in}
        
        \author{Alok Durgapal}
        \affil{Department of Physics 
        DSB Campus, Kumaun University, Nainital, 263002, India}
        \email{alokdurgapal@gmail.com}
        
        \author{Bhuwan C. Bhatt}
        \affil{Indian Institute of Astrophysics (IIA), Koramangala, Bangalore 560 034, India}
        \email{bcb@iiap.res.in}
        
         \author[0000-0002-6586-936X, sname='Verma']{Aayushi Verma}
        \affil{Aryabhatta Research Institute of Observational Sciences (ARIES),
        Manora Peak, Nainital 263 002, India}
        \email{aayushiverma@aries.res.in}
        
        \author[0009-0008-8490-8601, sname='Chand']{Tarak Chand}
        \affil{Aryabhatta Research Institute of Observational Sciences (ARIES),
        Manora Peak, Nainital 263 002, India}
        \email{tarakchand@aries.res.in}
        
         \author[0009-0004-4776-6780, sname='Gokhe']{Kartik Gokhe}
        \affil{Aryabhatta Research Institute of Observational Sciences (ARIES),
        Manora Peak, Nainital 263 002, India}
        \email{gokhekartik@aries.res.in}
        
        \author[0009-0001-4144-2281]{Mamta}
        \affil{Aryabhatta Research Institute of Observational Sciences (ARIES),
        Manora Peak, Nainital 263 002, India}
        \email{mamtaaries@aries.res.in}
        
        \author[0009-0009-6420-8058, sname='Chakraborty']{Manojit Chakraborty}
        \affil{Aryabhatta Research Institute of Observational Sciences (ARIES),
        Manora Peak, Nainital 263 002, India}
        \email{manojit@aries.res.in}





\begin{abstract}
To investigate the processes governing clustered star formation, we present a multi-wavelength study of the star-forming complex hosting clusters Waterloo~01  and [BDS2003]64 in Sh 2-208. Both clusters exhibit elongated morphologies, with radii of $\sim1.5$ and $\sim0.9$ pc, respectively, and are located at a common distance of $\sim3.5$ kpc. Wat01 displays hierarchical structure ($Q=0.6$) and signature of mass segregation ($\Gamma_{\rm MSR}=1.5\pm0.6$). Its relaxation time, comparable to or longer than its age, suggests that the observed segregation may have a primordial contribution. The mass function follows a single power law, $\Gamma=-1.62\pm0.15$, over $\sim0.7$--$19.6, M_\odot$. Molecular-line observations reveal distinct velocity components associated with S208 and Wat01, indicating different molecular environments within the complex. Wat01 is strongly shaped by massive-star feedback, with an evacuated cavity, surrounding shell, and compressed globules consistent with an expanding H\,{\sc ii} region; feedback may have promoted localized star formation, although a direct causal connection cannot be established. 
In contrast, \textit{Herschel} data uncover a hub-filament system (HFS) in S208, with the central hub associated with BDS64 and the H\,{\sc ii} region powered by `S1'. The filaments remain relatively well-preserved, indicating that gravity continues to play an important role in assembling dense gas toward the hub and in clustered star formation, even as feedback begins to affect the immediate surroundings. Overall, this complex illustrates the coexistence of feedback-processed and comparatively less-disrupted star-forming environments, highlighting the evolving interplay between stellar feedback and gravity in clustered star formation.

\end{abstract}


\section{Introduction} \label{intro}

Young stellar clusters (YSCs) are sites of star-forming activity, where the formation of massive stars and their subsequent influence on the natal molecular cloud can be examined \citep{2010ARA&A..48..431P,2019ARA&A..57..227K,
2020SSRv..216...69A}.
 A central challenge in star formation studies is to understand how the gravitational assembly of molecular gas and feedback from massive stars jointly regulate the formation and early evolution of stellar clusters
\citep{2019ARA&A..57..227K, 2024A&A...690A..94P, 2025ARA&A..63....1B}.
Hub--filament systems (HFSs) provide a direct connection between the large-scale structure of molecular clouds and the formation of YSCs and massive stars. In these systems, multiple filaments converge toward dense hubs, potentially assembling dense gas at sites of active clustered star formation.
\citep{2009ApJ...700.1609M,2018ARA&A..56...41M,2020A&A...642A..87K}. As massive stars form, their ionizing radiation and winds can subsequently disperse, compress, and restructure the surrounding molecular gas \citep{2019ARA&A..57..227K}. 
Regions at different evolutionary stages within the same molecular complex can therefore provide valuable insights into how the relative roles of gravity and stellar feedback change during clustered star formation.

As part of our multi-wavelength study of YSCs 
\citep{2016AJ....151..126S,2017MNRAS.467.2943S,2020ApJ...891...81P,2020ApJ...896...29K,2022ApJ...926...25P,2023ApJ...944..228M,2023JApA...44...52V,2023ApJ...953..145V,2023JApA...44...46S,2024MNRAS.527.9626P,2024AJ....168...98V,2024AJ....167..106S,2025ApJ...990..155V}, we investigate the star-forming complex G151.32+1.99. The complex contains the YSC Waterloo 01 (hereafter Wat01; \citealt{1979A&AS...38..197M}), which lies southwest of an embedded infrared cluster, [BDS2003]64 (hereafter BDS64), associated with the Sh 2-208 H\,{\sc ii} region (hereafter S208; \citealt{1959ApJS....4..257S, 2003A&A...404..223B}). The coexistence of these young stellar systems within the same molecular complex provides an opportunity to examine how different physical processes influence clustered star formation at different evolutionary stages.

\begin{figure*}
        \centering
        \hbox{
        \hspace{-0.5cm}
        \includegraphics[width=0.65\textwidth, angle=0]{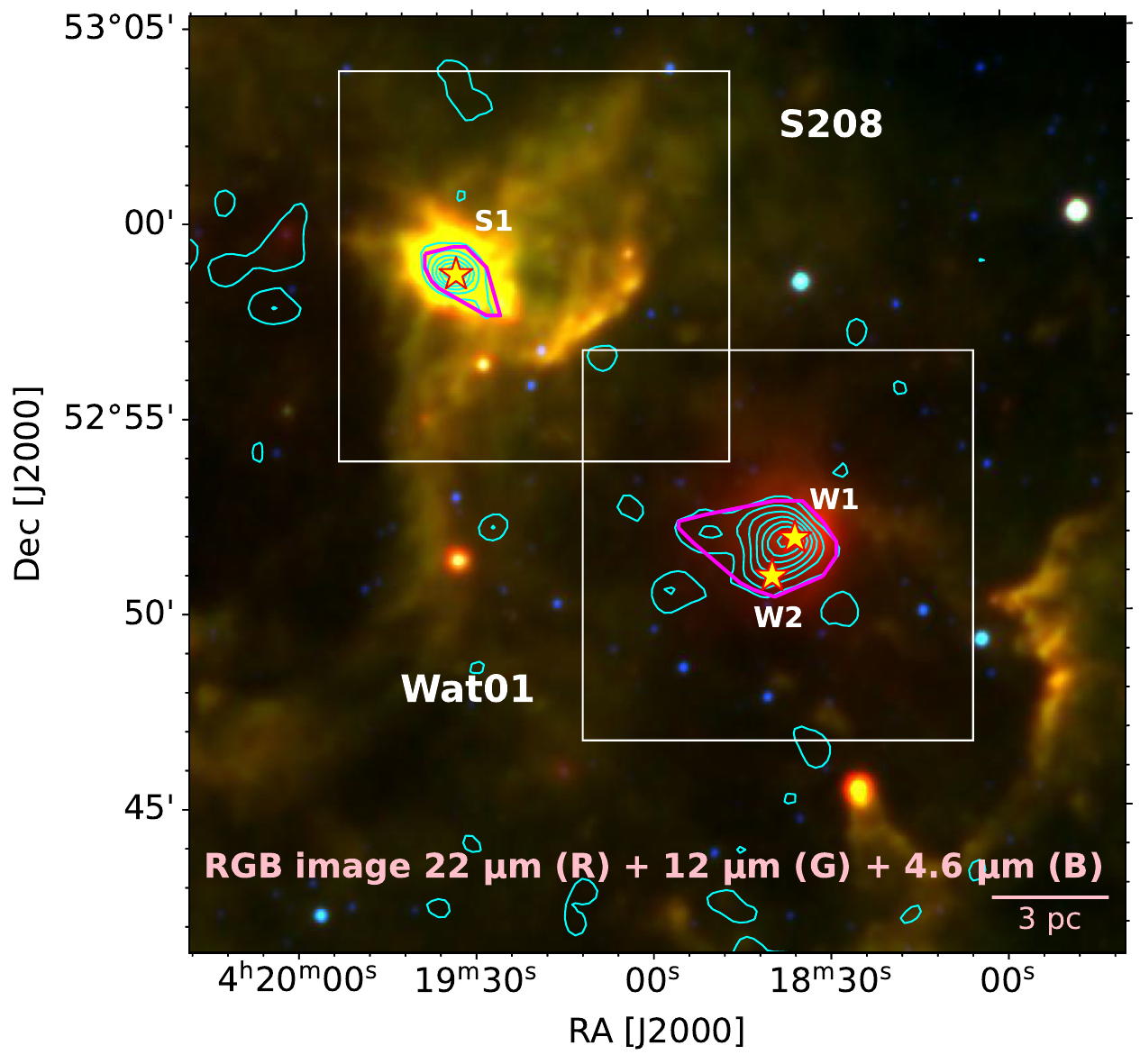}
        \hspace{-6.5cm}
        {\vbox
        {
        \includegraphics[width=0.32\textwidth, angle=0]{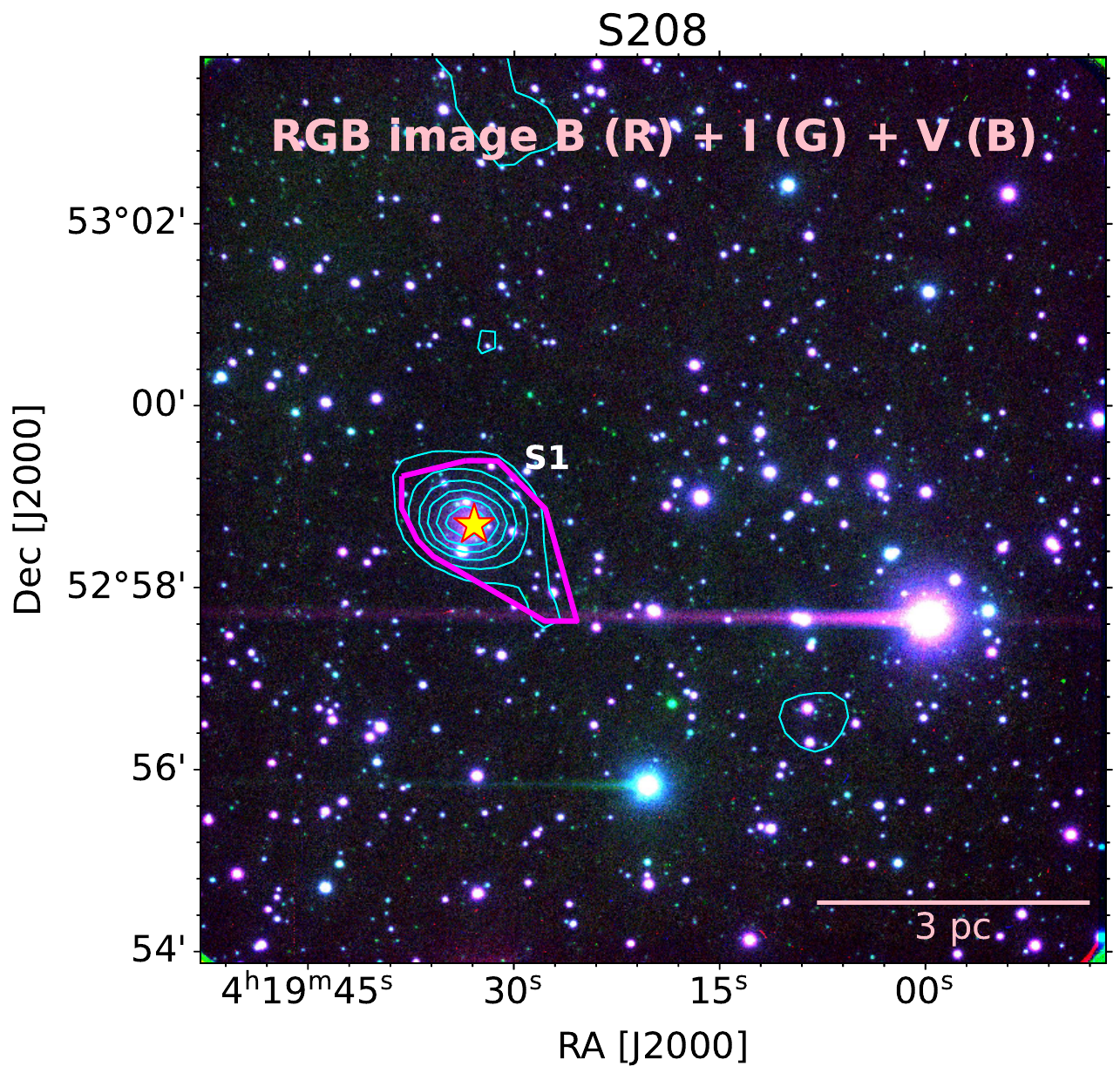}
        {\vbox{
        \vspace{0.6cm}
        \includegraphics[width=0.32\textwidth, angle=0]{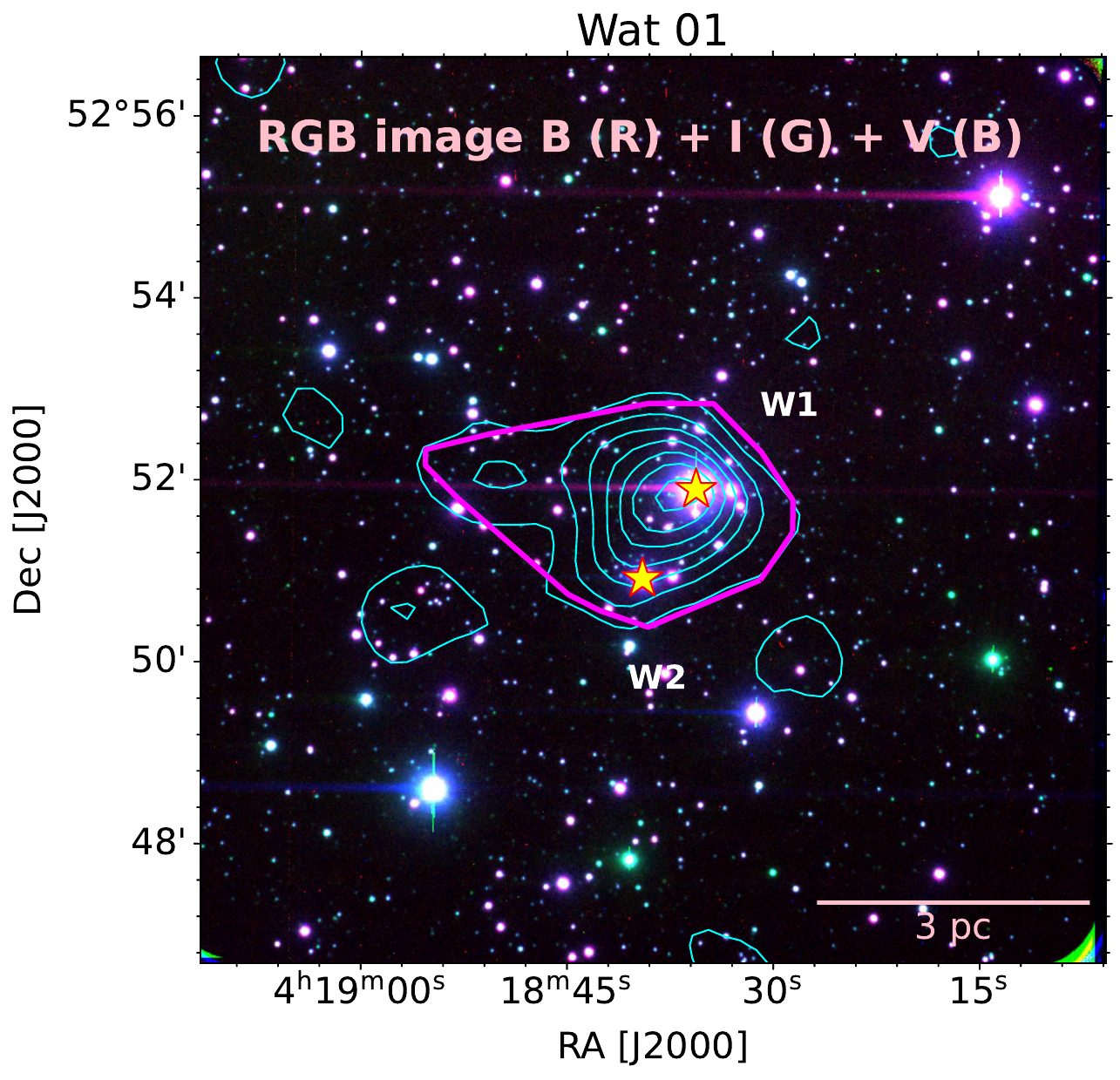}
        }}}}}
        \caption{\label{rgb32} {\it{Left panel}}: Color-composite image (Red: WISE 22$\mu$m, Green: WISE 12 $\mu$m and Blue: WISE 4.6 $\mu$m) of the $24^\prime\times24^\prime$  FOV of the star forming region G151.32+1.99  embracing S208 and Wat01 regions. This color image is overlaid with isodensity contours generated from the NIR catalog (2MASS+UKIDSS; see section~\ref{iso}).
        The lowest level for the isodensity contours is 3$\sigma$ above the mean stellar density (i.e., 20.5 stars arcmin$^{-2}$) with a step size of 1$\sigma$ (3.5 stars arcmin$^{-2}$). White boxes delineate the optically observed $10\times10$ arcmin$^2$ FOV from 2m HCT Hanle, and magenta curves are 
        the convex hull for the lowest density contour for both regions (see Section~\ref{opt}). Yellow asterisk symbols depict the location of massive stars in both regions. {\it{Right panels}}: Color-composite image (Red: B band, Green: I band, and Blue: V band; Optically observed from HCT Hanle) for S208 and Wat01 regions, shown as white boxes in the left panel. Overlaid contours and symbols are the same as those in the left panel.}
         \end{figure*}

Figure~\ref{rgb32} shows a $24^\prime\times24^\prime$ WISE color-composite image of the G151.32+1.99 complex. Wat01 shows a relatively evacuated central region surrounded by an arc-like PAH structure and hosts the massive stars ‘W1’ (O9V) and ‘W2’ (B1.5V). In contrast, S208 is associated with strong mid-infrared (MIR) emission and contains the ionizing source ‘S1’. The distinct morphologies of Wat01 and S208 indicate different physical conditions within the same molecular complex, possibly reflecting different stages of cloud evolution and interaction with massive stars.

In this work, we combine optical, infrared, molecular-line, and 1.4 GHz radio-continuum observations to investigate the stellar populations, molecular gas morphology and kinematics, and feedback processes in Wat01 and S208. We characterize the dynamical state of the clusters, examine the HFS structure in S208, and assess the impact of massive-star feedback on the molecular environment of Wat01. The contrasting properties of the two regions allow us to explore the relative roles of gravity and stellar feedback in regulating clustered star formation within a common molecular complex.

The paper is organized as follows. Section~\ref{obs} describes the observations and data sets. Section~\ref{s4} presents the stellar and molecular-gas properties of the regions. Section~\ref{dis} discusses the cluster dynamics, stellar feedback, and the hub--filament structure. Finally, Section~\ref{s6} summarizes our main conclusions.

    \section{Observation and data reduction}\label{obs}

        \subsection{Optical data}\label{opt}

        The broadband $UBV{(RI)}_c$ optical photometric data of the Wat01 centered at $\alpha_{J2000}$: 04$^h$18$^m$38.728$^s$, $\delta_{J2000}$: 
	   +52$^\circ$51$^\prime$40.24$^\prime$$^\prime$;
        $l=151^\circ.272$ and $b=1^\circ.789$, and S208 centered at $\alpha_{J2000}$: 04$^h$19$^m$19.76$^s$, $\delta_{J2000}$: 
	   +52$^\circ$58$^\prime$50$^\prime$$^\prime$;
        $l=151^\circ.262$ and $b=1^\circ.95$,
        were acquired by using the $2048\times 4096$ pixel$^2$ Hanle faint object 
	       spectrograph camera (HFOSC) mounted at
        2-m Himalayan Chandra Telescope (HCT), Hanle, Ladakh, India.
        The entire chip (Plate scale: 0.296 arcsec/pixel) covers a $\sim10^\prime\times10^\prime$ FOV on the sky. The observations were carried out in a $2\times2$ binning mode to improve the signal-to-noise ratio (SNR).
        The read-out noise and gain of the CCD are 4.8 $e^-$ and 1.22 $e^-$/ADU
        respectively. 
        The broad-band $UBV{(RI)}_c$ observations of Wat01 and S208  were  standardized by observing standard stars in the SA98 Landolt field \citep{1992AJ....104..340L} centered at
        $\alpha_{J2000}$: 06$^h$52$^m$12$^s$, $\delta_{J2000}$:
        -00$^\circ$19$^\prime$17$^\prime$$^\prime$.
        Many bias and flat frames were also taken during observations. Several short and deep (long) exposure frames were taken to observe both bright and faint stars in the field.
        The complete log of the observations is given in Table~\ref{log}.

        \begin{table}[htbp]
         \centering
        \caption{Log of optical observations with the 2m HCT, Hanle.}
        \label{log}
        \renewcommand{\arraystretch}{1.0}
        \begin{tabular}{@{}lll@{}}
        \toprule
        
        \textbf{Target/Date} & \textbf{Filter} & \textbf{Exposure $\times$ Frames (s)} \\
        \midrule

        \textbf{SA98} & & \\
       
        \textit{05 Nov 2005} & $U$ & $50\times4$ \\[-0.2em]
        
        & $B$ & $8\times2$, $10\times3$ \\[-0.2em]
       
        & $V$ & $3\times2$, $5\times4$ \\[-0.2em]
        
        & $R_c$ & $2\times2$, $3\times3$ \\[-0.2em]
        
        & $I_c$ & $2\times2$, $4\times3$, $5\times1$ \\
        

        \textbf{Waterloo 01} & & \\
       
        \textit{05 Nov 2005} & $U$ & $60\times1$, $120\times1$, $300\times3$ \\[-0.2em]
        
        & $B$ & $60\times1$, $180\times3$ \\[-0.2em]
        
        & $V$ & $10\times2$, $20\times1$, $120\times2$ \\[-0.2em]
       
        & $R_c$ & $2\times1$, $60\times1$, $120\times2$ \\[-0.2em]
       
        & $I_c$ & $2\times1$, $15\times1$, $60\times1$, $120\times1$, $300\times1$ \\
        

        \textbf{S208} & & \\
       
        \textit{05 Nov 2005} & $U$ & $300\times3$ \\[-0.2em]
       
        & $B$ & $30\times2$, $200\times3$ \\[-0.2em]
        
        & $V$ & $10\times1$, $120\times3$ \\[-0.2em]
       
        & $R_c$ & $5\times1$, $60\times3$ \\[-0.2em]
       
        & $I_c$ & $5\times1$, $60\times3$ \\
        
        \bottomrule
    \end{tabular}
    \end{table}

        The basic data reduction, including image cleaning,
        photometry and astrometry were performed using the standard procedure described in \citet{2020MNRAS.498.2309S, 2020ApJ...891...81P, 2020ApJ...896...29K}. 
        Calibration of the instrumental magnitudes to the standard system was done by using the procedures outlined by \citet{1992ASPC...25..297S}.
        The calibration equations derived by the least-squares linear regression are as follows:\\

        \begin{equation}
        \begin{split}
        u&= U + (3.096\pm0.027)-(0.009\pm0.017)(U-B)\\
        &\quad + (0.338\pm0.037)X_U
        \end{split}
        \end{equation}

        \begin{equation}
        \begin{split}
        b&= B + (1.023\pm0.012) -(0.001\pm0.005)(B-V)\\
        &\quad + (0.181\pm0.018)X_B
        \end{split}
        \end{equation}

        \begin{equation}
        \begin{split}
        v&= V + (0.626\pm0.004) -(0.085\pm0.007)(R-I_c)\\
        &\quad + (0.099\pm0.006)X_V
        \end{split}
        \end{equation}

        \begin{equation}
        \begin{split}
        r_c&= R_c + (0.563\pm0.009) -(0.003\pm0.004)(V-R_c)\\
        &\quad + (0.045\pm0.013)X_R
        \end{split}
        \end{equation}

        \begin{equation}
        \begin{split}
        i_c&= I_c + (0.831\pm0.006) -(0.065\pm0.009)(R-I_c)\\
        &\quad + (0.037\pm0.008)X_I 
        \end{split}
        \end {equation}

	    \noindent
        where $U,B,V,R_c$ and $I_c$ are the standard magnitudes and $u,b,v,r_c$ and $i_c$ are the instrumental aperture magnitudes,
        which are normalized per second of exposure time and $X_U$, $X_B$, $X_V$, $X_R$ and $X_I$ are the air-mass in respective filters.
        
      We have carried out a comparison of our calibrated $B, V$ magnitudes with the archival APASS\footnote{The AAVSO Photometric All-Sky Survey \citep{2015AAS...22533616H}, \url{https://www.aavso.org/apass}} photometric data and found that the difference is negligible and comparable to the photometric errors. 
       A total of 955 and 721 stars were identified with photometric error $<0.1$ mag in the $\sim10^\prime\times10^\prime$ FOV around  Wat01 and S208, respectively. The  $V$ and $I_{c}$ band detection limits (photometric error $<0.1$ mag) for our observations are found to be $\sim$22 mag and $\sim$20 mag, respectively.
      The astrometry of stars was done by using the Graphical Astronomy and Image Analysis (Gaia) 
       Tool\footnote{\url{http://star-www.dur.ac.uk/pdraper/gaia/gaia.html}} with rms noise of the order of $\sim0^{\prime\prime}.3$.

        \subsection{Completeness of the Photometric Data}\label{comp}
        Photometric data may be incomplete due to several factors, including nebulosity, stellar crowding, and instrument detection limits. Determining the mass corresponding to the photometric completeness limit is particularly important for deriving a reliable mass function (MF) slope. We estimated the completeness factor (CF) using artificial star experiments performed with the \texttt{ADDSTAR} task in \texttt{DAOPHOT II}, following the procedures described by \citet{1991A&A...250..324S,2008AJ....135.1934S, 2017ApJ...836...98J,2020MNRAS.498.2309S,2020ApJ...896...29K, 2023JApA...44...66K}. The optical data are  80\% complete down to $V\sim20.5$ mag, whereas the combined 2MASS and UKIDSS near-infrared (NIR) catalog is approximately 80\% complete down to $J\sim18.5$ mag, corresponding to a mass completeness limit of 0.7 M$_\odot$  for an adopted distance of 3.53 kpc and reddening, $E(B-V)=0.83$ mag. We therefore used NIR data for the MF analysis (for details, refer to Figure~\ref{masf} and Section~\ref{iso}).

        \subsection{Molecular Line Data}\label{mo}

         The Wat01 and S208 regions were observed in the $^{12}$CO ($J=1-0$) and $^{13}$CO ($J=1-0$) emission using the Five College Radio Astronomy Observatory (FCRAO) 14 m telescope in New Salem, Massachusetts, as part of the Extended Outer Galaxy Survey \citep[E--OGS;][]{2004ASPC..317...79B,1998ApJS..115..241H}, which extends the coverage of the FCRAO Outer Galaxy Survey \citep[OGS;][]{1998ApJS..115..241H} to Galactic longitude, $l=193^\circ$, over a Galactic latitude range of $-3^{\circ}$.5$\leq b \leq+5^{\circ}.5$. The FCRAO beam sizes are 45$^{\prime\prime}$ and 46$^{\prime\prime}$ for $^{12}$CO and $^{13}$CO, respectively. Both line data sets have a velocity resolution of 0.25 km\,s$^{-1}$. 
         Typical rms values for the spectra are 0.25 K for $^{12}$CO and 0.2 K for $^{13}$CO \citep{1996ApJ...463..630H}. This study focuses on emissions exceeding the $3\sigma$ threshold, where $\sigma$ denotes rms noise.

\begin{table*}[ht!]
    \centering
    \scriptsize
    \caption{List of surveys adopted in the present work 
    (from optical to radio wavelengths).}
    \label{surveys}
    \begin{tabular}{@{}lccl@{}}
    \toprule
   
    \textbf{Survey} & \textbf{Wavelength(s)} &
    \textbf{Resolution} & \textbf{References} \\
    \midrule

    Gaia DR3\textsuperscript{1}
    (magnitudes, parallax, and proper motion)
    & 330--1050\,nm & 0.4 mas & \citet{GaiaDR3_2023} \\

    2MASS\textsuperscript{2}
    & 1.25--2.17\,$\mu$m
    & $\sim2\farcs5$
    & \citet{2006AJ....131.1163S} \\

    UKIDSS GPS\textsuperscript{3}
    & 1.25--2.22\,$\mu$m
    & $\sim0\farcs8$
    & \citet{2007MNRAS.379.1599L} \\

    \textit{Spitzer} SEIP\textsuperscript{4}
    & 3.6, 4.5, 5.8, 8\,$\mu$m
    & $\sim2''$
    & \citet{https://doi.org/10.26131/irsa3}$^a$\\

    \textit{Spitzer} GLIMPSE360\textsuperscript{5}
    & 3.6, 4.5\,$\mu$m
    & $\sim2''$
    & \citet{2003PASP..115..953B} \\

    WISE\textsuperscript{6}
    & 3.4, 4.6, 12, 22\,$\mu$m
    & $\sim6\farcs1$--$\sim12''$
    & \citet{2010AJ....140.1868W} \\

    \textit{Herschel} Hi-GAL\textsuperscript{7}
    & 70, 160\,$\mu$m
    & $\sim5\farcs8$, $\sim12''$
    & \citet{2010PASP..122..314M} \\

    FCRAO OGS\textsuperscript{8}
    & 2.721, 2.603\,mm
    & $\sim50''$
    & \citet{1998ApJS..115..241H} \\

    NVSS\textsuperscript{9}
    & 21\,cm
    & $\sim45''$
    & \citet{1998AJ....115.1693C} \\
    \bottomrule
    \end{tabular}

    \vspace{0.2 cm}

    \parbox{\textwidth}{\scriptsize
    \textsuperscript{1}\url{https://www.cosmos.esa.int/web/Gaia/dr3};
    \textsuperscript{2}Two Micron All Sky Survey,
    \citet{https://doi.org/10.26131/irsa2};
    \textsuperscript{3}UKIRT Infrared Deep Sky Survey,
    \url{http://wsa.roe.ac.uk/};
    \textsuperscript{4}\textit{Spitzer} Enhanced Imaging Products,
    \citet{https://doi.org/10.26131/irsa3};
    \textsuperscript{5}Galactic Legacy Infrared Midplane Survey
    Extraordinaire, \citet{https://doi.org/10.26131/irsa214};
    \textsuperscript{6}\citet{https://doi.org/10.26131/irsa1};
    \textsuperscript{7}\url{http://archives.esac.esa.int/hsa/whsa/};
    \textsuperscript{8}See Section~\ref{mo};
    \textsuperscript{9}\url{https://www.cv.nrao.edu/nvss/postage.shtml};
    $^a$: \url{https://irsa.ipac.caltech.edu/data/SPITZER/Enhanced/SEIP/overview.html}.
    }
    \end{table*}

        \subsection{Archival data sets}

         Along with the observed optical data, we have also used archival datasets spanning optical to radio wavelengths to analyze the target regions.
         A concise overview of these data sets is provided in Table~\ref{surveys}.

\section{Results and Analysis}\label{s4}

\subsection{Physical State of the Clusters in the Region}\label{iso}

Wat01 and BDS64 (associated with S208) are young clusters located in regions containing significant gas and dust emission; therefore, their stellar populations are effectively traced at near-infrared (NIR) wavelengths. We used point-source catalogs from 2MASS and UKIDSS to investigate the stellar density distribution \citep[see also][]{2023JApA...44...66K}. Stellar surface-density maps were constructed using the nearest-neighbor (NN) method following \citet{2005ApJ...632..397G}. At each grid position, the distance to the $20^{\rm th}$ nearest star was determined using a grid spacing of $6^{\prime\prime}$, and the local stellar surface density was calculated as

\begin{equation}
\sigma(i,j) = \frac{N}{\pi r_N^2(i,j)},
\end{equation}

where $r_N(i,j)$ is the projected distance from the grid position $(i,j)$ to the $N^{\rm th}$ nearest star \citep{2009ApJS..184...18G}.
The resulting stellar-density contours are shown as cyan curves in Figure~\ref{rgb32}. The lowest contour corresponds to $3\sigma$ above the mean stellar density of 20.5 stars arcmin$^{-2}$, with subsequent contours separated by $1\sigma$ (3.6 stars arcmin$^{-2}$). Both regions show well-defined stellar density distributions, with elongated outer structures and more compact inner regions enclosed within approximately half of the peak density.

\begin{table*}[ht!]
        \centering
        \scriptsize
        \caption{\label{param} Structural parameters of Wat01 and BDS64 regions.}
        \begin{tabular}{@{}lcccccccccc@{}}
        \toprule
       
        \textbf{Region} & \textbf{Center RA} & \textbf{Center Dec} & \textbf{Peak Stellar Density} & \textbf{n$_{total}$}\textsuperscript{a} & \textbf{n$_{hull}$}\textsuperscript{b} & \textbf{A$_{hull}$}\textsuperscript{c} & \textbf{A$_{cluster}$}\textsuperscript{d} & \textbf{R$_{cluster}$}\textsuperscript{e} & \textbf{R$_{circ}$}\textsuperscript{f} & 
        \textbf{Aspect ratio} \\
       
        &  &  & (stars/pc$^{-2}$) &  &  & (pc)$^2$ & (pc)$^2$ & (pc) & (pc) & \\
        \midrule
        
        Wat01 & 04$^h$18$^m$38.73$^s$ & +52$^\circ$51$'$40.24$''$  &  $\sim$70 & 209 & 13 & 6.89  & 7.35  & 1.53 & 2.14 & 1.96 \\
       
         BDS64  & 04$^h$19$^m$19.76$^s$  & +52$^\circ$58$'$50$''$     &  $\sim$70 & 69  & 9  & 2.15  & 2.47  & 0.89 & 1.28 & 2.07 \\
        \bottomrule
        \end{tabular}

        \vspace{0.2 cm}

        \parbox{\textwidth}{\scriptsize
        \textsuperscript{a}Total number of stars within the convex hull;
        \textsuperscript{b}Number of vertices in the convex hull;
        \textsuperscript{c}Area of convex hull;
        \textsuperscript{d}Estimated cluster area;
        \textsuperscript{e}Estimated cluster radius;
        \textsuperscript{f}Radius of the smallest encompassing circle.}
    
        \end{table*}

Since the clusters exhibit distinctly elongated morphologies, we avoid assuming circular boundaries when estimating their sizes. Instead, we define the cluster area using the convex hull of the member stars enclosed within the lowest-density contour (magenta curves in Figure~\ref{rgb32}). The convex hull provides the minimum convex polygon enclosing the stellar distribution and is therefore better suited to irregular and elongated clusters than a circular approximation \citep{2006A&A...449..151S,2020MNRAS.498.2309S}. The cluster area, $A_{\rm cluster}$, is then estimated from the convex-hull area, $A_{\rm hull}$, after applying the appropriate geometrical correction :

        \begin{equation}
             A_{cluster} = {\frac{A_{hull}}{1-\frac{n_{hull}}{h_{total}}}}
         \end{equation}

        where $n_{hull}$ is the number of vertices in the convex hull and $n_{total}$ is the total number of data points (stars) within the convex hull. 
        The estimated values for $A_{hull}$, $A_{cluster}$, $R_{cluster}$, and $R_{circ}$ are listed in Table~\ref{param} for each region. The aspect ratio ${R^2_{circ}}/{R^2_{cluster}}$ of each cluster is also listed in the same table. 
        The cluster radius, $R_{cluster}$, is defined as the radius of the circle with the same area as $A_{cluster}$. 
        Whereas the circular radial size, $R_{circ}$, is the radius of the minimum area circle that encompasses the entire grouping, and it is defined as half of the largest distance between any two data points \citep[see details in][]{2006A&A...449..151S,2020MNRAS.498.2309S}. 

        \begin{figure*}[ht!]
        \centering
        \includegraphics[width=0.45\textwidth]{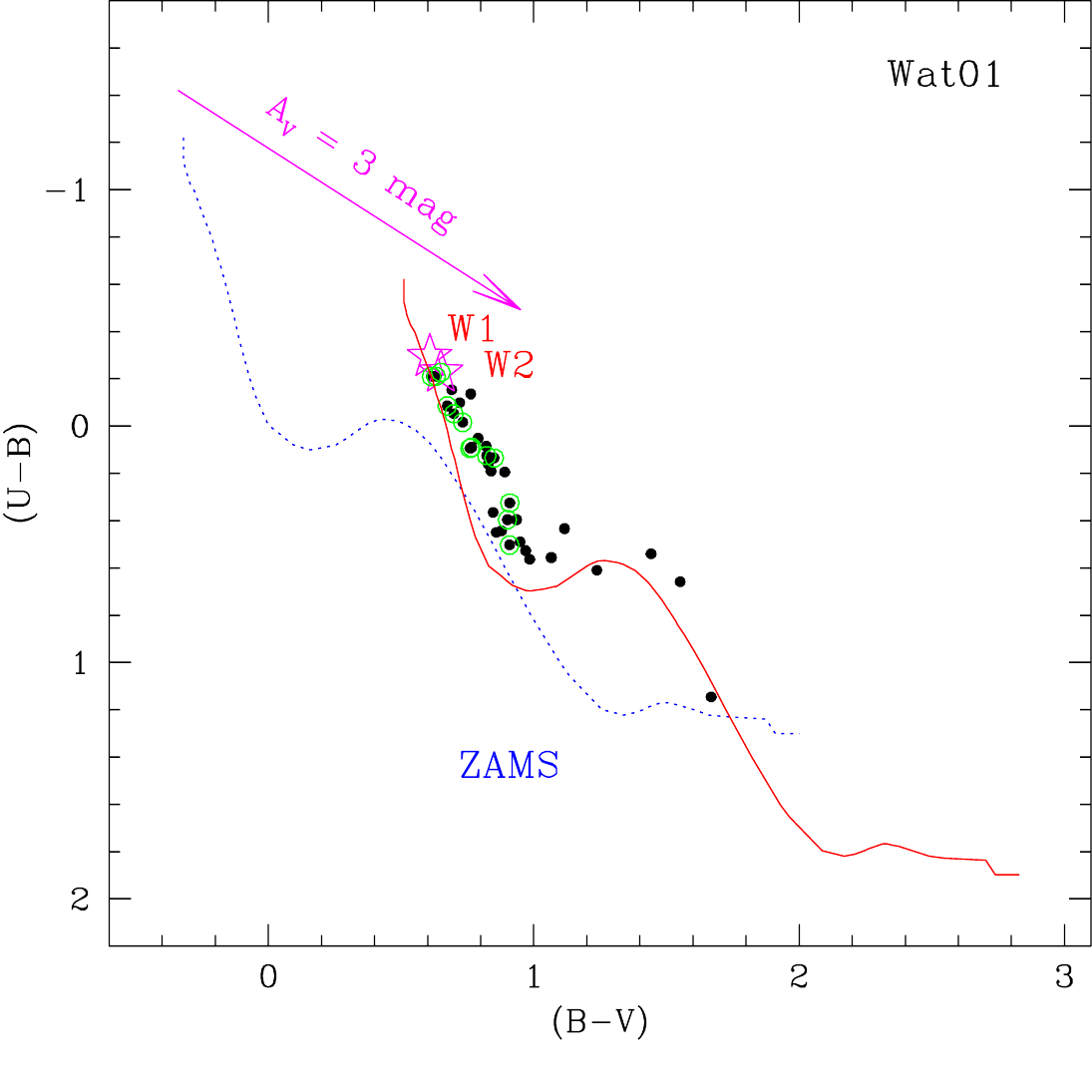}
        \includegraphics[width=0.45\textwidth]{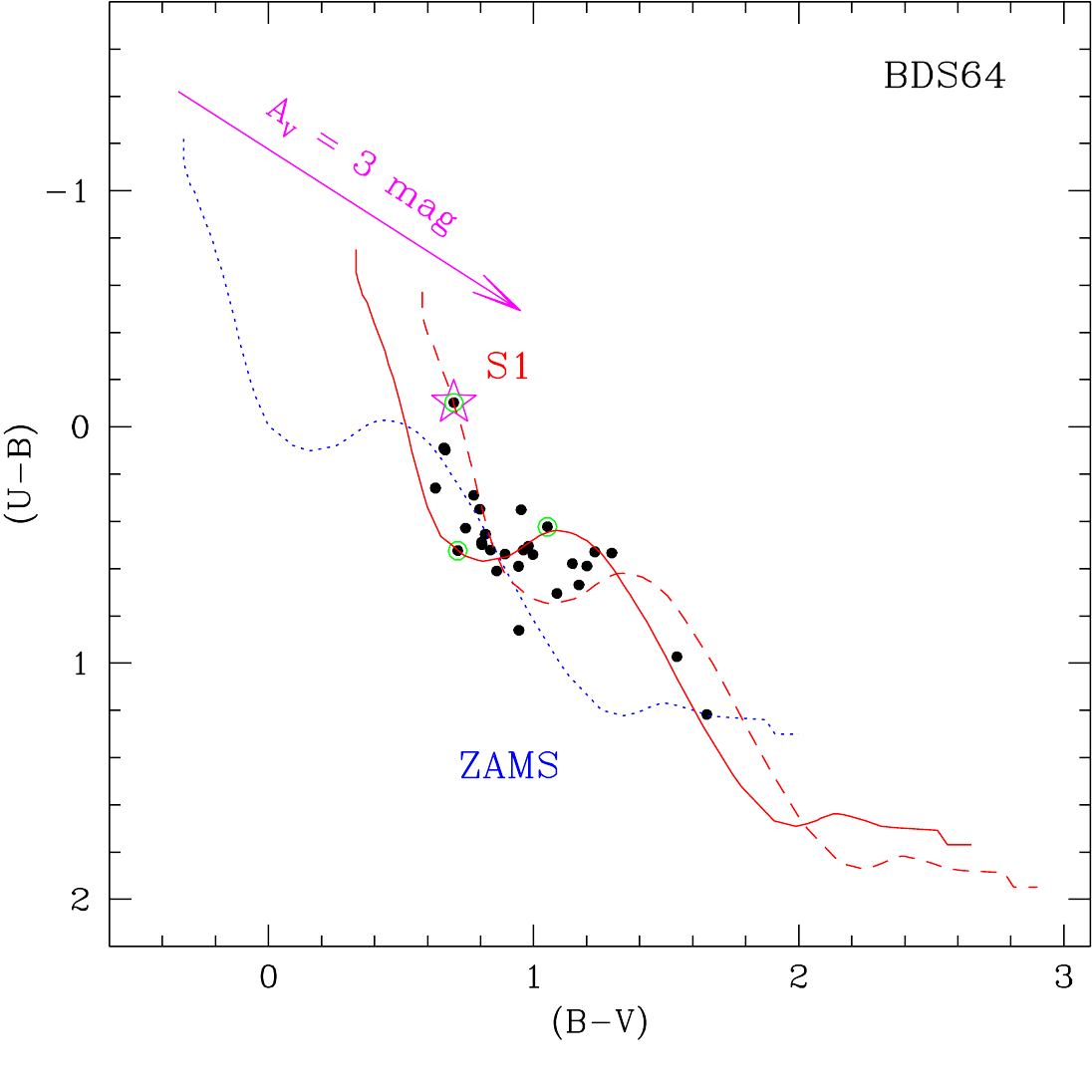}
        \caption{\label{cc} The $(U-B)$ vs.$(B-V)$ TCDs for stellar sources within hull boundaries of Wat01  and BDS64 (black dots). {\it Left panel}: The blue curve denotes the intrinsic Zero Age Main Sequence (ZAMS) locus (for $Z = 0.02$) from \cite{2013ApJS..208....9P}. The red solid curve shows this locus shifted along the reddening vector for the extinction of $E(B-V)=0.83$ mag.  Cluster members (P$_{\mu}>$80\%; for details see Appendix~\ref{mp}) are highlighted by green open circles, and magenta asterisks mark massive OB-type stars. {\it Right panel}: Same as the left panel, but for BDS64. The 
        solid and dashed red curves represent the intrinsic ZAMS reddened by $E(B-V)$=0.65 and 0.90 mag, respectively. The former corresponds to the adopted foreground extinction for the BDS64 region, while the latter represents the extinction towards the massive star ‘S1’.}
        \end{figure*}

    The color-magnitude diagrams (CMDs) and two-color diagrams (TCDs) of cluster's member stars can be utilized to accurately determine the distance, reddening (extinction), and age of the cluster \citep[see,][]{1994ApJS...90...31P, 2006AJ....132.1669S, 2020MNRAS.498.2309S, 2020ApJ...896...29K,2022ApJ...926...25P,2023JApA...44...66K, 2024AJ....167..106S, 2024AJ....168...98V}. 
    To constrain these parameters, we first investigate the extinction law toward Wat01 and BDS64 using the total-to-selective extinction ratio, $R_V = A_V/E(B-V)$, following the method described by  \citet{2003A&A...397..191P}.
    We employed ($V-\lambda$)\begingroup
    \renewcommand{\thefootnote}{\fnsymbol{footnote}}\footnote{$\lambda$ represents one of the wavelengths of the broadband filters \citep[($J, H, K, I_c, R, L $);][]{2003A&A...397..191P}.} \endgroup versus ($B-V$) TCDs to distinguish between the effect of normal extinction caused by the general ISM and the anomalous extinction occurring within the regions \citep{2003A&A...397..191P, 2020ApJ...896...29K}.  For TCDs, we consider all member stars within a 10$^\prime\times$10$^\prime$ FOV (P$_{\mu}>80\%$, see Appendix~\ref{mp}) of the Wat01 and BDS64 regions (see Figure~\ref{rgb32}), with optical and NIR observations, excluding young stellar objects (YSOs; see details in Appendix~\ref{yso}). 
    The slopes of the least-squares fit to the distribution of stars in the ($V-I_c$), ($V-J$), ($V-H$), and ($V-K$) versus ($B-V$) TCDs are found to be $0.99\pm0.23$, $1.74\pm0.45$, $2.21\pm0.33$, and $2.32\pm0.42$ for Wat01 and $1.11\pm0.21$, $1.88\pm0.53$, $2.40\pm0.41$, and $2.59\pm0.44$ for BDS64, respectively. 
    These values are in good agreement with those found for the general ISM \citep[1.10,1.96,2.42 and 2.60; see][]{2003A&A...397..191P}. Hence, we consider the normal reddening value (R$_V = 3.1$) for the Wat01 and BDS64 regions.

      \begin{figure}[ht!]
    \centering
    \includegraphics[width=0.48\textwidth]{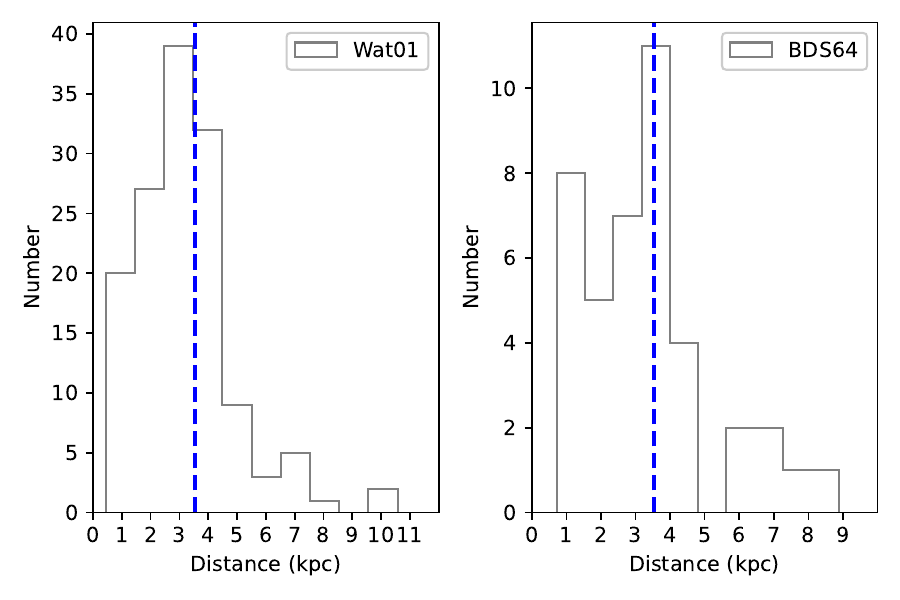}
    \caption{\label{hist} {Histogram showing the distribution of distances of stars inside the convex hull of Wat01 (left) and BDS64 (right) clusters. The Blue dotted line corresponds to the distance of 3.53 kpc, which is the estimated distance of both clusters in this study.}}
    \end{figure}

    The extinction toward both clusters was estimated using ($U-B$) versus ($B-V$) TCDs, as shown in Figure~\ref{cc}—the left panel for Wat01 and the right panel for BDS64. We shifted the intrinsic zero-age main sequence (ZAMS; blue dotted curve) locus from \citet{2013ApJS..208....9P} along the reddening vector\footnote{We adopt a reddening slope of $E(U-B)/E(B-V) = 0.72$, corresponding to $R_V \sim 3.1$, which is typical for the diffuse interstellar medium.} to align with the least-reddened member stars. The shifted ZAMS locus is shown as a continuous red curve. Using this approach, we estimate minimum color excesses of $E(B-V) = 0.83$ mag for Wat01 and $E(B-V) = 0.65$ mag for BDS64, corresponding to extinctions of $A_V=2.6$ and $2.0$ mag, respectively. The massive star in BDS64 exhibits a higher reddening of $E(B-V)=0.90$ mag, corresponding to $A_V=2.8$ mag, as indicated by the dashed red curve. The estimated uncertainty in the reddening is approximately 0.1 mag, following the method described in \citet{1994ApJS...90...31P}.
    We have performed a membership analysis using recent \textit{Gaia} DR3 data (see Appendix~\ref{mp}) and adopted the mean distance of member stars from \citet{2021AJ....161..147B} with parallax uncertainties $\leq 0.05$ mas. Based on this analysis, the distance to the Wat01 cluster is estimated to be $3.53 \pm 0.55$ kpc
    \citep[see also,][]{2023A&A...673A.114H}

    \begin{figure*}
    \centering
    \includegraphics[height=0.4\textheight]{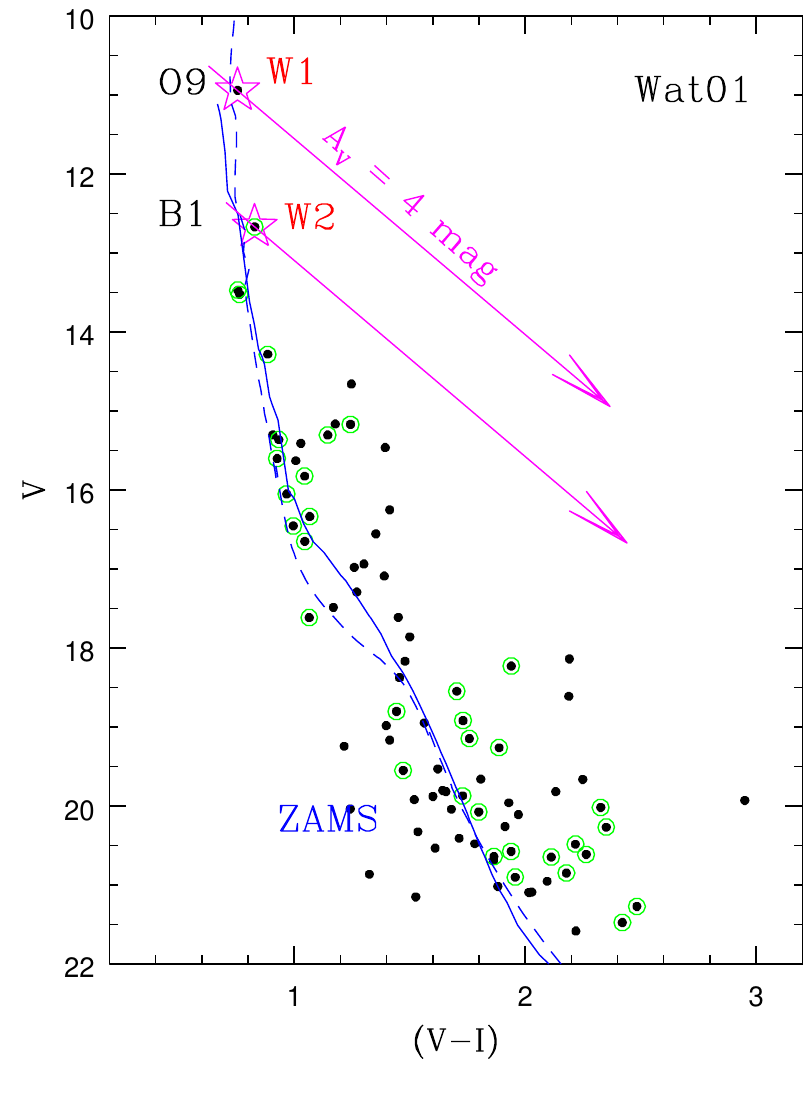}
    \includegraphics[height=0.4\textheight]{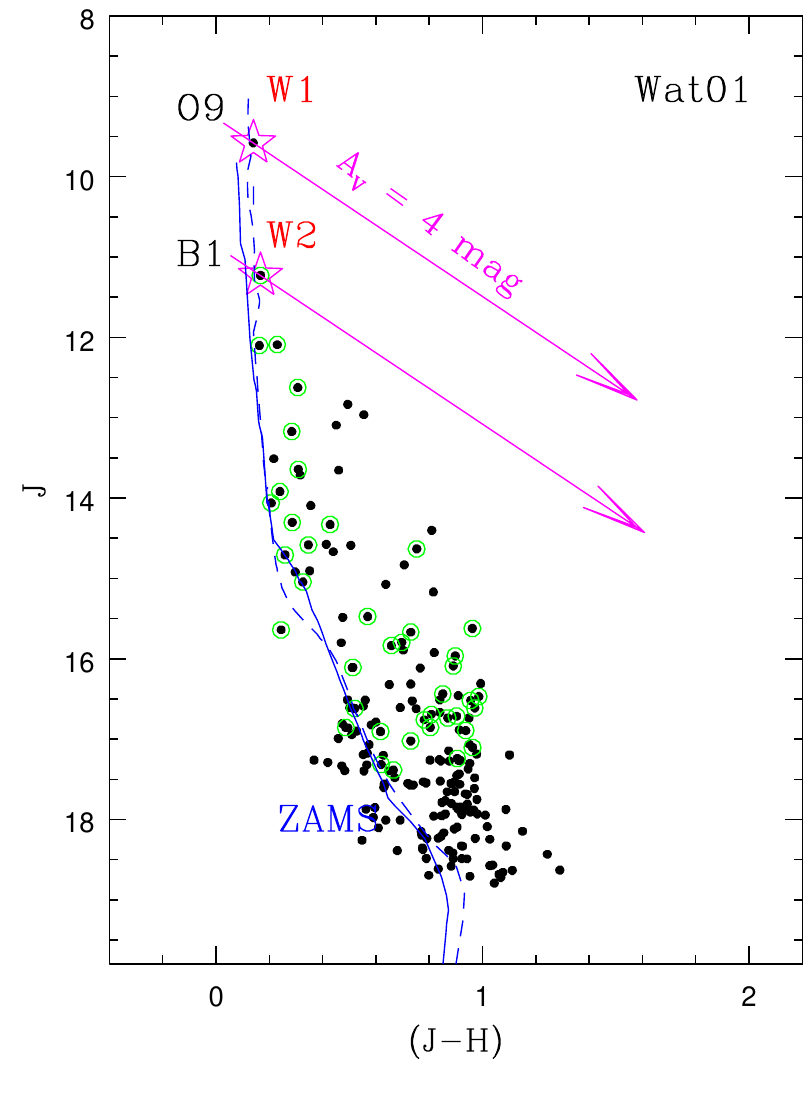}
    \includegraphics[height=0.4\textheight]{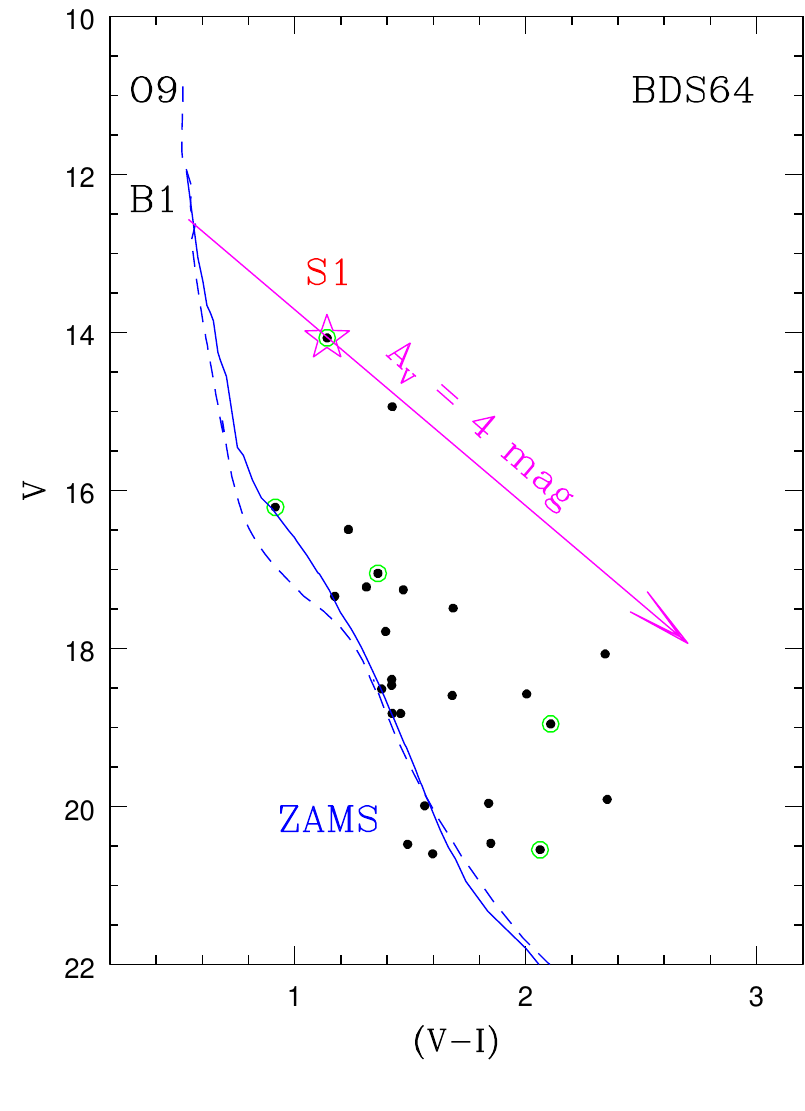}
    \includegraphics[height=0.4\textheight]{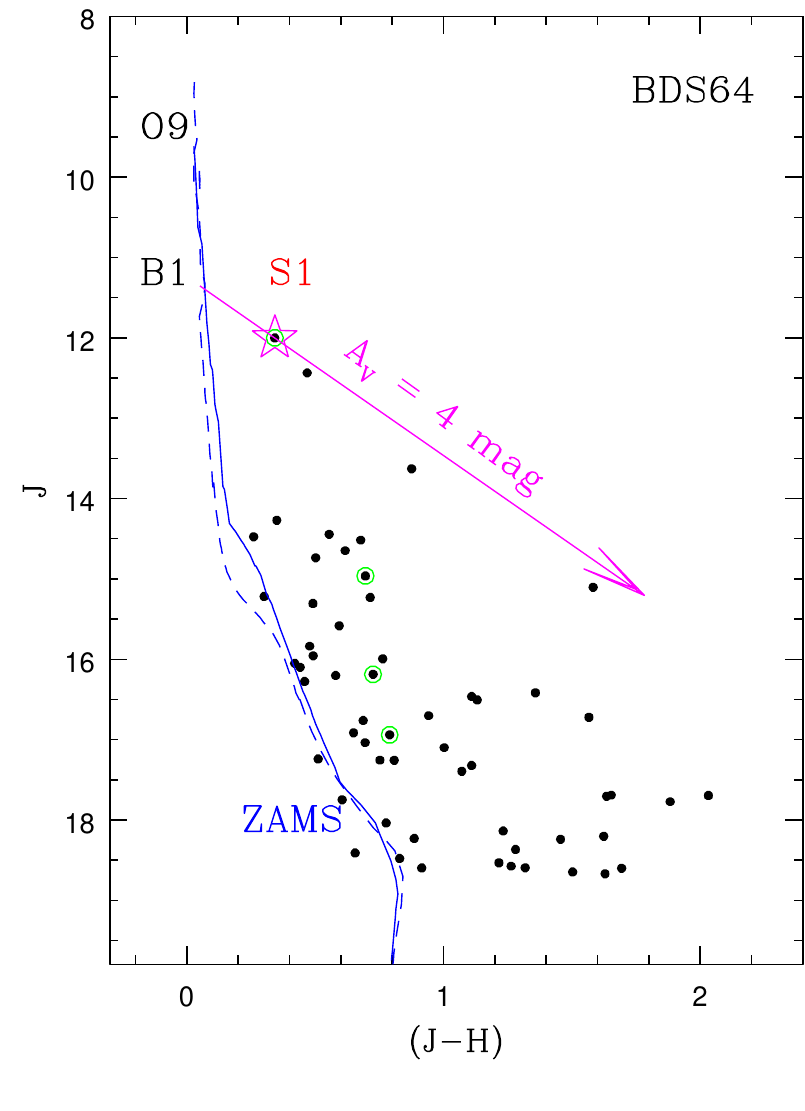}
    \caption{\label{cmd} Optical ($V$ vs. $(V-I)$)  and NIR ($J$ vs. $(J-H)$) color-magnitude diagrams for Wat01 and BDS64 cluster regions. Continuous and dotted blue curves represent the Zero Age Main Sequence (ZAMS) of \citet{2019MNRAS.485.5666P} and \citet{2013ApJS..208....9P}, respectively; all curves are corrected for the distance of $3.53$ kpc and extinction of $A_V=2.6$ mag for Wat01 and $A_V=2.0$ mag for BDS64 regions. Green open circles are the identified member stars from proper motion analysis (see details in Appendix~\ref{mp}), whereas magenta asterisk symbols are the location of massive OB-type stars in the respective clusters (see Section~\ref{intro}). Reddening vectors along the massive stars are represented by magenta arrows for $A_{V}=4$ mag in each cluster region.}
    \end{figure*}

In contrast to Wat01, BDS64 exhibits weaker stellar clustering and contains only a few optically detected member stars, making its distance determination from CMD analysis uncertain. Previous studies have reported distances ranging from $\sim3.9$ to $\sim10$ kpc \citep{1985A&AS...61..537L,2015ApJS..221...26A}. We therefore examine the distance distribution of all stars within the convex-hull regions of both clusters. As shown in Figure~\ref{hist}, the distance distributions of Wat01 and BDS64 peak at $\sim3.5$ kpc, suggesting that the two regions are likely located at a similar distance. We consequently adopt a common distance of $3.53\pm0.55$ kpc for both regions throughout this study.

    \begin{figure}[ht!]
        \centering
        \includegraphics[width=0.48\textwidth]{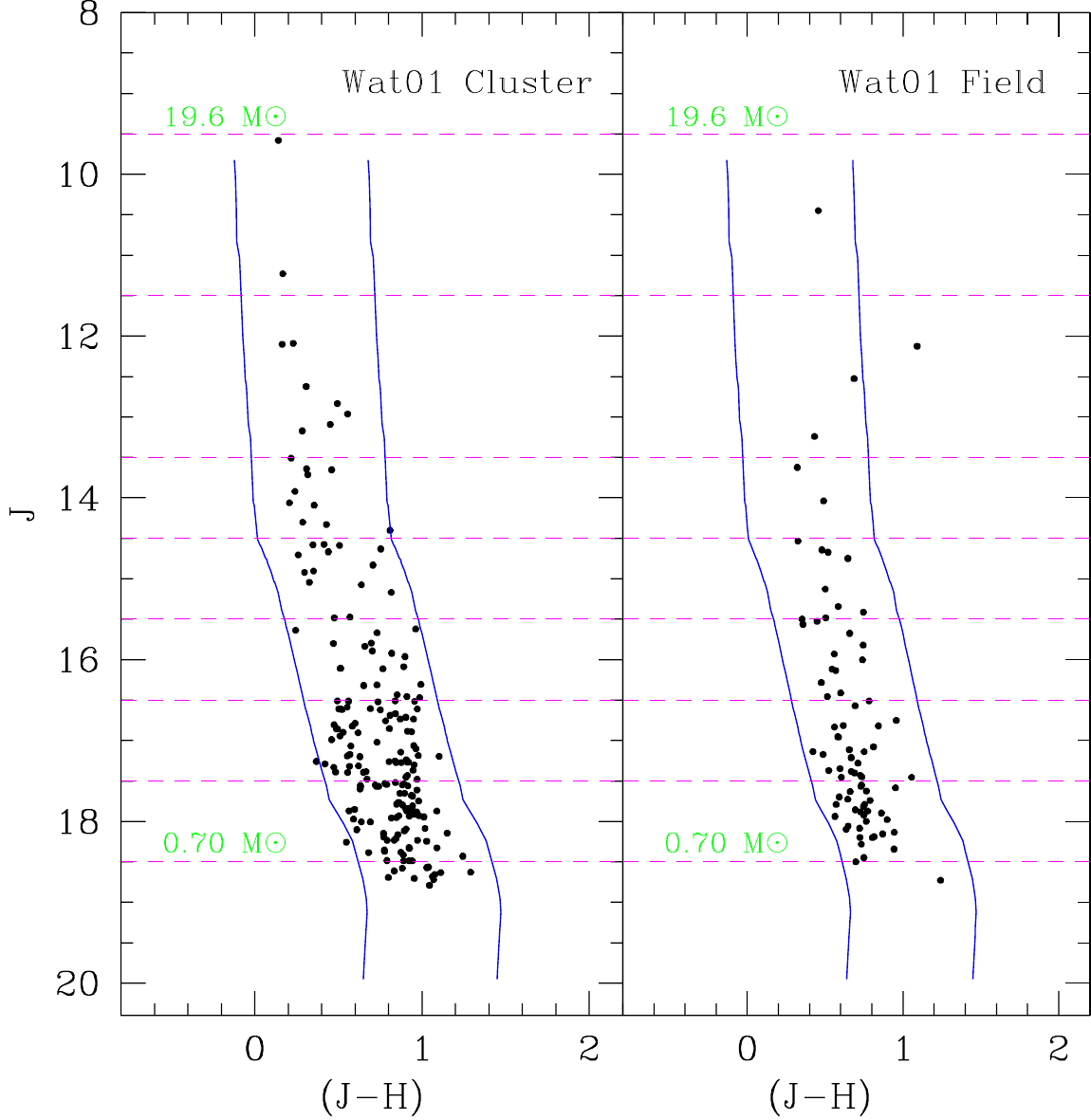}
         \includegraphics[width=0.48\textwidth]{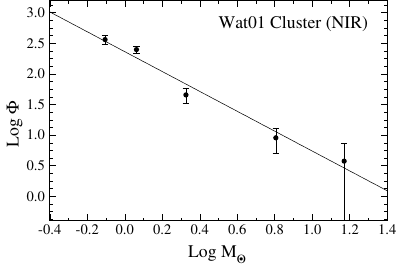}
        \caption{\label{masf} {\it Upper panel}: $J$ versus $(J-H) $ CMD for (a) stars within the cluster region, (b) stars within the field region of the same area as the cluster.
          The curves represents envelop of +0.6 mag (right curve; $A_{V} = 4.62$ mag ) and -0.2 mag (left curve; $A_{V} = 1.33$ mag) on the CMD from the isochrone of \cite{2013ApJS..208....9P}. 
          {\it Lower panel}: A plot of the mass function (MF) for the Wat01 region using the NIR (2MASS+UKIDSS) data. Log$\phi$ represents log(N/dlog m), and the error bars represent standard $\pm\sqrt{N}$ errors. The solid line shows a least-squares fit to the MF distribution.}
    \end{figure}
         
        To further validate the estimated distance, we applied the isochrone fitting method to both regions. For this, we used $V$ vs ($V-I_c$) and $J$ vs ($J-H$) CMDs, as shown in Figure~\ref{cmd}. 
        In these diagrams, stars inside the convex hull, probable members, and massive stars are represented by black dots, green circles, and magenta asterisks, respectively. The intrinsic ZAMS\footnote{To show fitting on the massive mass limit ($\sim$19.6$\odot$), we used two different isochrones in CMD fitting (see Figure~\ref{cmd}).}, 
        from  \citet[][blue dotted curves]{2019MNRAS.485.5666P}  and \citet[][blue solid curves]{2013ApJS..208....9P}  were adopted, corrected for extinction values of $A_V$=2.6 mag and $A_V$=2.0 mag for BDS64 and Wat01, respectively. For both regions, we have adopted a distance of 3.53 kpc.
         Isochrone fitting method \citep[see details in][]{2020ApJ...896...29K,2020MNRAS.498.2309S} confirms our estimated distance and extinction, as the ZAMS is closely aligned with the blue envelope of the stellar distribution in both CMDs. 
        In the Wat01 region, the probable massive stars `W1' and `W2' have estimated masses of 19.6 M$_\odot$ and 10 M$_\odot$ \citep{2013ApJS..208....9P}, corresponding to spectral types O9V and B1.5V, respectively. As `W1' is the brightest star in the Wat 01 region, the upper age limit of the cluster is estimated to be $\sim8.1$ Myr \citep{2004fost.book.....S}. Meanwhile, in the BDS64 region, the massive star `S1' has an estimated mass of 10 M$_\odot$ and corresponds to a spectral type of B1.5/B2V, with an upper age limit of $\sim12.6$ Myr \citep{2004fost.book.....S}.

        We have also used the deep NIR data from 2MASS and UKIDSS to derive MF slopes for the Wat01 cluster, extending to fainter magnitudes in a gas- and dust-rich environment. The MF is often expressed as a power law $N (\log m) \propto m^{\Gamma}$, where $\Gamma$ represents the slope of the distribution, and it's given as:
             
        \begin{equation}
         \Gamma = {\frac{d \log N (\log m)}{d \log m}}
        \end{equation}
        \noindent
        Here, $N (\log m)$ is the number of stars per unit logarithmic mass interval. 
        We constructed the NIR ($J$ vs. $(J-H)$) CMD of stars located within the convex hull of the Wat01 cluster and a nearby field region ($\alpha_{J2000}$: 04$^h$18$^m$26.6178$^s$, $\delta_{J2000}$: +52$^\circ$48$^\prime$26.272$^\prime$$^\prime$) of equal area (see Figure~\ref{masf} upper panel). 
        To minimize field star contamination, we defined an envelope around the cluster sequence of $+0.6$ mag and $-0.2$ mag in color (see Figure~\ref{masf}) and derived the cluster luminosity function (LF) by statistically subtracting field stars from cluster stars in different magnitude bins. Both the cluster and field star counts were corrected for data incompleteness (see Section~\ref{comp}).
        The completeness-corrected LF was then converted into an MF using the isochrones from \citet{2013ApJS..208....9P}, corrected for distance and extinction. The resulting MF distribution for the Wat01 cluster is shown in Figure~\ref{masf} (right panel), yielding a slope of $-1.62\pm0.15$ over the mass range $\sim0.70<M/M_{\odot}<19.6$.

         \begin{figure*}[ht!]
        \centering
        \includegraphics[width=1.0\textwidth]{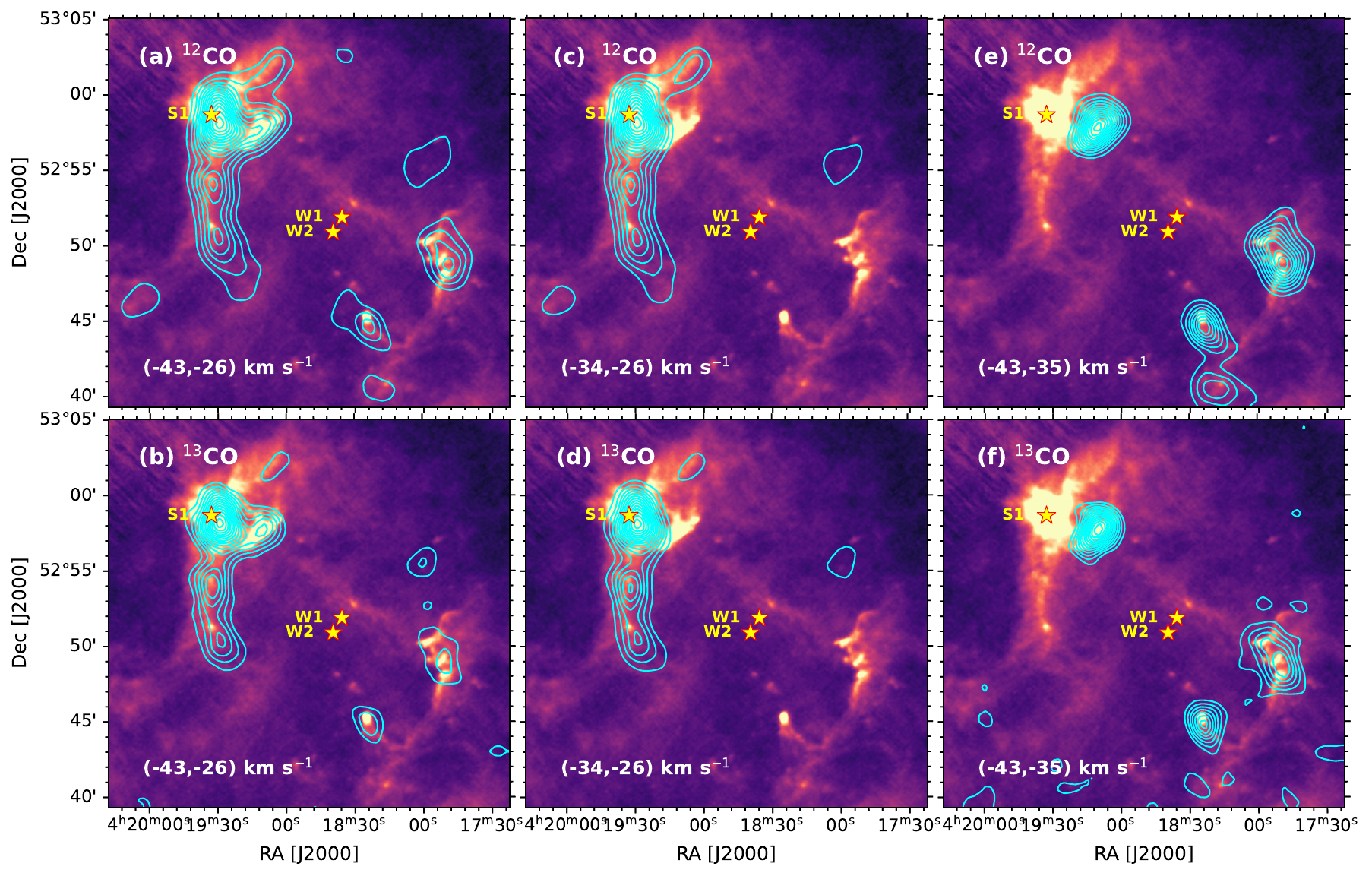}
        \caption{\textit{Herschel} 160 $\mu$m images overlaid with $^{12}$CO($J=1-0$) and $^{13}$CO($J=1-0$) emission contours at different velocities.}
        \label{hco}
        \end{figure*}

        \begin{figure*}[ht!]
        \centering
        \includegraphics[width=\textwidth]{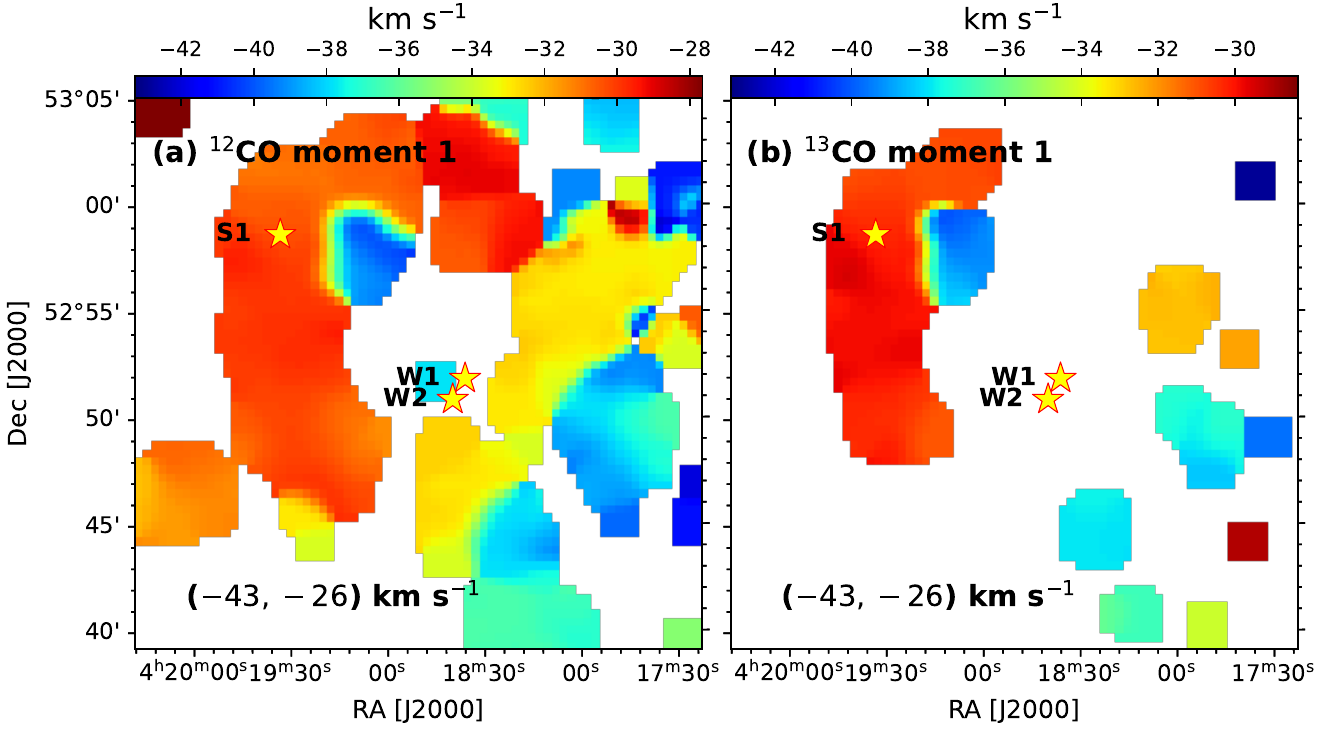}
         \includegraphics[width=\textwidth]{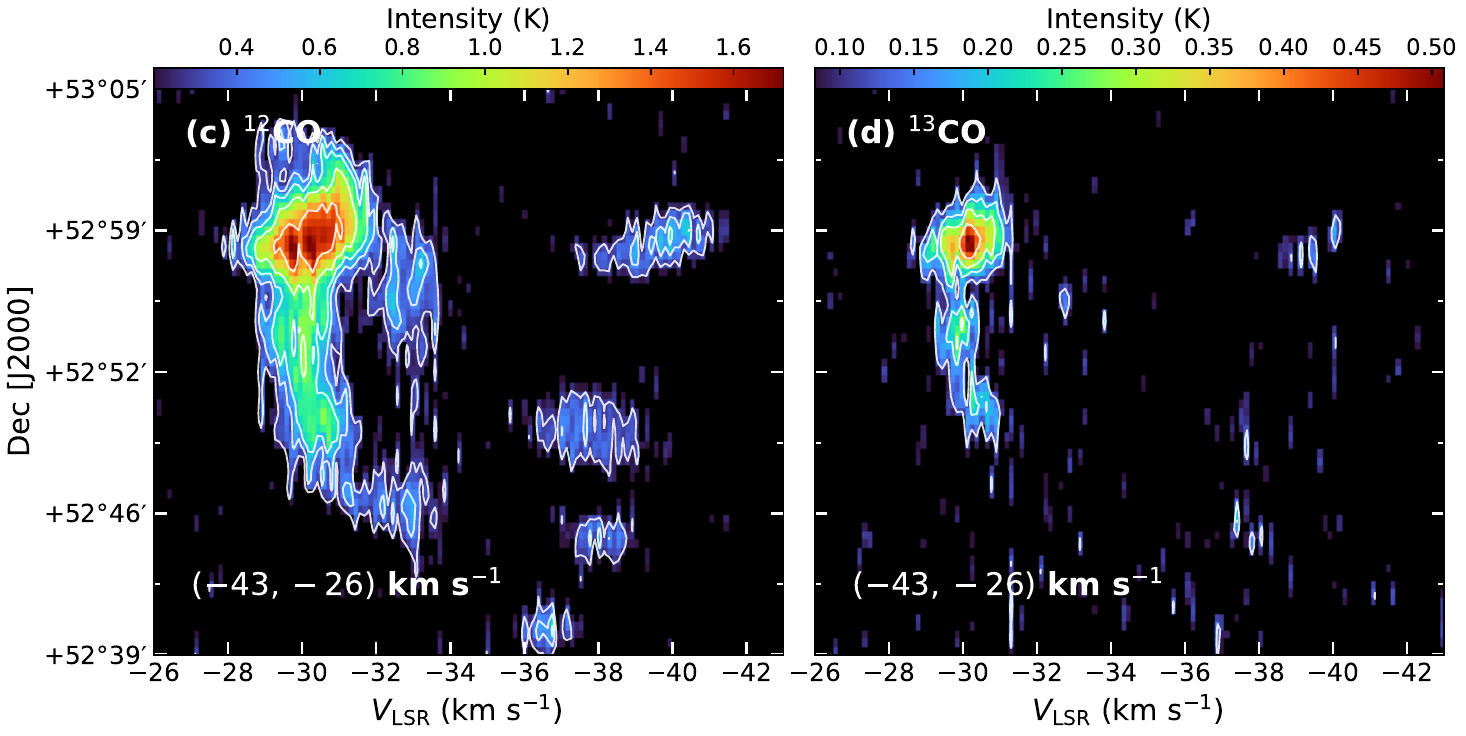} 
        \caption{\label{co} {{\it Panels (a) and (b)} represent the  moment-1 map (Intensity-weighted velocity) collapsed images for the $^{12}$CO($J=1-0$) and  $^{13}$CO($J=1-0$) cubes for a $24^\prime\times24^\prime$ region around the Wat01 and S208 regions. The asterisk symbols mark the location of massive stars in the cluster regions. {\it Panels (c) and (d)} show the declination velocity map for the velocity range [$-43, -26$] km\,s$^{-1}$, extracted from  $^{12}$CO($J=1-0$) and  $^{13}$CO($J=1-0$) data cubes, color-coded with intensity of molecular emission.}}
        \end{figure*}

    \subsection{Molecular gas and  its kinematics in the region}\label{mole}

    To investigate the morphology and kinematics of the molecular gas, we analyzed the molecular-line data cubes (see Section~\ref{mo}) over the velocity range $[-43,-26]$ km\,s$^{-1}$. As shown in Figure~\ref{hco}(a)–(b), the integrated $^{12}$CO and $^{13}$CO emission is concentrated in two distinct regions: (i) a northeastern filamentary structure associated with the massive star `S1', and (ii) a southwestern complex containing compact globule-like condensations oriented toward the massive stars `W1' and `W2'.
    To further examine the velocity structure, we decomposed the CO emission into narrower velocity intervals (see Figure~\ref{hco}(c)–(f)), revealing multiple kinematic components. The northeastern molecular cloud resolves into a linear filament peaking near `S1' and a compact globule coincident with an elongated cold-dust core. In contrast, the southwestern region consists of two compact globule-like condensations spatially associated with dust-emission peaks. These results indicate a clear morphological dichotomy: `S1' remains embedded within an extended filamentary cloud, whereas `W1' and `W2' are associated with compact clump-like condensations.
    In addition, diffuse molecular emission traces a ring- or bubble-like structure surrounding the Wat01 massive stars, with a noticeable deficit of molecular gas toward the central cavity, suggesting gas evacuation by stellar feedback. Conversely, `S1', associated with the S208 region, is located near the peak of the filamentary molecular emission. The distinct spatial and kinematic distributions of the molecular gas support a scenario in which feedback from the massive stars in Wat01 has significantly restructured the surrounding molecular cloud, producing the observed bubble morphology while leaving the S208 filament mostly unaffected.

    To probe the spatial and kinematic structure in more detail, we generated moment-1 (intensity-weighted velocity) maps from the $^{12}$CO($1\text{--}0$) and $^{13}$CO($1\text{--}0$) data cubes by collapsing them over the velocity range [$-43$, $-26$] km\,s$^{-1}$, following standard procedures \citep[see][]{2023ApJ...944..228M, 2024AJ....168...98V, 2025ApJ...990..155V}. The resulting maps, shown in Figure~\ref{co}(a)-(b), reveal a clear velocity segregation: the filamentary cloud associated with S208 (FC) is red-shifted, with peak velocities near $\sim-30$ km\,s$^{-1}$, while the clumpy structures linked to the Wat01 region (GB1–GB3) are blue-shifted around $\sim-38$ km\,s$^{-1}$. This offset of $\sim8$ km\,s$^{-1}$ signifies differential motion between the two components.

    To further explore the kinematic properties of the molecular gas, we constructed position--velocity (PV) diagrams for the $^{12}\mathrm{CO}(1\text{--}0)$ and $^{13}\mathrm{CO}(1\text{--}0)$ transitions, showing the variation in velocity as a function of declination ($\delta$), as presented in Figure~\ref{co}(c)--(d). The $^{12}\mathrm{CO}$ PV diagram reveals a group of prominent, isolated structures centered at $V_{\rm LSR}\sim-38~\mathrm{km\,s^{-1}}$, corresponding to GB1--GB3, together with a distinct redshifted filamentary feature centered at approximately $-30~\mathrm{km\,s^{-1}}$, associated with the filamentary cloud (FC) toward S208. The two components are clearly separated in velocity space, suggesting that they represent dynamically independent molecular structures with no direct physical connection evident in the present data.

    These features are more clearly resolved in the $^{12}\mathrm{CO}$ PV diagram than in the $^{13}\mathrm{CO}$ diagram, most likely because $^{12}\mathrm{CO}$ is more sensitive to extended and relatively diffuse molecular gas. The FC associated with S208 exhibits a velocity extent of approximately $5~\mathrm{km\,s^{-1}}$ around a systemic velocity of $V_{\rm LSR}\sim-30~\mathrm{km\,s^{-1}}$. This velocity spread may reflect turbulent and internal gas motions commonly observed in active star-forming environments.

To examine whether the velocity separation also reflects a line-of-sight distance separation, we estimated kinematic distances using the three methods of \citet{2018ApJ...856...52W} (Table~\ref{tab:kd}). The redshifted component associated with S208 ($V_{\rm LSR}=-30.253\pm0.040~\mathrm{km\,s^{-1}}$) has a Method-C distance of $3.26^{+0.65}_{-0.69}~\mathrm{kpc}$, consistent with the Gaia/CMD distance of $3.53~\mathrm{kpc}$ adopted for Wat01 and BDS64 (Section~\ref{iso}). In contrast, the blueshifted globules GB1--GB3 ($V_{\rm LSR}\sim-37.8$ to $-39.3~\mathrm{km\,s^{-1}}$) have nominal Method-C distances of $\sim4.1$--$4.8~\mathrm{kpc}$. These values suggest that the globules may be background structures projected along the same line of sight. However, their broad distance uncertainties do not conclusively establish a physical separation. Kinematic distances are sensitive to deviations from circular Galactic rotation, including peculiar motions, spiral-arm streaming, cloud-to-cloud motions, and local feedback-driven gas motions \citep{2009ApJ...700..137R, 2014ApJ...783..130R, 2018MNRAS.474.2028R, 2012MNRAS.424..377D}. Thus, the velocity estimates favor a separation between GB1--GB3 and the S208 complex, but cannot confirm it conclusively. No statistically significant $^{13}\mathrm{CO}$ emission is detected at the position of Wat01; hence, no molecular-gas kinematic distance is assigned to this cluster.

    \begin{table*}[ht!]
    \centering
    \scriptsize
    \caption{Systemic velocities and kinematic distances of the molecular components toward S208 and Wat01.}
    \label{tab:kd}
    
    {%
    \begin{tabular}{lccccccc}
    \toprule
    \textbf{Source} & \textbf{Velocity component}
    & \textbf{$l$}
    & \textbf{$b$}
    & \textbf{$V_{\rm LSR}$}
    & \textbf{$d_{\rm A}$}
    & \textbf{$d_{\rm B}$}
    & \textbf{$d_{\rm C}$} \\
    &
    & \textbf{(deg)}
    & \textbf{(deg)}
    & \textbf{(km\,s$^{-1}$)}
    & \textbf{(kpc)}
    & \textbf{(kpc)}
    & \textbf{(kpc)} \\
    \midrule
    S208  & Red  & 151.2873 & 1.9560 & $-30.253 \pm 0.040$ & 4.09 & 3.27 & $3.26^{+0.65}_{-0.69}$ \\
    GB1   & Blue & 151.2562 & 1.9163 & $-39.283 \pm 0.048$ & 5.91 & 4.67 & $4.82^{+0.77}_{-1.13}$ \\
    GB2   & Blue & 151.2201 & 1.6690 & $-37.804 \pm 0.081$ & 5.57 & 4.41 & $4.13^{+1.07}_{-0.70}$ \\
    GB3   & Blue & 151.3218 & 1.6819 & $-37.947 \pm 0.051$ & 5.63 & 4.46 & $4.29^{+0.99}_{-0.72}$ \\
    Wat01 & Blue & 151.2638 & 1.7867 & \multicolumn{4}{c}{No significant $^{13}$CO detection at the cluster position} \\
    \bottomrule
    \end{tabular}
    }

    \vspace{0.2 cm}
    \begin{minipage}{0.98\textwidth}
    \footnotesize
    The quoted $V_{\rm LSR}$ uncertainties combine the formal Gaussian-centroid uncertainty and the finite velocity-channel sampling term. Method A ($d_{\rm A}$) uses the \citet{1993A&A...275...67B} Galactic rotation curve and the IAU standard solar-motion parameters. Method B ($d_{\rm B}$) uses the
    \citet{2014ApJ...783..130R} rotation curve and updated solar-motion parameters. Method C ($d_{\rm C}$) is the Monte Carlo calculation of \citet{2018ApJ...856...52W}, based on 10,000 realizations; the asymmetric uncertainties enclose the 68.3\% confidence
    interval. At the exact Wat01 position, the fitted $^{13}$CO feature has only $S/N\simeq1.3$, and hence no reliable molecular-gas kinematic distance is reported. The independently derived distance to the Wat01 and BDS64 clusters is 3.53~kpc (for details see Section~\ref{iso} and Figure~\ref{cmd}).
    \end{minipage}
    \end{table*}

        \begin{figure*}[ht!]
        \centering
        \includegraphics[width=0.49\textwidth]{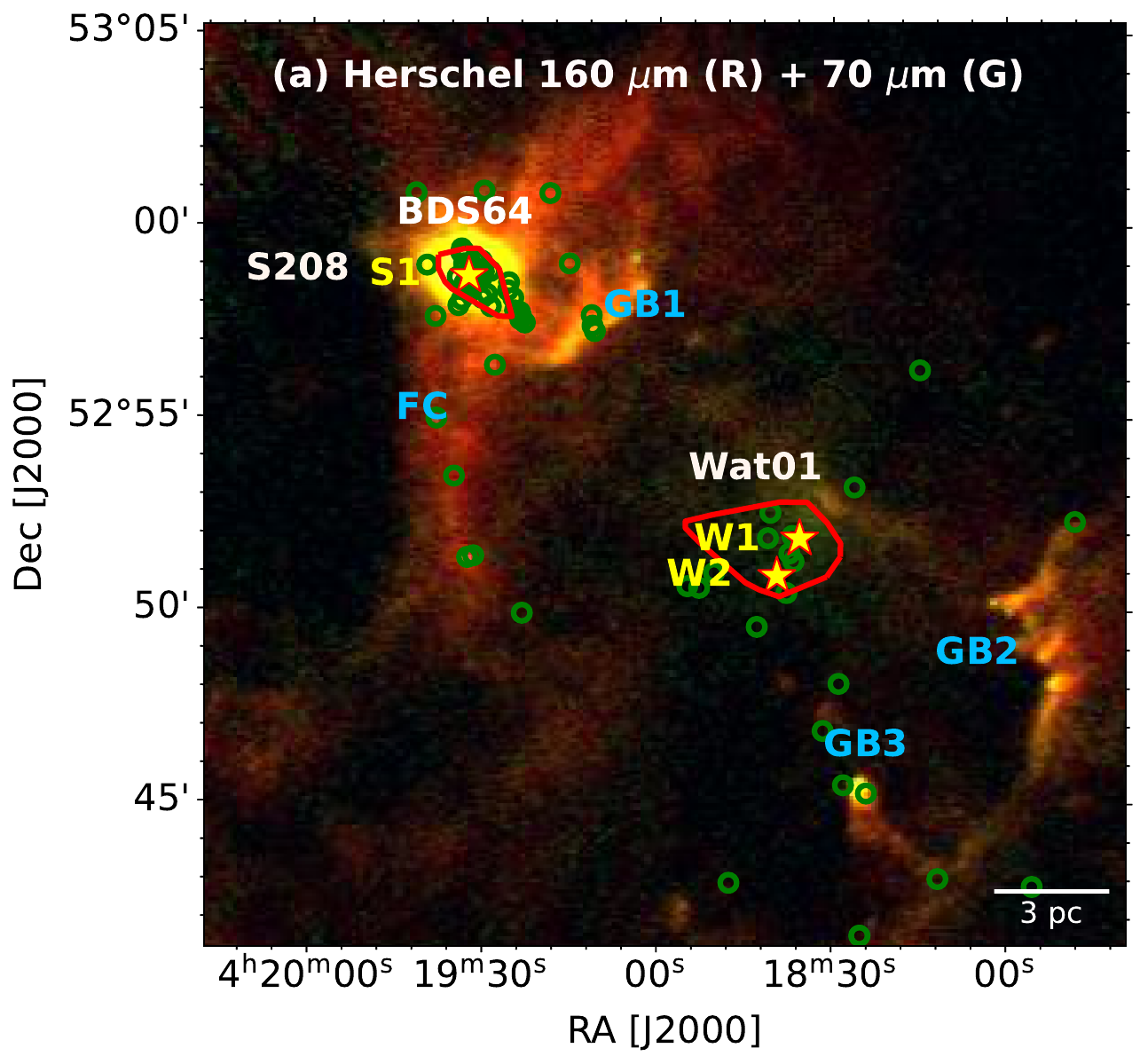}
        \includegraphics[width=0.49\textwidth]{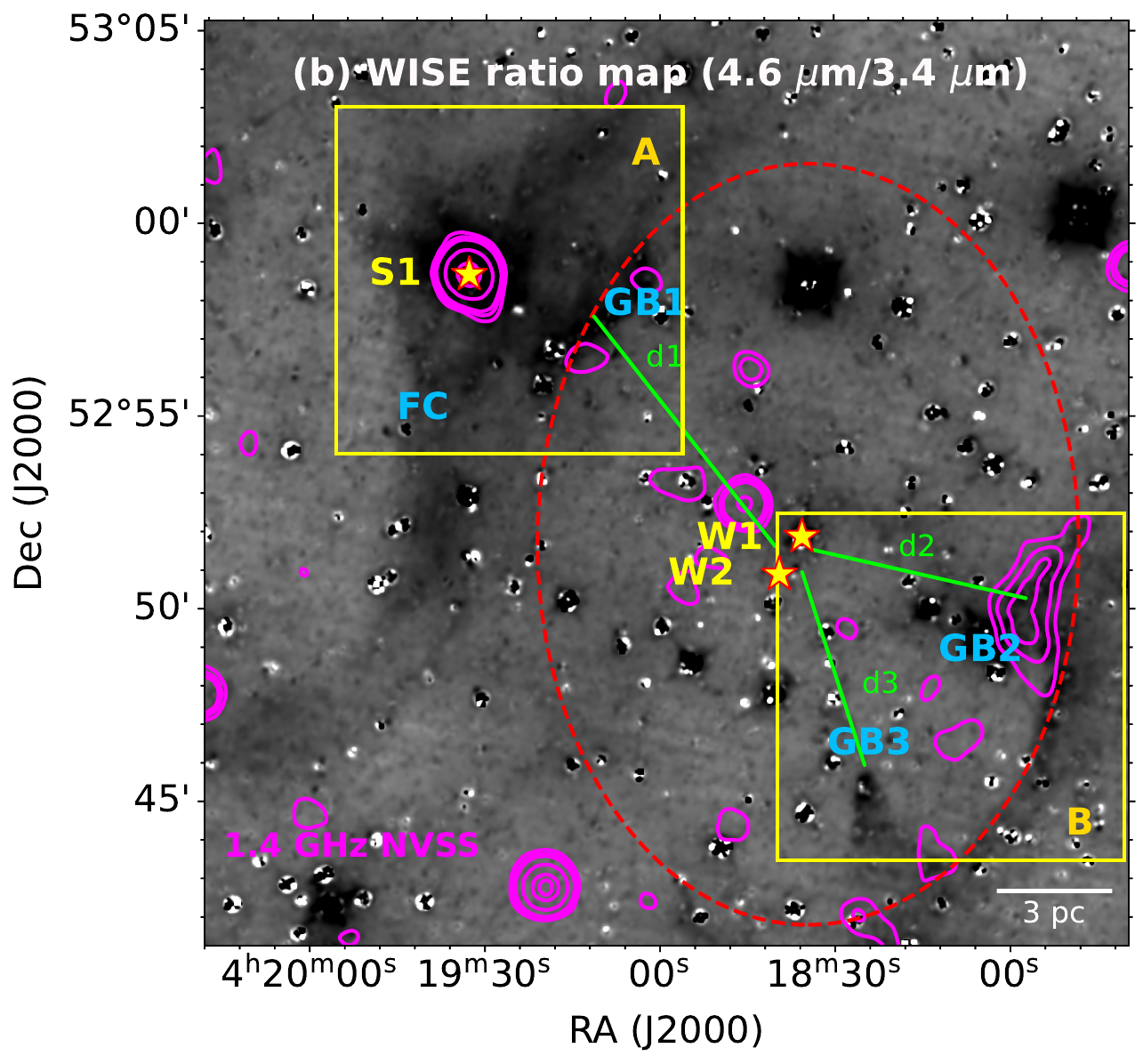}
         \includegraphics[width=0.49\textwidth]{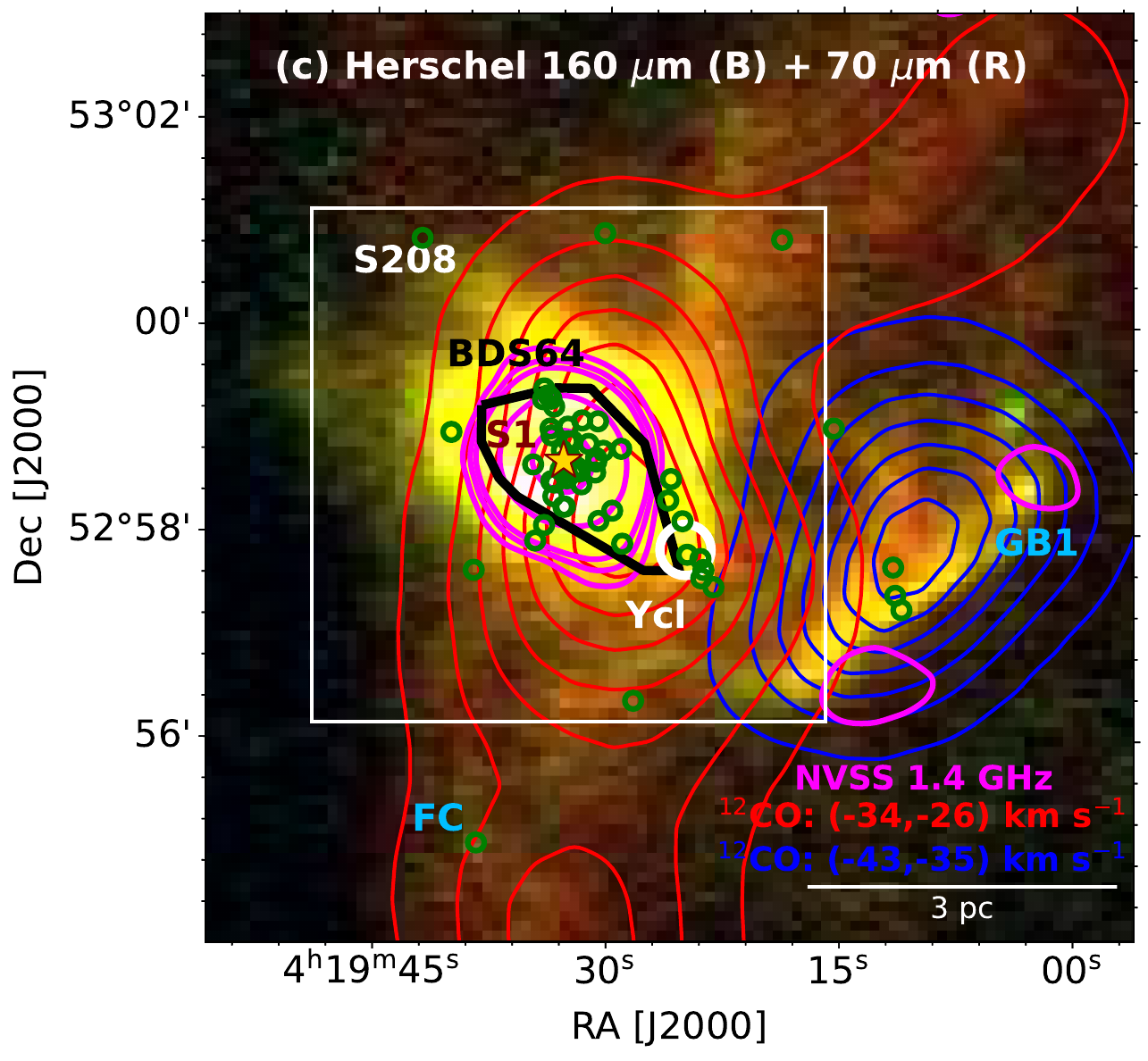}
         \includegraphics[width=0.49\textwidth]{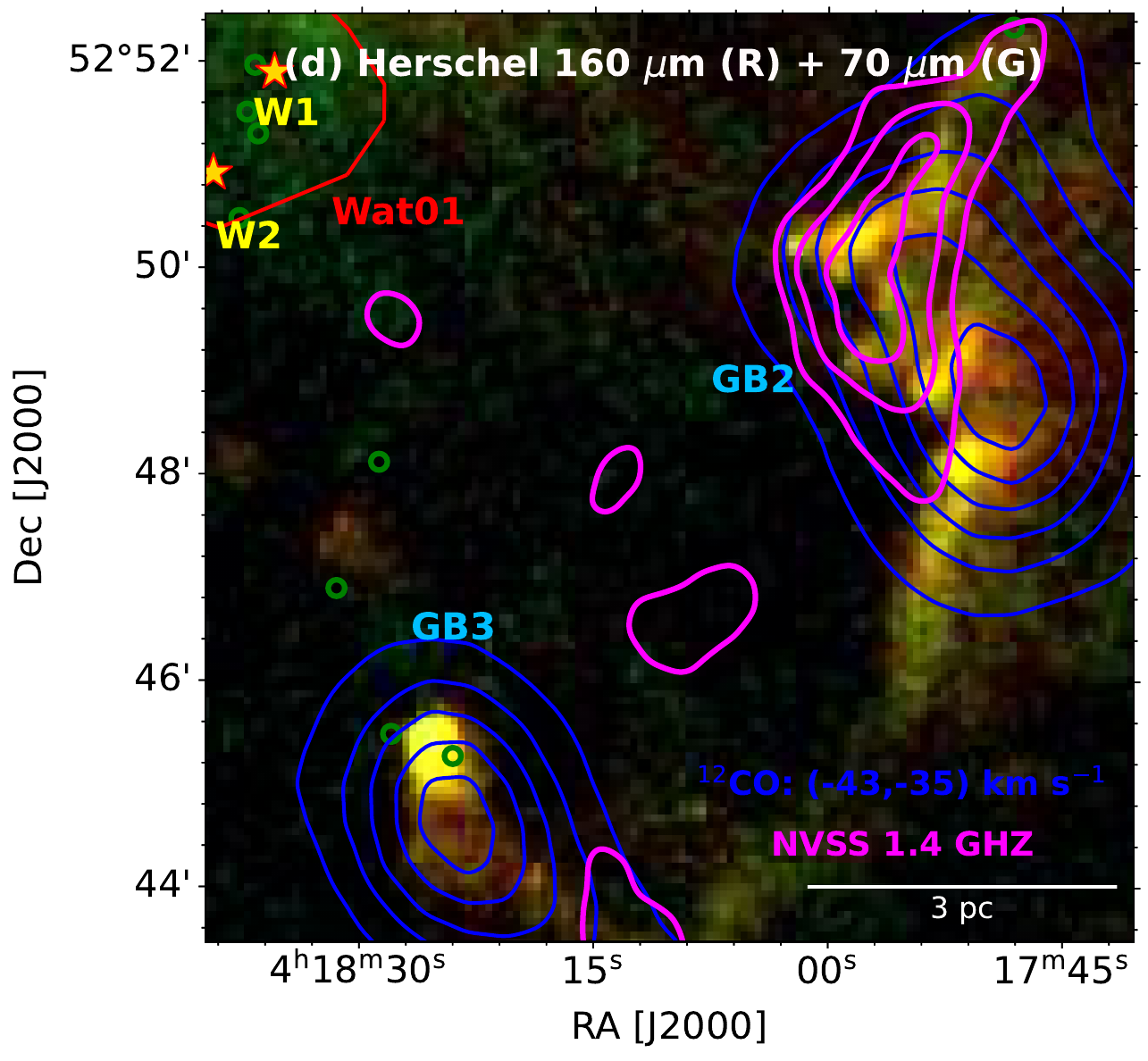}
        \caption{\label{akari}  {\it Panel (a):} \textit{Herschel} color-composite image of a $24^\prime\times24^\prime$ area containing Wat01 and S208, constructed using the 160 $\mu$m (red) and 70 $\mu$m (green) images. Green circles and yellow asterisks mark the YSOs and massive stars, respectively, while the maroon curves outline the convex-hull boundaries of Wat01 and BDS64. GB1--GB3 and FC denote the blue-shifted and red-shifted molecular peaks, respectively. {\it Panel (b):} WISE 4.6/3.4 $\mu$m ratio map. The yellow boxes mark the $9^\prime\times9^\prime$ regions A and B displayed in panels (c) and (d), respectively. The red dotted curve outlines the ring/bubble-like structure around Wat01 (Section~\ref{env}), and the green lines indicate the projected distances adopted for the pressure calculations (Appendix~\ref{pre}). The NVSS radio continuum emission is shown by magenta contours. The lowest NVSS contour is 1.35 mJy/beam with a 0.9 mJy/beam step size. {\it Panels (c) and (d):} Enlarged \textit{Herschel} views of regions A and B, respectively. Red and blue contours trace the $^{12}$CO emission integrated over [$-34$, $-26$] and [$-43$, $-35$] km\,s$^{-1}$, respectively; the magenta contours are the same as in panel (b). In panel (c), the white circle marks the embedded cluster Ycl identified by \citet{2016AJ....151..115Y}; BDS64, `S1', FC, and GB1 are also labeled. Panel (d) shows Wat01, `W1', `W2', GB2, and GB3. The symbols are the same as in panel (a).}
        \end{figure*}

\subsection{Multi-wavelength Morphology of the Region}\label{env} 
     
    We examine the multi-wavelength morphology of the Wat01--S208 complex using far-infrared (FIR), mid-infrared (MIR), and radio-continuum observations to trace cold and warm dust, PDRs, ionized gas, and embedded star-forming structures. Panel (a) of Figure~\ref{akari} shows a $24^\prime\times24^\prime$ color composite constructed from the \textit{Herschel} 160~$\mu$m and 70~$\mu$m bands. The Wat01 cluster (shown as maroon contours;  see Section~\ref{iso} for details) is surrounded by an arc-like distribution of cold dust, with the massive stars ‘W1’ and ‘W2’ located toward its interior. In contrast, S208 is associated with prominent 160~$\mu$m emission, indicating abundant cold dust. The massive star ‘S1’ is located at the peak of cold dust emission. The embedded cluster BDS64 (shown by maroon contours in the S208 region) lies within an elongated dust structure comprising the prominent filament FC, the globule GB1, and several YSOs (shown as green open circles; see details in Appendix~\ref{yso}). The Wat01 region itself is largely deficient in cold dust, although globule-like cold dust structures, GB2 and GB3, are present toward its southwest.

    Panel (b) Figure~\ref{akari} shows the WISE 4.6/3.4~$\mu$m ratio map \citep[for details see;][]{2017ApJ...834...22D, 2023ApJ...953..145V,2024AJ....167..106S}. The elongated structure comprising FC and GB1, together with GB2 and GB3 southwest of Wat01, exhibits strong PAH signatures (dark regions) that trace PDRs. The PAH-emitting material forms an arc-like structure around Wat01 that closely follows the cold dust distribution, giving the region an overall bubble-like morphology (red dotted ellipse). The NVSS 1.4~GHz emission traces diffuse ionized gas around the massive star ‘S1’ within the S208 region, consistent with ongoing ionization, and a distinct emission feature toward GB2. The latter is spatially associated with the bright rim of the globule and is oriented toward ‘W1’ and ‘W2’, consistent with an ionized boundary layer\footnote{A recombination layer that forms on the side of a molecular cloud facing an ionizing (massive) star, once the cloud’s surface becomes ionized.} (IBL) produced by external irradiation \citep{2020ApJ...905...61P,2022MNRAS.510..658P,2025ApJ...994..222C}.

    Panel (c) provides a zoomed view (a $9^\prime \times 9^\prime$ region ‘A’ as outlined in panel b of Figure~\ref{akari}) of the interface between the S208 and Wat01 regions using the \textit{Herschel} 160~$\mu$m (red) and 70~$\mu$m (green) images. The red and blue contours show the integrated $^{12}$CO(1--0) emission over [$-34$, $-26$]~km~s$^{-1}$ and [$-43$, $-35$]~km~s$^{-1}$, tracing FC and GB1, respectively (see Section~\ref{mole}). The extent of BDS64 is marked by the black convex hull, while the magenta contours trace the NVSS 1.4~GHz emission. The embedded cluster Ycl \citep{2016AJ....151..115Y} is marked by a white circle, and the massive star ‘S1’, located near the apex of the cold-dust ridge associated with FC, is shown by a yellow asterisk. Although FC and GB1 overlap in projection, their distinct velocity ranges do not provide direct evidence of a physical interaction. Several YSOs are associated with GB1, while the S208  region, in contrast, coincides with dense cold dust, the BDS64 cluster, and diffuse radio emission around the massive star ‘S1’. The embedded cluster Ycl is located near the edge of the dense FC component.

    Panel (d) shows a zoomed-in view of region ‘B’ (marked in panel b), using the \textit{Herschel} 160~$\mu$m (red) and 70~$\mu$m (green) images, highlighting the globules GB2 and GB3 southwest of Wat01. GB2 shows a bright radio-emitting rim facing the massive stars ‘W1’ and ‘W2’, while GB3 exhibits compact dense heads oriented towards Wat01 with diffuse tails extending away from it. The systematic orientation of these structures toward the massive stars, together with the associated PDR and radio continuum emission, indicates that they are strongly influenced by external stellar radiation. The presence of YSOs within the globules further suggests that star formation is ongoing in these feedback-affected structures.

    Overall, the multi-wavelength morphology reveals a clear contrast between the two regions: Wat01 is characterized by an evacuated, feedback-affected environment with compressed peripheral globules, whereas S208 retains a reservoir of cold, filamentary molecular material associated with ongoing embedded star formation.

    \begin{figure*}[ht!]
    \centering
    \includegraphics[width=0.48\textwidth]{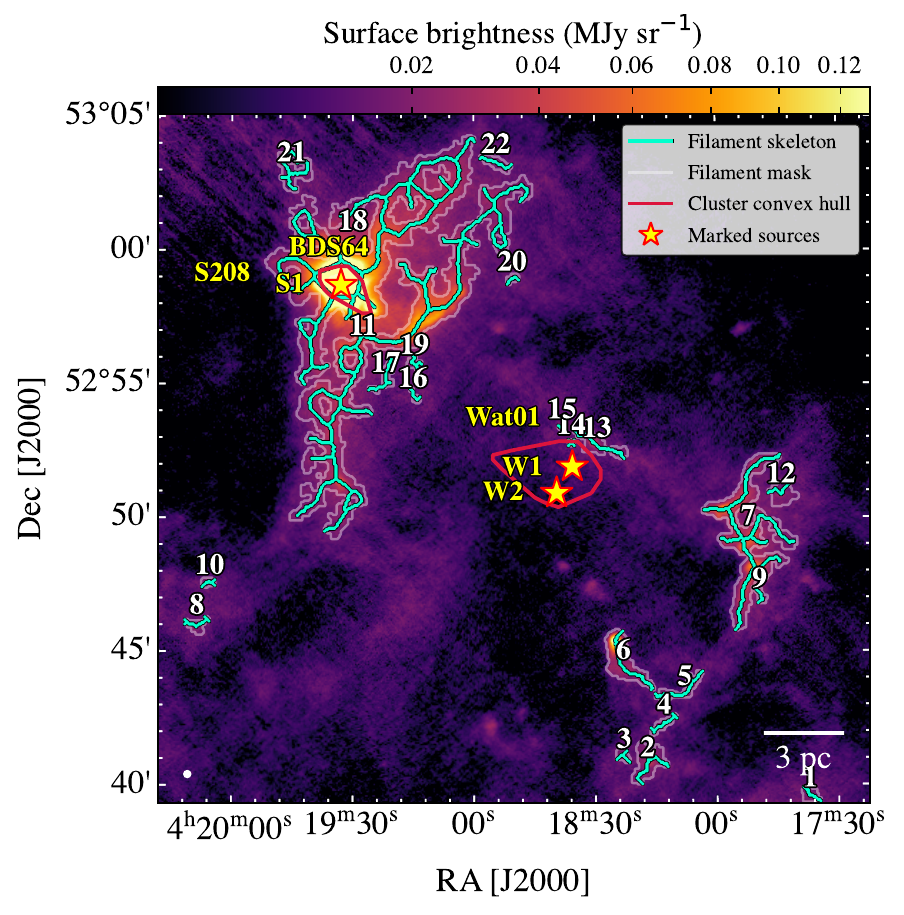}
    \includegraphics[width=0.48\textwidth]{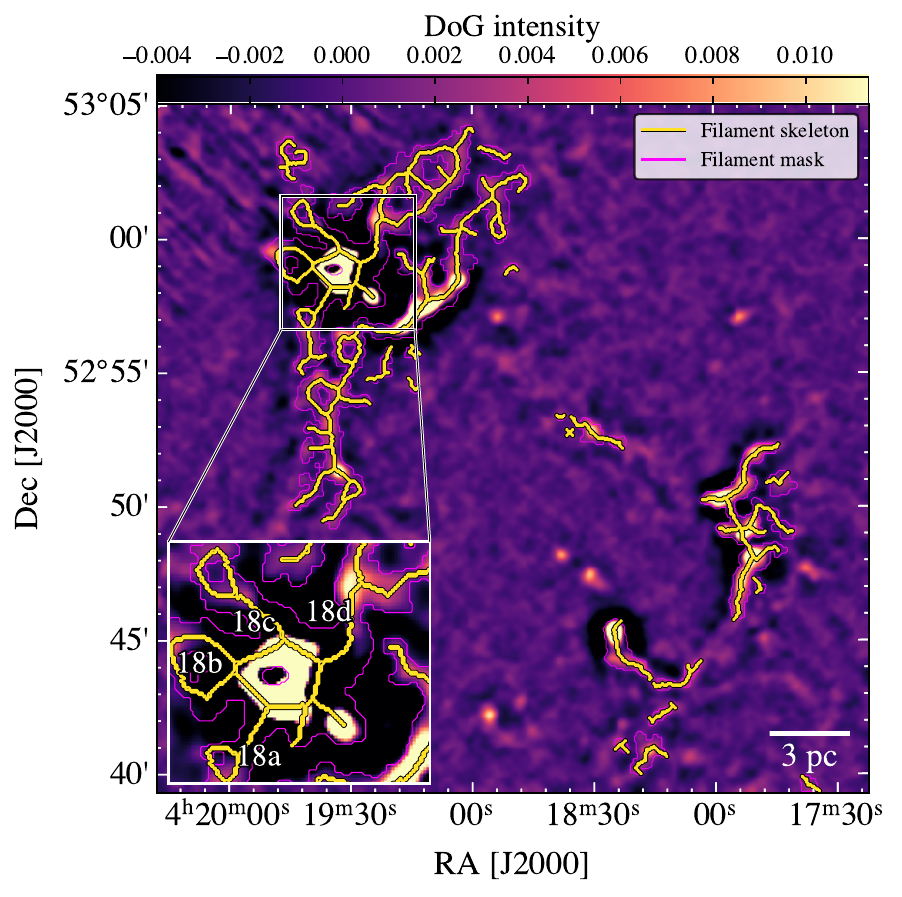}
    \caption{\label{fil} {\it Left panel}: \textit{Herschel} 160 $\mu$m emission map overlaid with the \texttt{FilFinder}-identified filamentary skeletons. The cyan curves mark the final adopted filament skeletons, and the numbered labels indicate the individual \texttt{FilFinder} filament IDs. The faint white contours show the corresponding mask boundary used for skeleton extraction. A beam size of 12 arcsec and a distance of 3.5 kpc were adopted; the beam and 3 pc scale bar are shown in the lower corners. In total, 22 candidate filamentary skeletons were identified using the final \texttt{FilFinder} parameters listed in Table~\ref{tab:ff_param}. The cluster extent is shown as red contours for Wat01 and S208, and yellow asterisks mark the position of massive stars `S1', `W1', and `W2'.
    {\it Right panel}: \textit{Herschel} 160 $\mu$m DoG image with the \texttt{FilFinder} filament skeleton overlaid. The DoG image was constructed as the difference between the map smoothed to effective resolutions of 20 arcsec and 60 arcsec, enhancing compact and intermediate-scale emission while suppressing diffuse large-scale background. Magenta contours show the filament mask, and yellow curves mark the extracted filament skeletons. The inset shows a zoomed view of the central S208 region, revealing a filamentary structure with HFS and four branches (18a-18d) converging towards the center. The scale bar corresponds to 3 pc at an assumed distance of 3.5 kpc.}
    \end{figure*}

\subsection{Morphological and Kinematic Evidence for a Hub--Filament System in S208}\label{f208}

    We extracted filamentary structures from the \textit{Herschel} 160~$\mu$m continuum image using the \texttt{FilFinder2D}\footnote{\url{https://fil-finder.readthedocs.io/en/latest/}} algorithm \citep{10.1093/mnras/stv1521}. The adopted parameters, selected to recover the principal filamentary network while minimizing fragmentation and spurious detections, are listed in Table~\ref{tab:ff_param}. The resulting skeletons are shown in the left panel of Figure~\ref{fil}. Filament~18 (hereafter, F18) forms the dominant filament in S208, with several secondary filamentary branches connected to it and converging toward the central high-emission region hosting the embedded cluster BDS64 and the massive star `S1' (see the left panel of Figure~\ref{fil} and panel (c) of Figure~\ref{akari}). The resulting branched morphology is characteristic of an HFS configuration. 

    We compared the continuum-derived skeletons with the $^{12}$CO and $^{13}$CO emission to examine their kinematic association. F18 and the upper part of filament~11 (hereafter, F11) coincide with the red-shifted molecular component FC ($V_{\rm LSR}=-34$ to $-26$~km~s$^{-1}$), whereas the lower part of F11, GB1, GB2, and GB3 are associated with the blue-shifted component ($V_{\rm LSR}=-43$ to $-35$~km~s$^{-1}$; Figures~\ref{akari} and \ref{fil}). Although F11 appears continuous in the \textit{Herschel} 160~$\mu$m  continuum image, the molecular-line observations demonstrate that it is composed of two kinematically distinct segments. The declination--velocity diagrams ( panels (c) and (d) of Figure~\ref{co}) show neither bridging emission nor a continuous velocity gradient between them, indicating that their apparent continuity is likely a consequence of projected continuum morphology along the line of sight.

    To test the correspondence between the extracted skeletons and dust-emission ridges, we constructed a Difference-of-Gaussians (DoG) image from the \textit{Herschel} 160~$\mu$m map by subtracting two smoothed images at different angular resolutions, $\mathrm{DoG}=I(20^{\prime\prime})-I(60^{\prime\prime})$, which enhances structures on intermediate spatial scales while suppressing the diffuse background. As shown in the right panel of Figure~\ref{fil}, the \texttt{FilFinder} skeletons closely follow the enhanced dust ridges. In particular, F18 and its secondary branches (18a-18d) converge toward the same central junction (see zoomed inset), providing additional morphological support for HFS structure in the S208 region. The morphological parameters of the F18 components are listed in Table~\ref{tab:f18}.

\begin{figure*}[ht!]
        \centering
        \includegraphics[width=0.9\textwidth]{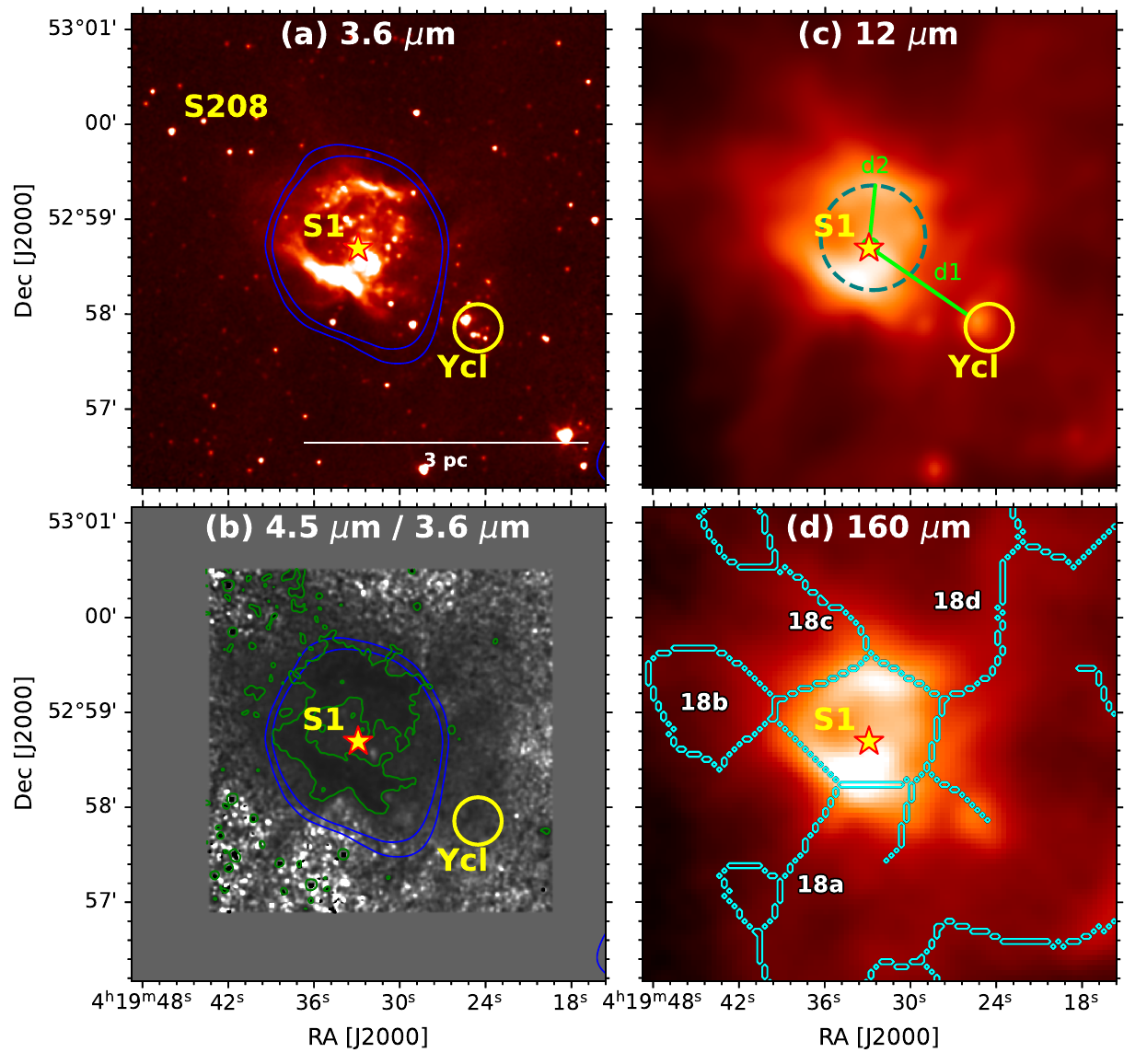}
        \caption{\label{all} {The zoomed-in (see; white box in panel (c) of Figure~\ref{akari} ) multi-wavelength picture of the S208 region ranging from MIR to FIR. Panel (a) shows the MIR 3.6 $\mu$m emission. Blue curves show the NVSS 1.4 GHz radio continuum emission. The lowest contour is at 1.35 mJy/beam with a 0.9 mJy/beam step size. The cyan circles show the distribution of identified YSOs. Panel (b) is a \textit{Spitzer} ratio map of the central region of S208 overlaid with a green PDR region. Panel (c) represents the WISE 12 $\mu$m image overlaid with projected distances, d1 and d2, from the massive star `S1' to the embedded cluster Ycl and the bubble boundary, respectively. Panel (d) shows the FIR emission representing the hub feature, highlighting the hub morphology with cyan filaments (for detail, see Section~\ref{f208}). The yellow asterisk indicates the location of the massive star ‘S1’ in all panels. }}
        \end{figure*}

    To further examine the S208 region and the associated filamentary cloud FC, Figure~\ref{all} provides a multi-wavelength zoomed-in view (white box in panel (c) of Figure~\ref{akari}) of the central S208 region, using \textit{Spitzer} 3.6~$\mu$m image (panel a), the 4.5/3.6~$\mu$m ratio map (panel b), WISE 12~$\mu$m image (panel c), and \textit{Herschel} 160~$\mu$m image (panel d). The 3.6~$\mu$m emission reveals a compact MIR bubble centered on ‘S1’, coincident with the radio-continuum emission (blue contours) and a PDR cavity (shown as green contours) in the 4.5/3.6~$\mu$m ratio map. The bubble has a radius of $\sim0.6$~pc (teal dashed circle in panel c) and contains most of the identified YSOs. The compact bubble and associated ionized emission indicate localized feedback (see Appendix~\ref{pre} and panel c of Figure~\ref{all}) around ‘S1’.

    The embedded cluster Ycl lies outside the bubble, close to the projected interface between FC and GB1 (see panel c in Figure~\ref{akari}). Faint filamentary extensions from the bubble become more prominent in the 12~$\mu$m and 160~$\mu$m images, where they connect spatially with the larger cold-dust network, shown as cyan (\texttt{FilFinder}- identified skeleton) filaments in panel (d). Although FC and GB1 are separated in velocity space, their projected interface coincides with Ycl. Thus, the S208 region contains a compact feedback-affected cavity embedded within a larger filamentary molecular network.

    \begin{table}
    \centering
    \caption{Adopted \texttt{FilFinder} parameters used for extracting candidate filamentary skeletons from the \textit{Herschel} 160~$\mu$m image. The beam FWHM was adopted as 12 arcsec, corresponding to $\sim$3.75 pixels.}
    \label{tab:ff_param}
    \begin{tabularx}{\linewidth}{@{}l c l@{}}
    \toprule
    
    \textbf{Parameter} &
    \textbf{Value} &
    \textbf{Physical scale } \\
    \midrule

    \texttt{Beam FWHM}           & 12 arcsec     & 1 beam ( $\sim$0.20 pc) \\

    \texttt{Distance}            & 3.5 kpc       & pixel scale: 0.054 pc pix$^{-1}$ \\

    \texttt{glob\_thresh}        & 0.015         & $\sim3\sigma$ robust background\textsuperscript{a} \\

    \texttt{adapt\_thresh}       & 30 pix        & 8 beams ($\sim$1.63 pc) \\

    \texttt{smooth\_size}        & 3 pix         & 0.8 beam \\

    \texttt{size\_thresh}        & 80 pix$^2$    & $\sim$5 beam areas \\

    \texttt{fill\_hole\_size}    & 50 pix$^2$    & $\sim$3 beam areas \\

    \texttt{branch\_thresh}      & 10 pix        & 2.7 beams ($\sim$0.54 pc) \\

    \texttt{skel\_thresh}        & 20 pix        & 5.3 beams ($\sim$1.09 pc) \\

    \texttt{flatten\_percent}    & 85            & $85^{th}$ percentile \\

    \bottomrule
    \end{tabularx}
    
    \vspace{2mm}
    
    \parbox{\columnwidth}{\footnotesize
    \textsuperscript{a}
    The adopted value of \texttt{glob\_thresh}=0.015 is approximately three times the robust background dispersion estimated from the median absolute deviation ($\sigma_{\rm robust}\sim0.005$ in image units). The \texttt{branch\_thresh} and \texttt{skel\_thresh} parameters were used during the pruning stage to remove short branches and insignificant
    skeleton structures from the extracted filament network.
    }
    
    \end{table}

\begin{table}[htbp]
    \centering
    \caption{Projected morphological parameters of the F18 filament components. The component spines were selected using DS9 regions and constrained to the FilFinder medial-axis skeleton. Widths are beam-deconvolved FWHM values from
    Gaussian-plus-background fits to the 160~$\mu$m radial intensity profiles. The first width uncertainty is the fitting uncertainty, and the second is the systematic uncertainty obtained by varying the radial-profile fitting
    radius. A distance of 3.5~kpc and a beam FWHM of 12~arcsec were adopted.}
    
    \label{tab:f18}
    \begin{tabular}{lccc}
    \toprule
   
    \textbf{Component} &
    \textbf{$L_{\rm fil}$ (pc)} &
    \textbf{$W_{\rm fil}$ (pc)} &
    \textbf{$L_{\rm fil}/W_{\rm fil}$} \\
    \midrule
   
    18a & 5.02 & $0.82 \pm 0.02_{\rm stat} \pm 0.02_{\rm sys}$ & 6.1 \\
   
    18b & 3.39 & $0.81 \pm 0.02_{\rm stat} \pm 0.01_{\rm sys}$ & 4.2 \\
   
    18c & 3.49 & $0.89 \pm 0.03_{\rm stat} \pm 0.11_{\rm sys}$ & 3.9 \\
   
    18d & 9.71 & $1.21 \pm 0.05_{\rm stat} \pm 0.11_{\rm sys}$ & 8.0 \\
    \bottomrule
    \end{tabular}

    \vspace{1mm}
    \begin{minipage}{0.96\linewidth}
    \footnotesize
    \textit{Notes.} $L_{\rm fil}$ is the projected curved length along the selected open spine, rather than the straight endpoint separation. $W_{\rm fil}$ is the Gaussian FWHM after deconvolution from the adopted beam. 
    \end{minipage}
    \end{table}

\section{Discussion} \label{dis}

\subsection{Structural and Dynamical Evolution of the Clusters} \label{ds}

The Wat01 cluster has an elongated morphology, a radius of $\sim1.5$ pc, and a peak stellar density of $\sim70$ stars pc$^{-2}$. We estimate a distance of $3.53\pm0.55$ kpc and a reddening of $E(B-V)=0.83$ mag. Fifty-two probable members are identified, including the massive stars ‘W1’ (O9V) and ‘W2’ (B1.5V). The main-sequence lifetime of ‘W1’ provides an upper limit to the cluster age of $\sim8.1$ Myr \citep{2023JApA...44...66K}.

The spatial distribution of the members gives $Q=0.6$, indicating a significant degree of hierarchical substructure \citep{2004MNRAS.348..589C,2009MNRAS.392..341C}. This suggests that Wat01 has not yet evolved into a centrally concentrated configuration. At the same time, the MST analysis of mass segregation yields $\Gamma_{\rm MSR}=1.5\pm0.6$, indicating a tendency for the massive stars to be more centrally concentrated than the lower-mass population \citep{2009MNRAS.395.1449A,2011A&A...532A.119O,2018MNRAS.473..849D}. The estimated dynamical relaxation time, $T_E\sim8.7$ Myr, is comparable to the upper age limit of the cluster. Thus, although dynamical evolution may have contributed to the observed segregation, the cluster may not yet have undergone sufficient relaxation for dynamical processes to be its sole origin. A primordial contribution to the mass segregation is therefore plausible.

The MF of Wat01 is described by a single power law with $\Gamma=-1.62\pm0.15$ down to $\sim0.7\,M_\odot$, steeper than the Salpeter slope of $-1.35$ \citep{1955ApJ...121..161S}. In contrast to our previous studies \citep{2020ApJ...896...29K,2020MNRAS.498.2309S,2024MNRAS.527.9626P}, we do not find a statistically significant turnover near $\sim1\,M_\odot$. The steep MF and absence of a clear turnover may reflect the unevolved dynamical state of Wat01, although observational incompleteness, unresolved binaries, and residual field contamination can also affect the measured MF. The present results, therefore, suggest that the stellar mass distribution of Wat01 may still retain considerable information about its formation conditions.

BDS64, embedded within S208, has a smaller radius of $\sim0.9$ pc and lies at the same distance within the uncertainties. The B1.5V star ‘S1’ provides an upper age limit of $\sim12.6$ Myr. However, the compact H\,{\sc ii} region, strong infrared emission, and embedded stellar population indicate ongoing star formation and suggest that this upper limit does not necessarily represent the current age of the active star-forming population. The high extinction and incompleteness of Gaia DR3 prevent a robust determination of the spatial structure, MF, mass segregation, and relaxation state of BDS64.

\subsection{Feedback-regulated Evolution of the Wat01 Molecular Cloud}

The molecular gas around Wat01 appears to have evolved under the influence of feedback from its massive stars (see Section~\ref{env}). The 1.4~GHz radio continuum emission associated with the cluster is surrounded by an arc-like structure traced by PAH emission, cold dust, and molecular gas (see Figure~\ref{akari}), consistent with an H\,{\sc ii} region interacting with its surrounding material \citep{2010A&A...523A...6D,2019ARA&A..57..227K}. The molecular emission is weak toward the cluster center, whereas compact molecular globules (GB1--GB3) are observed along its periphery at $V_{\rm LSR}\sim-38~\mathrm{km\,s^{-1}}$. This morphology suggests that low-density molecular material may have been dispersed from the central region, while denser structures survived along the boundary of the H\,{\sc ii} region (see Figures~\ref{hco} and \ref{akari}). Similar feedback-shaped cloud structures have been reported around other Galactic H\,{\sc ii} regions \citep{2009A&A...496..177D,2012A&A...542A..10A,2022ApJ...926...25P}.

The pressure analysis (see Appendix~\ref{pre}) provides further constraints on the nature of this interaction. The ionized-gas pressure, $P_{\mathrm{HII}}$, exceeds both the radiation ($P_{\mathrm{rad}}$) and stellar-wind ($P_{\mathrm{wind}}$)  pressures (see Table~\ref{fp}), indicating that thermal pressure from the H\,{\sc ii} region is the dominant feedback component acting on the surrounding gas. The peripheral globules GB1--GB3 possess IBL features, cometary morphologies, enhanced PDR emission, and dense molecular material preferentially oriented toward ‘W1’ and ‘W2’. Their association with YSOs makes them prominent sites of feedback-influenced star formation. These properties are consistent with radiation-driven implosion (RDI), in which an expanding ionization front compresses pre-existing dense structures. However, the present observations do not uniquely establish that the YSOs were formed as a direct consequence of the feedback. 

Taken together, Wat01 appears to be in a stage where massive-star feedback has significantly altered the parental cloud while dense molecular structures survive along the periphery. The region therefore demonstrates the dual role of massive-star feedback: dispersing low-density molecular material while simultaneously compressing selected dense structures that can continue to form stars.

\subsection{Evolution of the Hub--Filament System in S208} 
    
    The molecular environment of S208 exhibits a markedly different morphology. Molecular-line observations reveal a prominent filamentary structure associated with the region. The \textit{Herschel} 160 $\mu$m emission, together with the \texttt{FilFinder} skeletons and DoG analysis (Figure~9), reveals a central hub-like region surrounded by four parsec-scale filaments (see panel (d) in Figure~10 and Table~\ref{tab:f18}). This central hub is associated with BDS64 and the H\,{\sc ii} region powered by the massive star ‘S1’. Infrared emission at 3.6 and 4.5 $\mu$m also suggests the presence of a PDR surrounding ‘S1’. Such a morphology is broadly consistent with the HFS paradigm, in which filamentary structures converge toward a dense central hub. Numerical simulations suggest that HFSs can arise from large-scale converging flows that generate filamentary networks and channel gas toward dense hubs, connecting cloud-scale gas dynamics to massive-star and cluster formation \citep{semadeni07,smith16,nozaki26}. These converging motions may result from turbulent compression, global gravitational collapse, or the self-gravity of filamentary networks \citep{2009ApJ...700.1609M,2018ARA&A..56...41M,padoan20,semadeni19}. A detailed investigation of the origin and kinematics of the filamentary structures in S208, however, is beyond the scope of the present study.

    Although the present observations do not directly demonstrate mass inflow along the filaments, the observed morphology is consistent with a gravitationally organized, intermediate-stage HFS in which the dense gas reservoir remains available for ongoing star formation. Unlike Wat01, the molecular cloud around S208 is comparatively less disrupted, with ‘S1’ still embedded within the dense filamentary material. The associated PDR and compact MIR bubble (see Figure~\ref{all}) indicate localized stellar feedback, while the persistence of the larger-scale filamentary network suggests that feedback has not yet disrupted the molecular reservoir. The spatial coincidence of the dense hub with BDS64, ‘S1’, and numerous YSOs further identifies S208 as an active site of clustered star formation.

\subsection{Contrasting Star Formation Environments in the Wat01--S208 Complex}

   Wat01 and S208 are located at a similar distance of $\sim3.5$ kpc and are likely to be associated with the same star-forming complex, yet their molecular and stellar environments differ considerably. Wat01 is characterized by an evacuated central region, a shell-like molecular morphology, and pressure-dominated expansion driven by ionized gas. S208, in contrast, retains an interconnected filamentary network, an embedded cluster, and a massive star. These differences demonstrate that distinct physical conditions can coexist within a common star-forming environment.

    The blue-shifted molecular component associated with Wat01 is fragmented into globules and appears strongly affected by stellar feedback, whereas the red-shifted component associated with S208 retains a well-defined filamentary morphology. The two components are separated in velocity by $\sim8$ km\,s$^{-1}$, and their distinct kinematic structures do not by themselves establish that they belong to a single physically interacting cloud. Nevertheless, their contrasting properties provide a useful comparison between a feedback-processed environment and a less disrupted hub--filament system \citep{2022MNRAS.514.6038Z, 2023MNRAS.522.3719L, 2024MNRAS.528.1460R}.

    The embedded cluster Ycl, located near the interface between FC and GB1, may represent an intermediate case. Its spatial coincidence with dense molecular material, together with the pressure exerted by feedback from ‘S1’ on the surrounding gas, suggests that both gravitational assembly and localized feedback may influence its formation. However, the available observations do not uniquely distinguish between spontaneous and feedback-triggered star formation at this location.

    Overall, the Wat01--S208 complex demonstrates that gravity and stellar feedback operate concurrently during clustered star formation, although their relative importance may vary across neighboring regions. Wat01 represents an environment in which massive stars have dispersed and compressed the natal molecular material, whereas S208 contains a comparatively intact HFS with a dense gas reservoir available for ongoing clustered star formation. Rather than implying a strictly sequential evolutionary connection between the two regions, their properties highlight the diversity of evolutionary states and physical mechanisms that can coexist within a single molecular complex.

\section{Summary and Conclusions}\label{s6}

To investigate the underlying physical processes, we have carried out a multi-wavelength study of the Wat01--S208 complex. Our main conclusions are as follows:

\begin{itemize}

\item The Wat01 and BDS64 clusters display elongated morphologies, with characteristic radii of $\sim1.53$ and $\sim0.88$ pc, respectively, and are located at a common distance of $3.53\pm0.55$ kpc. Using \textit{Gaia} DR3, we identify 52 probable members of Wat01. The estimated reddening values are $(E(B-V)\sim0.83$ and $\sim0.65$ mag toward Wat01 and BDS64, respectively, corresponding to visual extinctions of $A_V\sim2.6$ mag and $\sim2.0$ mag for $R_V=3.1$.

\item Wat01 shows significant hierarchical structure, characterized by $Q=0.6$, along with marginal tendancy towards mass segregation, with $\Gamma_{\rm MSR}=1.5\pm0.6$. Since its relaxation time is comparable to or longer than its age, the observed mass segregation may have a primordial component. The MF is well described by a single power law, with $\Gamma=-1.62\pm0.15$ over the mass range $\sim$0.7--19.6 $M_\odot$, and shows no significant turnover near $\sim$1 $M_\odot$.

\item The molecular gas exhibits two distinct kinematic components: a filamentary cloud hosting S208 at $V_{\rm LSR}\sim-34$ to $-26$ km\,s$^{-1}$ and a collection of compact globules (GB1--GB3) associated with Wat01 at $V_{\rm LSR}\sim-43$ to $-35$ km\,s$^{-1}$. These distinct velocity components indicate that Wat01 and S208 occupy physically and kinematically different molecular environments within the broader star-forming complex.

\item  Wat01 has been strongly affected by massive-star feedback, as indicated by its evacuated central cavity, encompassing shell, and peripheral globules, which are consistent with an expanding H\,{\sc ii} region. The dominance of ionized-gas pressure and the morphology of GB1--GB3 support a scenario in which stellar feedback has compressed the surrounding molecular gas and may have triggered localized star formation.
However, a direct causal connection cannot be established from the present observations alone.

\item In contrast, our investigation of S208 using \textit{Herschel} FIR observations reveals a prominent HFS, whose central hub is associated with BDS64 and the H\,{\sc ii}  region ionized by the massive star ‘S1’. The F18a–F18d components have projected lengths of 3.39–9.71 pc, beam-deconvolved widths of 0.81–1.21 pc, and aspect ratios of 3.9–8.0, consistent with their elongated morphology and convergence toward a common hub. The relatively well-preserved filamentary network suggests that gravity continues to play an important role in assembling dense gas and sustaining clustered star formation, while localized stellar feedback is beginning to influence the surrounding cloud. The Ycl cluster near the GB1--FC interface may represent a region where gravitational assembly and stellar feedback jointly contribute to ongoing star formation.

\end{itemize}

In conclusion, the Wat01--S208 complex provides an example of coexisting star-forming environments at different evolutionary stages, in which feedback has extensively reshaped Wat01 while the filamentary structure of S208 remains comparatively coherent. These results highlight the evolving interplay between stellar feedback and gravity in regulating gas assembly and clustered star formation.

\section*{Acknowledgments}

We sincerely thank the anonymous referee for constructive comments. 
This work is based on observations obtained with the 2-m Himalayan Chandra Telescope (HCT) at the Indian Astronomical Observatory (Hanle) and uses archival data from UKIDSS, 2MASS, the \textit{Spitzer} Space Telescope, WISE, the \textit{Herschel} space telescope, and the FCRAO Outer Galaxy CO Survey. We thank Prof. Mark Heyer for providing the FCRAO Outer Galaxy Survey data. A.V. acknowledges the financial support of DST- INSPIRE (No. DST/INSPIRE Fellowship/2019/IF190550). D.K.O. acknowledges support from the Department of Atomic Energy, Government of India, under Project No. RTI 4012. M.C. acknowledges the financial support of DST-INSPIRE (No. DST/INSPIRE Fellowship/2022/IF220587). The research work at the Physical Research Laboratory is funded by the Department of Space, Government of India.






\bibliography{wat01}{}

@ARTICLE{semadeni19,
       author = {{V{\'a}zquez-Semadeni}, Enrique and {Palau}, Aina and {Ballesteros-Paredes}, Javier and {G{\'o}mez}, Gilberto C. and {Zamora-Avil{\'e}s}, Manuel},
        title = "{Global hierarchical collapse in molecular clouds. Towards a comprehensive scenario}",
      journal = {\mnras},
         year = 2019,
        month = dec,
       volume = {490},
       number = {3},
        pages = {3061-3097},
          doi = {10.1093/mnras/stz2736},
archivePrefix = {arXiv},
       eprint = {1903.11247},
 primaryClass = {astro-ph.GA},
       adsurl = {https://ui.adsabs.harvard.edu/abs/2019MNRAS.490.3061V}
}

@ARTICLE{semadeni07,
       author = {{V{\'a}zquez-Semadeni}, Enrique and {G{\'o}mez}, Gilberto C. and {Jappsen}, A. Katharina and {Ballesteros-Paredes}, Javier and {Gonz{\'a}lez}, Ricardo F. and {Klessen}, Ralf S.},
        title = "{Molecular Cloud Evolution. II. From Cloud Formation to the Early Stages of Star Formation in Decaying Conditions}",
      journal = {\apj},
         year = 2007,
        month = mar,
       volume = {657},
       number = {2},
        pages = {870-883},
          doi = {10.1086/510771},
archivePrefix = {arXiv},
       eprint = {astro-ph/0608375},
 primaryClass = {astro-ph},
       adsurl = {https://ui.adsabs.harvard.edu/abs/2007ApJ...657..870V}
}

@ARTICLE{smith16,
       author = {{Smith}, Rowan J. and {Glover}, Simon C.~O. and {Klessen}, Ralf S. and {Fuller}, Gary A.},
        title = "{On the nature of star-forming filaments - II. Subfilaments and velocities}",
      journal = {\mnras},
         year = 2016,
        month = feb,
       volume = {455},
       number = {4},
        pages = {3640-3655},
          doi = {10.1093/mnras/stv2559},
archivePrefix = {arXiv},
       eprint = {1509.03321},
 primaryClass = {astro-ph.GA},
       adsurl = {https://ui.adsabs.harvard.edu/abs/2016MNRAS.455.3640S}
}

@ARTICLE{nozaki26,
       author = {{Nozaki}, Shingo and {Inutsuka}, Shu-ichiro},
        title = "{An Origin of Radially Aligned Filaments in Hub-filament Systems}",
      journal = {\apjl},
         year = 2026,
        month = mar,
       volume = {1000},
       number = {1},
          eid = {L31},
        pages = {L31},
          doi = {10.3847/2041-8213/ae4c84},
archivePrefix = {arXiv},
       eprint = {2603.03063},
 primaryClass = {astro-ph.GA},
       adsurl = {https://ui.adsabs.harvard.edu/abs/2026ApJ..1000L..31N}
}

@ARTICLE{padoan20,
	author = {{Padoan}, Paolo and {Pan}, Liubin and {Juvela}, Mika and {Haugb{\o}lle}, Troels and {Nordlund}, {\r{A}}ke},
	title = "{The Origin of Massive Stars: The Inertial-inflow Model}",
	journal = {\apj},
	year = 2020,
	month = sep,
	volume = {900},
	number = {1},
	eid = {82},
	pages = {82},
	doi = {10.3847/1538-4357/abaa47},
	archivePrefix = {arXiv},
	eprint = {1911.04465},
	primaryClass = {astro-ph.GA},
	adsurl = {https://ui.adsabs.harvard.edu/abs/2020ApJ...900...82P}
}

@ARTICLE{2008AJ....135.1934S,
   author = {{Sharma}, S. and {Pandey}, A.~K. and {Ogura}, K. and {Aoki}, T. and 
	{Pandey}, K. and {Sandhu}, T.~S. and {Sagar}, R.},
    title = "{Mass Functions and Photometric Binaries in Nine Open Clusters}",
  journal = {\aj},
archivePrefix = "arXiv",
   eprint = {0803.0122},
     year = 2008,
    month = may,
   volume = 135,
    pages = {1934-1945},
      doi = {10.1088/0004-6256/135/5/1934},
   adsurl = {http://adsabs.harvard.edu/abs/2008AJ....135.1934S}
}

@ARTICLE{1991A&A...250..324S,
   author = {{Sagar}, R. and {Richtler}, T.},
    title = "{Mass functions of five young Large Magellanic Cloud star clusters}",
  journal = {\aap},
     year = 1991,
    month = oct,
   volume = 250,
    pages = {324-339},
   adsurl = {http://adsabs.harvard.edu/abs/1991A\%26A...250..324S}
}

@ARTICLE{1955ApJ...121..161S,
   author = {{Salpeter}, E.~E.},
    title = "{The Luminosity Function and Stellar Evolution.}",
  journal = {\apj},
     year = 1955,
    month = jan,
   volume = 121,
    pages = {161},
      doi = {10.1086/145971},
   adsurl = {http://adsabs.harvard.edu/abs/1955ApJ...121..161S}
}

@ARTICLE{2003A&A...397..191P,
   author = {{Pandey}, A.~K. and {Upadhyay}, K. and {Nakada}, Y. and {Ogura}, K.
	},
    title = "{Interstellar extinction in the open clusters towards galactic longitude around 130$^{deg}$}",
  journal = {\aap},
   eprint = {astro-ph/0210680},
     year = 2003,
    month = jan,
   volume = 397,
    pages = {191-200},
      doi = {10.1051/0004-6361:20021509},
   adsurl = {http://adsabs.harvard.edu/abs/2003A\%26A...397..191P}
}

@ARTICLE{1994ApJS...90...31P,
   author = {{Phelps}, R.~L. and {Janes}, K.~A.},
    title = "{Young open clusters as probes of the star formation process. 1: an atlas of open cluster photometry}",
  journal = {\apjs},
     year = 1994,
    month = jan,
   volume = 90,
    pages = {31-82},
      doi = {10.1086/191857},
   adsurl = {http://adsabs.harvard.edu/abs/1994ApJS...90...31P}
}

@ARTICLE{2009ApJS..184...18G,
   author = {{Gutermuth}, R.~A. and {Megeath}, S.~T. and {Myers}, P.~C. and 
	{Allen}, L.~E. and {Pipher}, J.~L. and {Fazio}, G.~G.},
    title = "{A Spitzer Survey of Young Stellar Clusters Within One Kiloparsec of the Sun: Cluster Core Extraction and Basic Structural Analysis}",
  journal = {\apjs},
archivePrefix = "arXiv",
   eprint = {0906.0201},
 primaryClass = "astro-ph.SR",
     year = 2009,
    month = sep,
   volume = 184,
    pages = {18-83},
      doi = {10.1088/0067-0049/184/1/18},
   adsurl = {http://adsabs.harvard.edu/abs/2009ApJS..184...18G}
}

@ARTICLE{1992AJ....104..340L,
   author = {{Landolt}, A.~U.},
    title = "{UBVRI photometric standard stars in the magnitude range 11.5-16.0 around the celestial equator}",
  journal = {\aj},
     year = 1992,
    month = jul,
   volume = 104,
    pages = {340-371},
      doi = {10.1086/116242},
   adsurl = {http://adsabs.harvard.edu/abs/1992AJ....104..340L}
}

@INPROCEEDINGS{1992ASPC...25..297S,
   author = {{Stetson}, P.~B.},
    title = "{More Experiments with DAOPHOT II and WF/PC Images}",
booktitle = {Astronomical Data Analysis Software and Systems I},
     year = 1992,
   series = {Astronomical Society of the Pacific Conference Series},
   volume = 25,
   editor = {{Worrall}, D.~M. and {Biemesderfer}, C. and {Barnes}, J.},
    pages = {297},
   adsurl = {http://adsabs.harvard.edu/abs/1992ASPC...25..297S}
}

@ARTICLE{2004ApJ...608..797O,
   author = {{Ojha}, D.~K. and {Tamura}, M. and {Nakajima}, Y. and {Fukagawa}, M. and 
	{Sugitani}, K. and {Nagashima}, C. and {Nagayama}, T. and {Nagata}, T. and 
	{Sato}, S. and {Pickles}, A.~J. and {Ogura}, K.},
    title = "{Deep Near-Infrared Observations of the W3 Main Star-forming Region}",
  journal = {\apj},
   eprint = {astro-ph/0403139},
     year = 2004,
    month = jun,
   volume = 608,
    pages = {797-808},
      doi = {10.1086/420876},
   adsurl = {http://adsabs.harvard.edu/abs/2004ApJ...608..797O}
}

@ARTICLE{2014ApJ...791..131K,
   author = {{Koenig}, X.~P. and {Leisawitz}, D.~T.},
    title = "{A Classification Scheme for Young Stellar Objects Using the Wide-field Infrared Survey Explorer AllWISE Catalog: Revealing Low-density Star Formation in the Outer Galaxy}",
  journal = {\apj},
archivePrefix = "arXiv",
   eprint = {1407.2262},
     year = 2014,
    month = aug,
   volume = 791,
      eid = {131},
    pages = {131},
      doi = {10.1088/0004-637X/791/2/131},
   adsurl = {http://adsabs.harvard.edu/abs/2014ApJ...791..131K}
}

@ARTICLE{1998A&AS..133..387B,
   author = {{Balaguer-N{\'u}nez}, L. and {Tian}, K.~P. and {Zhao}, J.~L.
	},
    title = "{Determination of proper motions and membership of the open clusters NGC 1817 and NGC 1807}",
  journal = {\aaps},
     year = 1998,
    month = dec,
   volume = 133,
    pages = {387-394},
      doi = {10.1051/aas:1998324},
   adsurl = {http://adsabs.harvard.edu/abs/1998A\%26AS..133..387B}
}

@ARTICLE{2009A&A...493..959B,
   author = {{Bellini}, A. and {Piotto}, G. and {Bedin}, L.~R. and {Anderson}, J. and 
	{Platais}, I. and {Momany}, Y. and {Moretti}, A. and {Milone}, A.~P. and 
	{Ortolani}, S.},
    title = "{Ground-based CCD astrometry with wide field imagers. III. WFI@2.2m proper-motion catalog of the globular cluster {$\omega$} Centauri}",
  journal = {\aap},
archivePrefix = "arXiv",
   eprint = {0810.1914},
     year = 2009,
    month = jan,
   volume = 493,
    pages = {959-978},
      doi = {10.1051/0004-6361:200810880},
   adsurl = {http://adsabs.harvard.edu/abs/2009A\%26A...493..959B}
}

@ARTICLE{2012A&A...543A..87S,
   author = {{Sariya}, D.~P. and {Yadav}, R.~K.~S. and {Bellini}, A.},
    title = "{Proper motions and membership probabilities of stars in the region of globular cluster NGC 6809}",
  journal = {\aap},
archivePrefix = "arXiv",
   eprint = {1206.1157},
 primaryClass = "astro-ph.SR",
     year = 2012,
    month = jul,
   volume = 543,
      eid = {A87},
    pages = {A87},
      doi = {10.1051/0004-6361/201219306},
   adsurl = {http://adsabs.harvard.edu/abs/2012A\%26A...543A..87S}
}

@ARTICLE{2015A&A...584A..59S,
   author = {{Sariya}, D.~P. and {Yadav}, R.~K.~S.},
    title = "{Proper motions and membership probabilities of stars in the region of globular cluster NGC 6366}",
  journal = {\aap},
     year = 2015,
    month = dec,
   volume = 584,
      eid = {A59},
    pages = {A59},
      doi = {10.1051/0004-6361/201526688},
   adsurl = {http://adsabs.harvard.edu/abs/2015A\%26A...584A..59S}
}

@ARTICLE{2017AJ....153..134S,
   author = {{Sariya}, D.~P. and {Jiang}, I.-G. and {Yadav}, R.~K.~S.},
    title = "{A Proper Motions Study of the Globular Cluster NGC 3201}",
  journal = {\aj},
archivePrefix = "arXiv",
   eprint = {1701.04859},
 primaryClass = "astro-ph.SR",
     year = 2017,
    month = mar,
   volume = 153,
      eid = {134},
    pages = {134},
      doi = {10.3847/1538-3881/aa5be6},
   adsurl = {http://adsabs.harvard.edu/abs/2017AJ....153..134S}
}

@ARTICLE{2013MNRAS.430.3350Y,
   author = {{Yadav}, R.~K.~S. and {Sariya}, D.~P. and {Sagar}, R.},
    title = "{Proper motions and membership probabilities of stars in the region of open cluster NGC 3766}",
  journal = {\mnras},
archivePrefix = "arXiv",
   eprint = {1302.2558},
     year = 2013,
    month = apr,
   volume = 430,
    pages = {3350-3358},
      doi = {10.1093/mnras/stt136},
   adsurl = {http://adsabs.harvard.edu/abs/2013MNRAS.430.3350Y}
}

@ARTICLE{1989AJ.....98..227G,
   author = {{Girard}, T.~M. and {Grundy}, W.~M. and {Lopez}, C.~E. and {van Altena}, W.~F.
	},
    title = "{Relative proper motions and the stellar velocity dispersion of the open cluster M67}",
  journal = {\aj},
     year = 1989,
    month = jul,
   volume = 98,
    pages = {227-243},
      doi = {10.1086/115139},
   adsurl = {http://adsabs.harvard.edu/abs/1989AJ.....98..227G}
}

@ARTICLE{2017MNRAS.467.2943S,
   author = {{Sharma}, S. and {Pandey}, A.~K. and {Ojha}, D.~K. and {Bhatt}, H. and 
	{Ogura}, K. and {Kobayashi}, N. and {Yadav}, R. and {Pandey}, J.~C.
	},
    title = "{Stellar contents and star formation in the NGC 7538 region}",
  journal = {\mnras},
archivePrefix = "arXiv",
   eprint = {1701.00975},
 primaryClass = "astro-ph.SR",
     year = 2017,
    month = may,
   volume = 467,
    pages = {2943-2965},
      doi = {10.1093/mnras/stx014},
   adsurl = {http://adsabs.harvard.edu/abs/2017MNRAS.467.2943S}
}

@ARTICLE{2006AJ....132.1669S,
   author = {{Sharma}, S. and {Pandey}, A.~K. and {Ogura}, K. and {Mito}, H. and 
	{Tarusawa}, K. and {Sagar}, R.},
    title = "{Wide-Field CCD Photometry around Nine Open Clusters}",
  journal = {\aj},
   eprint = {astro-ph/0607538},
     year = 2006,
    month = oct,
   volume = 132,
    pages = {1669-1691},
      doi = {10.1086/507094},
   adsurl = {http://adsabs.harvard.edu/abs/2006AJ....132.1669S}
}

@ARTICLE{2005ApJ...632..397G,
   author = {{Gutermuth}, R.~A. and {Megeath}, S.~T. and {Pipher}, J.~L. and
        {Williams}, J.~P. and {Allen}, L.~E. and {Myers}, P.~C. and
        {Raines}, S.~N.},
    title = "{The Initial Configuration of Young Stellar Clusters: A K-Band Number Counts Analysis of the Surface Density of Stars}",
  journal = {\apj},
   eprint = {arXiv:astro-ph/0410750},
     year = 2005,
    month = oct,
   volume = 632,
    pages = {397-420},
      doi = {10.1086/432460},
   adsurl = {http://adsabs.harvard.edu/abs/2005ApJ...632..397G}
}

@ARTICLE{2007MNRAS.379.1599L,
   author = {{Lawrence}, A. and {Warren}, S.~J. and {Almaini}, O. and {Edge}, A.~C. and 
	{Hambly}, N.~C. and {Jameson}, R.~F. and {Lucas}, P. and {Casali}, M. and 
	{Adamson}, A. and {Dye}, S. and {Emerson}, J.~P. and {Foucaud}, S. and 
	{Hewett}, P. and {Hirst}, P. and {Hodgkin}, S.~T. and {Irwin}, M.~J. and 
	{Lodieu}, N. and {McMahon}, R.~G. and {Simpson}, C. and {Smail}, I. and 
	{Mortlock}, D. and {Folger}, M.},
    title = "{The UKIRT Infrared Deep Sky Survey (UKIDSS)}",
  journal = {\mnras},
   eprint = {astro-ph/0604426},
     year = 2007,
    month = aug,
   volume = 379,
    pages = {1599-1617},
      doi = {10.1111/j.1365-2966.2007.12040.x},
   adsurl = {http://adsabs.harvard.edu/abs/2007MNRAS.379.1599L}
}

@ARTICLE{2006AJ....131.1163S,
   author = {{Skrutskie}, M.~F. and {Cutri}, R.~M. and {Stiening}, R. and 
	{Weinberg}, M.~D. and {Schneider}, S. and {Carpenter}, J.~M. and 
	{Beichman}, C. and {Capps}, R. and {Chester}, T. and {Elias}, J. and 
	{Huchra}, J. and {Liebert}, J. and {Lonsdale}, C. and {Monet}, D.~G. and 
	{Price}, S. and {Seitzer}, P. and {Jarrett}, T. and {Kirkpatrick}, J.~D. and 
	{Gizis}, J.~E. and {Howard}, E. and {Evans}, T. and {Fowler}, J. and 
	{Fullmer}, L. and {Hurt}, R. and {Light}, R. and {Kopan}, E.~L. and 
	{Marsh}, K.~A. and {McCallon}, H.~L. and {Tam}, R. and {Van Dyk}, S. and 
	{Wheelock}, S.},
    title = "{The Two Micron All Sky Survey (2MASS)}",
  journal = {\aj},
     year = 2006,
    month = feb,
   volume = 131,
    pages = {1163-1183},
      doi = {10.1086/498708},
   adsurl = {http://adsabs.harvard.edu/abs/2006AJ....131.1163S}
}

@ARTICLE{2010AJ....140.1868W,
   author = {{Wright}, E.~L. and {Eisenhardt}, P.~R.~M. and {Mainzer}, A.~K. and 
	{Ressler}, M.~E. and {Cutri}, R.~M. and {Jarrett}, T. and {Kirkpatrick}, J.~D. and 
	{Padgett}, D. and {McMillan}, R.~S. and {Skrutskie}, M. and 
	{Stanford}, S.~A. and {Cohen}, M. and {Walker}, R.~G. and {Mather}, J.~C. and 
	{Leisawitz}, D. and {Gautier}, III, T.~N. and {McLean}, I. and 
	{Benford}, D. and {Lonsdale}, C.~J. and {Blain}, A. and {Mendez}, B. and 
	{Irace}, W.~R. and {Duval}, V. and {Liu}, F. and {Royer}, D. and 
	{Heinrichsen}, I. and {Howard}, J. and {Shannon}, M. and {Kendall}, M. and 
	{Walsh}, A.~L. and {Larsen}, M. and {Cardon}, J.~G. and {Schick}, S. and 
	{Schwalm}, M. and {Abid}, M. and {Fabinsky}, B. and {Naes}, L. and 
	{Tsai}, C.-W.},
    title = "{The Wide-field Infrared Survey Explorer (WISE): Mission Description and Initial On-orbit Performance}",
  journal = {\aj},
archivePrefix = "arXiv",
   eprint = {1008.0031},
 primaryClass = "astro-ph.IM",
     year = 2010,
    month = dec,
   volume = 140,
    pages = {1868-1881},
      doi = {10.1088/0004-6256/140/6/1868},
   adsurl = {http://adsabs.harvard.edu/abs/2010AJ....140.1868W}
}

@ARTICLE{2016AJ....151..126S,
   author = {{Sharma}, S. and {Pandey}, A.~K. and {Borissova}, J. and {Ojha}, D.~K. and
        {Ivanov}, V.~D. and {Ogura}, K. and {Kobayashi}, N. and {Kurtev}, R. and
        {Gopinathan}, M. and {Kesh Yadav}, R.},
    title = "{Structural Studies of Eight Bright Rimmed Clouds in the Southern Hemisphere}",
  journal = {\aj},
archivePrefix = "arXiv",
   eprint = {1602.07798},
 primaryClass = "astro-ph.SR",
     year = 2016,
    month = may,
   volume = 151,
      eid = {126},
    pages = {126},
      doi = {10.3847/0004-6256/151/5/126},
   adsurl = {https://ui.adsabs.harvard.edu/abs/2016AJ....151..126S}
}

@ARTICLE{2012ApJ...758L..28B,
       author = {{Bressert}, E. and {Ginsburg}, A. and {Bally}, J. and {Battersby}, C. and
         {Longmore}, S. and {Testi}, L.},
        title = "{How to Find Young Massive Cluster Progenitors}",
      journal = {\apjl},
         year = "2012",
        month = "Oct",
       volume = {758},
       number = {2},
          eid = {L28},
        pages = {L28},
          doi = {10.1088/2041-8205/758/2/L28},
archivePrefix = {arXiv},
       eprint = {1208.3472},
 primaryClass = {astro-ph.GA},
       adsurl = {https://ui.adsabs.harvard.edu/abs/2012ApJ...758L..28B}
}

@ARTICLE{2009A&A...498..837M,
       author = {{Marcolino}, W.~L.~F. and {Bouret}, J. -C. and {Martins}, F. and
         {Hillier}, D.~J. and {Lanz}, T. and {Escolano}, C.},
        title = "{Analysis of Galactic late-type O dwarfs: more constraints on the weak wind problem}",
      journal = {\aap},
         year = "2009",
        month = "May",
       volume = {498},
       number = {3},
        pages = {837-852},
          doi = {10.1051/0004-6361/200811289},
archivePrefix = {arXiv},
       eprint = {0902.1833},
 primaryClass = {astro-ph.SR},
       adsurl = {https://ui.adsabs.harvard.edu/abs/2009A&A...498..837M}
}

@ARTICLE{2017ApJ...834...22D,
       author = {{Dewangan}, L.~K. and {Ojha}, D.~K. and {Zinchenko}, I. and
         {Janardhan}, P. and {Luna}, A.},
        title = "{Multiwavelength Study of the Star Formation in the S237 H II Region}",
      journal = {\apj},
         year = "2017",
        month = "Jan",
       volume = {834},
       number = {1},
          eid = {22},
        pages = {22},
          doi = {10.3847/1538-4357/834/1/22},
archivePrefix = {arXiv},
       eprint = {1610.08428},
 primaryClass = {astro-ph.GA},
       adsurl = {https://ui.adsabs.harvard.edu/abs/2017ApJ...834...22D}
}

@ARTICLE{2010PASP..122..314M,
       author = {{Molinari}, S. and {Swinyard}, B. and {Bally}, J. and {Barlow}, M. and
         {Bernard}, J. -P. and {Martin}, P. and {Moore}, T. and
         {Noriega-Crespo}, A. and {Plume}, R. and {Testi}, L.},
        title = "{Hi-GAL: The Herschel Infrared Galactic Plane Survey}",
      journal = {\pasp},
         year = "2010",
        month = "Mar",
       volume = {122},
       number = {889},
        pages = {314},
          doi = {10.1086/651314},
archivePrefix = {arXiv},
       eprint = {1001.2106},
 primaryClass = {astro-ph.GA},
       adsurl = {https://ui.adsabs.harvard.edu/abs/2010PASP..122..314M}
}

@ARTICLE{2003PASP..115..953B,
       author = {{Benjamin}, Robert A. and {Churchwell}, E. and {Babler}, Brian L. and
         {Bania}, T.~M. and {Clemens}, Dan P. and {Cohen}, Martin and
         {Dickey}, John M. and {Indebetouw}, R{\'e}my and {Jackson}, James M. and
         {Kobulnicky}, Henry A.},
        title = "{GLIMPSE. I. An SIRTF Legacy Project to Map the Inner Galaxy}",
      journal = {\pasp},
         year = "2003",
        month = "Aug",
       volume = {115},
       number = {810},
        pages = {953-964},
          doi = {10.1086/376696},
archivePrefix = {arXiv},
       eprint = {astro-ph/0306274},
 primaryClass = {astro-ph},
       adsurl = {https://ui.adsabs.harvard.edu/abs/2003PASP..115..953B}
}

@INPROCEEDINGS{2004ASPC..317...79B,
       author = {{Brunt}, C.},
        title = "{A Large Scale Survey of $^{12}$CO (J=1rightarrow 0) and $^{13}$CO (J=1rightarrow 0) at High Latitudes in the Outer Galaxy}",
    booktitle = {Milky Way Surveys: The Structure and Evolution of our Galaxy},
         year = "2004",
       editor = {{Clemens}, Dan and {Shah}, Ronak and {Brainerd}, Teresa},
       series = {Astronomical Society of the Pacific Conference Series},
       volume = {317},
        month = "Dec",
        pages = {79},
       adsurl = {https://ui.adsabs.harvard.edu/abs/2004ASPC..317...79B}
}

@ARTICLE{1998AJ....115.1693C,
       author = {{Condon}, J.~J. and {Cotton}, W.~D. and {Greisen}, E.~W. and
         {Yin}, Q.~F. and {Perley}, R.~A. and {Taylor}, G.~B. and
         {Broderick}, J.~J.},
        title = "{The NRAO VLA Sky Survey}",
      journal = {\aj},
         year = "1998",
        month = "May",
       volume = {115},
       number = {5},
        pages = {1693-1716},
          doi = {10.1086/300337},
       adsurl = {https://ui.adsabs.harvard.edu/abs/1998AJ....115.1693C}
}

@ARTICLE{1996ApJ...463..630H,
       author = {{Heyer}, Mark H. and {Carpenter}, John M. and {Ladd}, E.~F.},
        title = "{Giant Molecular Cloud Complexes with Optical H II Regions: 12CO and 13CO Observations and Global Cloud Properties}",
      journal = {\apj},
         year = "1996",
        month = "Jun",
       volume = {463},
        pages = {630},
          doi = {10.1086/177277},
       adsurl = {https://ui.adsabs.harvard.edu/abs/1996ApJ...463..630H}
}

@ARTICLE{1998ApJS..115..241H,
       author = {{Heyer}, Mark H. and {Brunt}, Christopher and {Snell}, Ronald L. and
         {Howe}, John E. and {Schloerb}, F. Peter and {Carpenter}, John M.},
        title = "{The Five College Radio Astronomy Observatory CO Survey of the Outer Galaxy}",
      journal = {\apjs},
         year = "1998",
        month = "Apr",
       volume = {115},
       number = {2},
        pages = {241-258},
          doi = {10.1086/313086},
       adsurl = {https://ui.adsabs.harvard.edu/abs/1998ApJS..115..241H}
}

@ARTICLE{2009ApJ...700.1609M,
       author = {{Myers}, Philip C.},
        title = "{Filamentary Structure of Star-forming Complexes}",
      journal = {\apj},
         year = "2009",
        month = "Aug",
       volume = {700},
       number = {2},
        pages = {1609-1625},
          doi = {10.1088/0004-637X/700/2/1609},
archivePrefix = {arXiv},
       eprint = {0906.2005},
 primaryClass = {astro-ph.GA},
       adsurl = {https://ui.adsabs.harvard.edu/abs/2009ApJ...700.1609M}
}

@ARTICLE{1973AJ.....78..929P,
       author = {{Panagia}, Nino},
        title = "{Some Physical parameters of early-type stars}",
      journal = {\aj},
         year = "1973",
        month = "Nov",
       volume = {78},
        pages = {929-934},
          doi = {10.1086/111498},
       adsurl = {https://ui.adsabs.harvard.edu/abs/1973AJ.....78..929P}
}

@ARTICLE{2017ApJ...836...98J,
       author = {{Jose}, Jessy and {Herczeg}, Gregory J. and {Samal}, Manash R. and
         {Fang}, Qiliang and {Panwar}, Neelam},
        title = "{The Low-mass Population in the Young Cluster Stock 8: Stellar Properties and Initial Mass Function}",
      journal = {\apj},
         year = "2017",
        month = "Feb",
       volume = {836},
       number = {1},
          eid = {98},
        pages = {98},
          doi = {10.3847/1538-4357/836/1/98},
archivePrefix = {arXiv},
       eprint = {1612.00697},
 primaryClass = {astro-ph.SR},
	       adsurl = {https://ui.adsabs.harvard.edu/abs/2017ApJ...836...98J}
}

@ARTICLE{2010A&A...523A...6D,
	       author = {{Deharveng}, L. and {Schuller}, F. and {Anderson}, L.~D. and
		 {Zavagno}, A. and {Wyrowski}, F. and {Menten}, K.~M. and
		 {Bronfman}, L. and {Testi}, L. and {Walmsley}, C.~M. and {Wienen}, M.},
		title = "{A gallery of bubbles. The nature of the bubbles observed by Spitzer and what ATLASGAL tells us about the surrounding neutral material}",
	      journal = {\aap},
		 year = "2010",
		month = "Nov",
	       volume = {523},
		  eid = {A6},
		pages = {A6},
		  doi = {10.1051/0004-6361/201014422},
	archivePrefix = {arXiv},
	       eprint = {1008.0926},
	 primaryClass = {astro-ph.GA},
	       adsurl = {https://ui.adsabs.harvard.edu/abs/2010A&A...523A...6D}
}

@ARTICLE{2019MNRAS.485.5666P,
       author = {{Pastorelli}, Giada and {Marigo}, Paola and {Girardi}, L{\'e}o and
         {Chen}, Yang and {Rubele}, Stefano and {Trabucchi}, Michele and
         {Aringer}, Bernhard and {Bladh}, Sara and {Bressan}, Alessandro and
         {Montalb{\'a}n}, Josefina and {Boyer}, Martha L. and
         {Dalcanton}, Julianne J. and {Eriksson}, Kjell and
         {Groenewegen}, Martin A.~T. and {H{\"o}fner}, Susanne and
         {Lebzelter}, Thomas and {Nanni}, Ambra and {Rosenfield}, Philip and
         {Wood}, Peter R. and {Cioni}, Maria-Rosa L.},
        title = "{Constraining the thermally pulsing asymptotic giant branch phase with resolved stellar populations in the Small Magellanic Cloud}",
      journal = {\mnras},
         year = 2019,
        month = jun,
       volume = {485},
       number = {4},
        pages = {5666-5692},
          doi = {10.1093/mnras/stz725},
archivePrefix = {arXiv},
       eprint = {1903.04499},
 primaryClass = {astro-ph.SR},
       adsurl = {https://ui.adsabs.harvard.edu/abs/2019MNRAS.485.5666P}
}

@ARTICLE{2020ApJ...891...81P,
       author = {{Pandey}, Rakesh and {Sharma}, Saurabh and {Panwar}, Neelam and
         {Dewangan}, Lokesh K. and {Ojha}, Devendra K. and {Bisen}, D.~P. and
         {Sinha}, Tirthendu and {Ghosh}, Arpan and {Pandey}, Anil K.},
        title = "{Stellar Cores in the Sh 2-305 H II Region}",
      journal = {\apj},
         year = 2020,
        month = mar,
       volume = {891},
       number = {1},
          eid = {81},
        pages = {81},
          doi = {10.3847/1538-4357/ab6dc7},
archivePrefix = {arXiv},
       eprint = {2001.04722},
 primaryClass = {astro-ph.SR},
       adsurl = {https://ui.adsabs.harvard.edu/abs/2020ApJ...891...81P}
}

@ARTICLE{2013ApJS..208....9P,
       author = {{Pecaut}, Mark J. and {Mamajek}, Eric E.},
        title = "{Intrinsic Colors, Temperatures, and Bolometric Corrections of Pre-main-sequence Stars}",
      journal = {\apjs},
         year = 2013,
        month = sep,
       volume = {208},
       number = {1},
          eid = {9},
        pages = {9},
          doi = {10.1088/0067-0049/208/1/9},
archivePrefix = {arXiv},
       eprint = {1307.2657},
 primaryClass = {astro-ph.SR},
       adsurl = {https://ui.adsabs.harvard.edu/abs/2013ApJS..208....9P}
}

@ARTICLE{2020MNRAS.498.2309S,
       author = {{Sharma}, Saurabh and {Ghosh}, Arpan and {Ojha}, D.~K. and {Pandey}, R. and {Sinha}, T. and {Pandey}, A.~K. and {Ghosh}, S.~K. and {Panwar}, N. and {Pandey}, S.~B.},
        title = "{The disintegrating old open cluster Czernik 3}",
      journal = {\mnras},
         year = 2020,
        month = oct,
       volume = {498},
       number = {2},
        pages = {2309-2322},
          doi = {10.1093/mnras/staa2412},
archivePrefix = {arXiv},
       eprint = {2008.04102},
 primaryClass = {astro-ph.GA},
       adsurl = {https://ui.adsabs.harvard.edu/abs/2020MNRAS.498.2309S}
}

@ARTICLE{2021A&A...649A...1G,
       author = {{Gaia Collaboration} and {Brown}, A.~G.~A. and {Vallenari}, A. and {Prusti}, T. and {de Bruijne}, J.~H.~J. and {Babusiaux}, C. and {Biermann}, M. and {Creevey}, O.~L. and {Evans}, D.~W. and {Eyer}, L. and {Hutton}, A. and {Jansen}, F. and {Jordi}, C. and {Klioner}, S.~A. and {Lammers}, U. and {Lindegren}, L. and {Luri}, X. and {Mignard}, F. and {Panem}, C. and {Pourbaix}, D. and {Randich}, S. and {Sartoretti}, P. and {Soubiran}, C. and {Walton}, N.~A. and {Arenou}, F. and {Bailer-Jones}, C.~A.~L. and {Bastian}, U. and {Cropper}, M. and {Drimmel}, R. and {Katz}, D. and {Lattanzi}, M.~G. and {van Leeuwen}, F. and {Bakker}, J. and {Cacciari}, C. and {Casta{\~n}eda}, J. and {De Angeli}, F. and {Ducourant}, C. and {Fabricius}, C. and {Fouesneau}, M. and {Fr{\'e}mat}, Y. and {Guerra}, R. and {Guerrier}, A. and {Guiraud}, J. and {Jean-Antoine Piccolo}, A. and {Masana}, E. and {Messineo}, R. and {Mowlavi}, N. and {Nicolas}, C. and {Nienartowicz}, K. and {Pailler}, F. and {Panuzzo}, P. and {Riclet}, F. and {Roux}, W. and {Seabroke}, G.~M. and {Sordo}, R. and {Tanga}, P. and {Th{\'e}venin}, F. and {Gracia-Abril}, G. and {Portell}, J. and {Teyssier}, D. and {Altmann}, M. and {Andrae}, R. and {Bellas-Velidis}, I. and {Benson}, K. and {Berthier}, J. and {Blomme}, R. and {Brugaletta}, E. and {Burgess}, P.~W. and {Busso}, G. and {Carry}, B. and {Cellino}, A. and {Cheek}, N. and {Clementini}, G. and {Damerdji}, Y. and {Davidson}, M. and {Delchambre}, L. and {Dell'Oro}, A. and {Fern{\'a}ndez-Hern{\'a}ndez}, J. and {Galluccio}, L. and {Garc{\'\i}a-Lario}, P. and {Garcia-Reinaldos}, M. and {Gonz{\'a}lez-N{\'u}{\~n}ez}, J. and {Gosset}, E. and {Haigron}, R. and {Halbwachs}, J. -L. and {Hambly}, N.~C. and {Harrison}, D.~L. and {Hatzidimitriou}, D. and {Heiter}, U. and {Hern{\'a}ndez}, J. and {Hestroffer}, D. and {Hodgkin}, S.~T. and {Holl}, B. and {Jan{\ss}en}, K. and {Jevardat de Fombelle}, G. and {Jordan}, S. and {Krone-Martins}, A. and {Lanzafame}, A.~C. and {L{\"o}ffler}, W. and {Lorca}, A. and {Manteiga}, M. and {Marchal}, O. and {Marrese}, P.~M. and {Moitinho}, A. and {Mora}, A. and {Muinonen}, K. and {Osborne}, P. and {Pancino}, E. and {Pauwels}, T. and {Petit}, J. -M. and {Recio-Blanco}, A. and {Richards}, P.~J. and {Riello}, M. and {Rimoldini}, L. and {Robin}, A.~C. and {Roegiers}, T. and {Rybizki}, J. and {Sarro}, L.~M. and {Siopis}, C. and {Smith}, M. and {Sozzetti}, A. and {Ulla}, A. and {Utrilla}, E. and {van Leeuwen}, M. and {van Reeven}, W. and {Abbas}, U. and {Abreu Aramburu}, A. and {Accart}, S. and {Aerts}, C. and {Aguado}, J.~J. and {Ajaj}, M. and {Altavilla}, G. and {{\'A}lvarez}, M.~A. and {{\'A}lvarez Cid-Fuentes}, J. and {Alves}, J. and {Anderson}, R.~I. and {Anglada Varela}, E. and {Antoja}, T. and {Audard}, M. and {Baines}, D. and {Baker}, S.~G. and {Balaguer-N{\'u}{\~n}ez}, L. and {Balbinot}, E. and {Balog}, Z. and {Barache}, C. and {Barbato}, D. and {Barros}, M. and {Barstow}, M.~A. and {Bartolom{\'e}}, S. and {Bassilana}, J. -L. and {Bauchet}, N. and {Baudesson-Stella}, A. and {Becciani}, U. and {Bellazzini}, M. and {Bernet}, M. and {Bertone}, S. and {Bianchi}, L. and {Blanco-Cuaresma}, S. and {Boch}, T. and {Bombrun}, A. and {Bossini}, D. and {Bouquillon}, S. and {Bragaglia}, A. and {Bramante}, L. and {Breedt}, E. and {Bressan}, A. and {Brouillet}, N. and {Bucciarelli}, B. and {Burlacu}, A. and {Busonero}, D. and {Butkevich}, A.~G. and {Buzzi}, R. and {Caffau}, E. and {Cancelliere}, R. and {C{\'a}novas}, H. and {Cantat-Gaudin}, T. and {Carballo}, R. and {Carlucci}, T. and {Carnerero}, M.~I. and {Carrasco}, J.~M. and {Casamiquela}, L. and {Castellani}, M. and {Castro-Ginard}, A. and {Castro Sampol}, P. and {Chaoul}, L. and {Charlot}, P. and {Chemin}, L. and {Chiavassa}, A. and {Cioni}, M. -R.~L. and {Comoretto}, G. and {Cooper}, W.~J. and {Cornez}, T. and {Cowell}, S. and {Crifo}, F. and {Crosta}, M. and {Crowley}, C. and {Dafonte}, C. and {Dapergolas}, A. and {David}, M. and {David}, P. and {de Laverny}, P. and {De Luise}, F. and {De March}, R. and {De Ridder}, J. and {de Souza}, R. and {de Teodoro}, P. and {de Torres}, A. and {del Peloso}, E.~F. and {del Pozo}, E. and {Delbo}, M. and {Delgado}, A. and {Delgado}, H.~E. and {Delisle}, J. -B. and {Di Matteo}, P. and {Diakite}, S. and {Diener}, C. and {Distefano}, E. and {Dolding}, C. and {Eappachen}, D. and {Edvardsson}, B. and {Enke}, H. and {Esquej}, P. and {Fabre}, C. and {Fabrizio}, M. and {Faigler}, S. and {Fedorets}, G. and {Fernique}, P. and {Fienga}, A. and {Figueras}, F. and {Fouron}, C. and {Fragkoudi}, F. and {Fraile}, E. and {Franke}, F. and {Gai}, M. and {Garabato}, D. and {Garcia-Gutierrez}, A. and {Garc{\'\i}a-Torres}, M. and {Garofalo}, A. and {Gavras}, P. and {Gerlach}, E. and {Geyer}, R. and {Giacobbe}, P. and {Gilmore}, G. and {Girona}, S. and {Giuffrida}, G. and {Gomel}, R. and {Gomez}, A. and {Gonzalez-Santamaria}, I. and {Gonz{\'a}lez-Vidal}, J.~J. and {Granvik}, M. and {Guti{\'e}rrez-S{\'a}nchez}, R. and {Guy}, L.~P. and {Hauser}, M. and {Haywood}, M. and {Helmi}, A. and {Hidalgo}, S.~L. and {Hilger}, T. and {H{\l}adczuk}, N. and {Hobbs}, D. and {Holland}, G. and {Huckle}, H.~E. and {Jasniewicz}, G. and {Jonker}, P.~G. and {Juaristi Campillo}, J. and {Julbe}, F. and {Karbevska}, L. and {Kervella}, P. and {Khanna}, S. and {Kochoska}, A. and {Kontizas}, M. and {Kordopatis}, G. and {Korn}, A.~J. and {Kostrzewa-Rutkowska}, Z. and {Kruszy{\'n}ska}, K. and {Lambert}, S. and {Lanza}, A.~F. and {Lasne}, Y. and {Le Campion}, J. -F. and {Le Fustec}, Y. and {Lebreton}, Y. and {Lebzelter}, T. and {Leccia}, S. and {Leclerc}, N. and {Lecoeur-Taibi}, I. and {Liao}, S. and {Licata}, E. and {Lindstr{\o}m}, E.~P. and {Lister}, T.~A. and {Livanou}, E. and {Lobel}, A. and {Madrero Pardo}, P. and {Managau}, S. and {Mann}, R.~G. and {Marchant}, J.~M. and {Marconi}, M. and {Marcos Santos}, M.~M.~S. and {Marinoni}, S. and {Marocco}, F. and {Marshall}, D.~J. and {Martin Polo}, L. and {Mart{\'\i}n-Fleitas}, J.~M. and {Masip}, A. and {Massari}, D. and {Mastrobuono-Battisti}, A. and {Mazeh}, T. and {McMillan}, P.~J. and {Messina}, S. and {Michalik}, D. and {Millar}, N.~R. and {Mints}, A. and {Molina}, D. and {Molinaro}, R. and {Moln{\'a}r}, L. and {Montegriffo}, P. and {Mor}, R. and {Morbidelli}, R. and {Morel}, T. and {Morris}, D. and {Mulone}, A.~F. and {Munoz}, D. and {Muraveva}, T. and {Murphy}, C.~P. and {Musella}, I. and {Noval}, L. and {Ord{\'e}novic}, C. and {Orr{\`u}}, G. and {Osinde}, J. and {Pagani}, C. and {Pagano}, I. and {Palaversa}, L. and {Palicio}, P.~A. and {Panahi}, A. and {Pawlak}, M. and {Pe{\~n}alosa Esteller}, X. and {Penttil{\"a}}, A. and {Piersimoni}, A.~M. and {Pineau}, F. -X. and {Plachy}, E. and {Plum}, G. and {Poggio}, E. and {Poretti}, E. and {Poujoulet}, E. and {Pr{\v{s}}a}, A. and {Pulone}, L. and {Racero}, E. and {Ragaini}, S. and {Rainer}, M. and {Raiteri}, C.~M. and {Rambaux}, N. and {Ramos}, P. and {Ramos-Lerate}, M. and {Re Fiorentin}, P. and {Regibo}, S. and {Reyl{\'e}}, C. and {Ripepi}, V. and {Riva}, A. and {Rixon}, G. and {Robichon}, N. and {Robin}, C. and {Roelens}, M. and {Rohrbasser}, L. and {Romero-G{\'o}mez}, M. and {Rowell}, N. and {Royer}, F. and {Rybicki}, K.~A. and {Sadowski}, G. and {Sagrist{\`a} Sell{\'e}s}, A. and {Sahlmann}, J. and {Salgado}, J. and {Salguero}, E. and {Samaras}, N. and {Sanchez Gimenez}, V. and {Sanna}, N. and {Santove{\~n}a}, R. and {Sarasso}, M. and {Schultheis}, M. and {Sciacca}, E. and {Segol}, M. and {Segovia}, J.~C. and {S{\'e}gransan}, D. and {Semeux}, D. and {Shahaf}, S. and {Siddiqui}, H.~I. and {Siebert}, A. and {Siltala}, L. and {Slezak}, E. and {Smart}, R.~L. and {Solano}, E. and {Solitro}, F. and {Souami}, D. and {Souchay}, J. and {Spagna}, A. and {Spoto}, F. and {Steele}, I.~A. and {Steidelm{\"u}ller}, H. and {Stephenson}, C.~A. and {S{\"u}veges}, M. and {Szabados}, L. and {Szegedi-Elek}, E. and {Taris}, F. and {Tauran}, G. and {Taylor}, M.~B. and {Teixeira}, R. and {Thuillot}, W. and {Tonello}, N. and {Torra}, F. and {Torra}, J. and {Turon}, C. and {Unger}, N. and {Vaillant}, M. and {van Dillen}, E. and {Vanel}, O. and {Vecchiato}, A. and {Viala}, Y. and {Vicente}, D. and {Voutsinas}, S. and {Weiler}, M. and {Wevers}, T. and {Wyrzykowski}, {\L}. and {Yoldas}, A. and {Yvard}, P. and {Zhao}, H. and {Zorec}, J. and {Zucker}, S. and {Zurbach}, C. and {Zwitter}, T.},
        title = "{Gaia Early Data Release 3. Summary of the contents and survey properties}",
      journal = {\aap},
         year = 2021,
        month = may,
       volume = {649},
          eid = {A1},
        pages = {A1},
          doi = {10.1051/0004-6361/202039657},
archivePrefix = {arXiv},
       eprint = {2012.01533},
 primaryClass = {astro-ph.GA},
       adsurl = {https://ui.adsabs.harvard.edu/abs/2021A&A...649A...1G}
}

@ARTICLE{2021AJ....161..147B,
       author = {{Bailer-Jones}, C.~A.~L. and {Rybizki}, J. and {Fouesneau}, M. and {Demleitner}, M. and {Andrae}, R.},
        title = "{Estimating Distances from Parallaxes. V. Geometric and Photogeometric Distances to 1.47 Billion Stars in Gaia Early Data Release 3}",
      journal = {\aj},
         year = 2021,
        month = mar,
       volume = {161},
       number = {3},
          eid = {147},
        pages = {147},
          doi = {10.3847/1538-3881/abd806},
archivePrefix = {arXiv},
       eprint = {2012.05220},
 primaryClass = {astro-ph.SR},
       adsurl = {https://ui.adsabs.harvard.edu/abs/2021AJ....161..147B}
}

@ARTICLE{2016AJ....151..115Y,
       author = {{Yasui}, Chikako and {Kobayashi}, Naoto and {Saito}, Masao and {Izumi}, Natsuko},
        title = "{Low-metallicity Young Clusters in the Outer Galaxy. II. Sh 2-208}",
      journal = {\aj},
         year = 2016,
        month = may,
       volume = {151},
       number = {5},
          eid = {115},
        pages = {115},
          doi = {10.3847/0004-6256/151/5/115},
archivePrefix = {arXiv},
       eprint = {1604.00144},
 primaryClass = {astro-ph.GA},
       adsurl = {https://ui.adsabs.harvard.edu/abs/2016AJ....151..115Y}
}

@ARTICLE{1979A&AS...38..197M,
       author = {{Moffat}, A.~F.~J. and {Fitzgerald}, M.~P. and {Jackson}, P.~D.},
        title = "{The rotation and structure of the Galaxy beyond the solar circle. I. Photometry and spectroscopy of 276 stars in 45 H II regions and other young stellar groups toward the galactic anticentre.}",
      journal = {\aaps},
         year = 1979,
        month = nov,
       volume = {38},
        pages = {197-225},
       adsurl = {https://ui.adsabs.harvard.edu/abs/1979A&AS...38..197M}
}

@ARTICLE{2003A&A...404..223B,
       author = {{Bica}, E. and {Dutra}, C.~M. and {Soares}, J. and {Barbuy}, B.},
        title = "{New infrared star clusters in the Northern and Equatorial Milky Way with 2MASS}",
      journal = {\aap},
         year = 2003,
        month = jun,
       volume = {404},
        pages = {223-232},
          doi = {10.1051/0004-6361:20030486},
archivePrefix = {arXiv},
       eprint = {astro-ph/0304379},
 primaryClass = {astro-ph},
       adsurl = {https://ui.adsabs.harvard.edu/abs/2003A&A...404..223B}
}

@ARTICLE{1985A&AS...61..537L,
       author = {{Lahulla}, J.~F.},
        title = "{UBVRI photometry of stars in several HII regions.}",
      journal = {\aaps},
         year = 1985,
        month = sep,
       volume = {61},
        pages = {537-545},
       adsurl = {https://ui.adsabs.harvard.edu/abs/1985A&AS...61..537L}
}

@ARTICLE{2015ApJS..221...26A,
       author = {{Anderson}, L.~D. and {Armentrout}, W.~P. and {Johnstone}, B.~M. and {Bania}, T.~M. and {Balser}, Dana S. and {Wenger}, Trey V. and {Cunningham}, V.},
        title = "{Finding Distant Galactic HII Regions}",
      journal = {\apjs},
         year = 2015,
        month = dec,
       volume = {221},
       number = {2},
          eid = {26},
        pages = {26},
          doi = {10.1088/0067-0049/221/2/26},
archivePrefix = {arXiv},
       eprint = {1510.07347},
 primaryClass = {astro-ph.GA},
       adsurl = {https://ui.adsabs.harvard.edu/abs/2015ApJS..221...26A}
}

@ARTICLE{1959ApJS....4..257S,
       author = {{Sharpless}, Stewart},
        title = "{A Catalogue of H II Regions.}",
      journal = {\apjs},
         year = 1959,
        month = dec,
       volume = {4},
        pages = {257},
          doi = {10.1086/190049},
       adsurl = {https://ui.adsabs.harvard.edu/abs/1959ApJS....4..257S}
}

@ARTICLE{2018MNRAS.473..849D,
       author = {{Dib}, Sami and {Schmeja}, Stefan and {Parker}, Richard J.},
        title = "{Structure and mass segregation in Galactic stellar clusters}",
      journal = {\mnras},
         year = 2018,
        month = jan,
       volume = {473},
       number = {1},
        pages = {849-859},
          doi = {10.1093/mnras/stx2413},
archivePrefix = {arXiv},
       eprint = {1707.00744},
 primaryClass = {astro-ph.GA},
       adsurl = {https://ui.adsabs.harvard.edu/abs/2018MNRAS.473..849D}
}

@ARTICLE{2022ApJ...926...25P,
       author = {{Pandey}, Rakesh and {Sharma}, Saurabh and {Dewangan}, Lokesh K. and {Ojha}, Devendra K. and {Panwar}, Neelam and {Das}, Swagat and {Bisen}, D.~P. and {Ghosh}, Arpan and {Sinha}, Tirthendu},
        title = "{Sh 2-301: A Blistered H II Region Undergoing Star Formation}",
      journal = {\apj},
         year = 2022,
        month = feb,
       volume = {926},
       number = {1},
          eid = {25},
        pages = {25},
          doi = {10.3847/1538-4357/ac41c3},
archivePrefix = {arXiv},
       eprint = {2112.01539},
 primaryClass = {astro-ph.GA},
       adsurl = {https://ui.adsabs.harvard.edu/abs/2022ApJ...926...25P}
}

@ARTICLE{2020ApJ...896...29K,
       author = {{Kaur}, Harmeen and {Sharma}, Saurabh and {Dewangan}, Lokesh K. and {Ojha}, Devendra K. and {Durgapal}, Alok and {Panwar}, Neelam},
        title = "{Unveiling the Physical Conditions in NGC 6910}",
      journal = {\apj},
         year = 2020,
        month = jun,
       volume = {896},
       number = {1},
          eid = {29},
        pages = {29},
          doi = {10.3847/1538-4357/ab9122},
archivePrefix = {arXiv},
       eprint = {2004.10500},
 primaryClass = {astro-ph.SR},
       adsurl = {https://ui.adsabs.harvard.edu/abs/2020ApJ...896...29K}
}

@ARTICLE{2009MNRAS.395.1449A,
       author = {{Allison}, Richard J. and {Goodwin}, Simon P. and {Parker}, Richard J. and {Portegies Zwart}, Simon F. and {de Grijs}, Richard and {Kouwenhoven}, M.~B.~N.},
        title = "{Using the minimum spanning tree to trace mass segregation}",
      journal = {\mnras},
         year = 2009,
        month = may,
       volume = {395},
       number = {3},
        pages = {1449-1454},
          doi = {10.1111/j.1365-2966.2009.14508.x},
archivePrefix = {arXiv},
       eprint = {0901.2047},
 primaryClass = {astro-ph.GA},
       adsurl = {https://ui.adsabs.harvard.edu/abs/2009MNRAS.395.1449A}
}

@ARTICLE{2006A&A...449..151S,
       author = {{Schmeja}, S. and {Klessen}, R.~S.},
        title = "{Evolving structures of star-forming clusters}",
      journal = {\aap},
         year = 2006,
        month = apr,
       volume = {449},
       number = {1},
        pages = {151-159},
          doi = {10.1051/0004-6361:20054464},
archivePrefix = {arXiv},
       eprint = {astro-ph/0511448},
 primaryClass = {astro-ph},
       adsurl = {https://ui.adsabs.harvard.edu/abs/2006A&A...449..151S}
}

@misc{https://doi.org/10.26131/irsa3,
  doi = {10.26131/IRSA3},
  url = {https://catcopy.ipac.caltech.edu/dois/doi.php?id=10.26131/IRSA3},
  author = {{Capak; Peter}},
  title = {Spitzer Enhanced Imaging Products (SEIP) Source List},
  publisher = {IPAC},
  year = {2019}
}

@misc{https://doi.org/10.26131/irsa214,
  doi = {10.26131/IRSA214},
  url = {https://catcopy.ipac.caltech.edu/dois/doi.php?id=10.26131/IRSA214},
  author = {{GLIMPSE Team}},
  title = {GLIMPSE 360 Catalog},
  publisher = {IPAC},
  year = {2020}
}

@misc{https://doi.org/10.26131/irsa2,
  doi = {10.26131/IRSA2},
  url = {https://catcopy.ipac.caltech.edu/dois/doi.php?id=10.26131/IRSA2},
  author = {{Skrutskie},  M. F. and {Cutri},  R. M. and {Stiening,  R.; Weinberg,  M. D.; Schneider,  S.; Carpenter,  J. M.; Beichman,  C.; Capps,  R.; Chester,  T.; Elias,  J.; Huchra,  J.; Liebert,  J.; Lonsdale,  C.; Monet,  D. G.; Price,  S.; Seitzer,  P.; Jarrett,  T.; Kirkpatrick,  J. D.; Gizis,  J. E.; Howard,  E.; Evans,  T.; Fowler,  J.; Fullmer,  L.; Hurt,  R.; Light,  R.; Kopan,  E. L.; Marsh,  K. A.; McCallon,  H. L.; Tam,  R.; Van Dyk,  S.; Wheelock,  S.}},
  title = {2MASS All-Sky Point Source Catalog},
  publisher = {IPAC},
  year = {2003}
}

@misc{https://doi.org/10.26131/irsa1,
  doi = {10.26131/IRSA1},
  url = {https://catcopy.ipac.caltech.edu/dois/doi.php?id=10.26131/IRSA1},
  author = {{Wright},  Edward L. and {Eisenhardt},  Peter R. M. and {Mainzer,  Amy K.; Ressler,  Michael E.; Cutri,  Roc M.; Jarrett,  Thomas; Kirkpatrick,  J. Davy; Padgett,  Deborah; McMillan,  Robert S.; Skrutskie,  Michael; Stanford,  S. A.; Cohen,  Martin; Walker,  Russell G.; Mather,  John C.; Leisawitz,  David; Gautier,  Thomas N.,  III; McLean,  Ian; Benford,  Dominic; Lonsdale,  Carol J.; Blain,  Andrew; Mendez,  Bryan; Irace,  William R.; Duval,  Valerie; Liu,  Fengchuan; Royer,  Don; Heinrichsen,  Ingolf; Howard,  Joan; Shannon,  Mark; Kendall,  Martha; Walsh,  Amy L.; Larsen,  Mark; Cardon,  Joel G.; Schick,  Scott; Schwalm,  Mark; Abid,  Mohamed; Fabinsky,  Beth; Naes,  Larry; Tsai,  ChaoWei}},
  title = {AllWISE Source Catalog},
  publisher = {IPAC},
  year = {2019}
}

@INPROCEEDINGS{2015AAS...22533616H,
       author = {{Henden}, Arne A. and {Levine}, Stephen and {Terrell}, Dirk and {Welch}, Douglas L.},
        title = "{APASS - The Latest Data Release}",
    booktitle = {American Astronomical Society Meeting Abstracts \#225},
         year = 2015,
       series = {American Astronomical Society Meeting Abstracts},
       volume = {225},
        month = jan,
          eid = {336.16},
        pages = {336.16},
       adsurl = {https://ui.adsabs.harvard.edu/abs/2015AAS...22533616H}
}

@ARTICLE{2023A&A...673A.114H,
       author = {{Hunt}, Emily L. and {Reffert}, Sabine},
        title = "{Improving the open cluster census. II. An all-sky cluster catalogue with Gaia DR3}",
      journal = {\aap},
         year = 2023,
        month = may,
       volume = {673},
          eid = {A114},
        pages = {A114},
          doi = {10.1051/0004-6361/202346285},
archivePrefix = {arXiv},
       eprint = {2303.13424},
 primaryClass = {astro-ph.GA},
       adsurl = {https://ui.adsabs.harvard.edu/abs/2023A&A...673A.114H}
}

@ARTICLE{2004MNRAS.348..589C,
       author = {{Cartwright}, Annabel and {Whitworth}, Anthony P.},
        title = "{The statistical analysis of star clusters}",
      journal = {\mnras},
         year = 2004,
        month = feb,
       volume = {348},
       number = {2},
        pages = {589-598},
          doi = {10.1111/j.1365-2966.2004.07360.x},
       adsurl = {https://ui.adsabs.harvard.edu/abs/2004MNRAS.348..589C}
}

@ARTICLE{2009MNRAS.392..341C,
       author = {{Cartwright}, Annabel and {Whitworth}, Anthony P.},
        title = "{Analysing the structure of elongated star clusters}",
      journal = {\mnras},
         year = 2009,
        month = jan,
       volume = {392},
       number = {1},
        pages = {341-345},
          doi = {10.1111/j.1365-2966.2008.14055.x},
       adsurl = {https://ui.adsabs.harvard.edu/abs/2009MNRAS.392..341C}
}

@ARTICLE{2011A&A...532A.119O,
       author = {{Olczak}, C. and {Spurzem}, R. and {Henning}, Th.},
        title = "{A highly efficient measure of mass segregation in star clusters}",
      journal = {\aap},
         year = 2011,
        month = aug,
       volume = {532},
          eid = {A119},
        pages = {A119},
          doi = {10.1051/0004-6361/201116902},
archivePrefix = {arXiv},
       eprint = {1107.1842},
 primaryClass = {astro-ph.GA},
       adsurl = {https://ui.adsabs.harvard.edu/abs/2011A&A...532A.119O}
}

@BOOK{2004fost.book.....S,
       author = {{Stahler}, Steven W. and {Palla}, Francesco},
        title = "{The Formation of Stars}",
         year = 2004,
       adsurl = {https://ui.adsabs.harvard.edu/abs/2004fost.book.....S}
}

@ARTICLE{2023JApA...44...66K,
       author = {{Kaur}, Harmeen and {Sharma}, Saurabh and {Durgapal}, Alok and {Dewangan}, Lokesh K. and {Verma}, Aayushi and {Panwar}, Neelam and {Pandey}, Rakesh and {Ghosh}, Arpan},
        title = "{Structural analysis of open cluster Bochum 2}",
      journal = {Journal of Astrophysics and Astronomy},
         year = 2023,
        month = dec,
       volume = {44},
       number = {2},
          eid = {66},
        pages = {66},
          doi = {10.1007/s12036-023-09953-9},
archivePrefix = {arXiv},
       eprint = {2305.01212},
 primaryClass = {astro-ph.SR},
       adsurl = {https://ui.adsabs.harvard.edu/abs/2023JApA...44...66K}
}

@ARTICLE{2023JApA...44...46S,
       author = {{Sharma}, Saurabh and {Dewangan}, Lokesh and {Panwar}, Neelam and {Kaur}, Harmeen and {Ojha}, Devendra K. and {Yadav}, Ramkesh and {Verma}, Aayushi and {Baug}, Tapas and {Sinha}, Tirthendu and {Pandey}, Rakesh and {Ghosh}, Arpan and {Chand}, Tarak},
        title = "{Teutsch 76: A deep near-infrared study}",
      journal = {Journal of Astrophysics and Astronomy},
         year = 2023,
        month = jun,
       volume = {44},
       number = {1},
          eid = {46},
        pages = {46},
          doi = {10.1007/s12036-023-09936-w},
archivePrefix = {arXiv},
       eprint = {2302.04516},
 primaryClass = {astro-ph.SR},
       adsurl = {https://ui.adsabs.harvard.edu/abs/2023JApA...44...46S}
}

@ARTICLE{2020ApJ...905...61P,
       author = {{Panwar}, Neelam and {Sharma}, Saurabh and {Ojha}, D.~K. and {Baug}, T. and {Dewangan}, L.~K. and {Bhatt}, B.~C. and {Pandey}, Rakesh},
        title = "{Star Formation and Evolution of Blister-type H II Region Sh2-112}",
      journal = {\apj},
         year = 2020,
        month = dec,
       volume = {905},
       number = {1},
          eid = {61},
        pages = {61},
          doi = {10.3847/1538-4357/abc42e},
archivePrefix = {arXiv},
       eprint = {2010.11065},
 primaryClass = {astro-ph.SR},
       adsurl = {https://ui.adsabs.harvard.edu/abs/2020ApJ...905...61P}
}

@ARTICLE{2023ApJ...944..228M,
       author = {{Mallick}, Kshitiz K. and {Dewangan}, Lokesh K. and {Ojha}, Devendra K. and {Baug}, Tapas and {Zinchenko}, Igor I.},
        title = "{Structure and Kinematics of Sh2-138-A Distant Hub-filament System in the Outer Galactic Plane}",
      journal = {\apj},
         year = 2023,
        month = feb,
       volume = {944},
       number = {2},
          eid = {228},
        pages = {228},
          doi = {10.3847/1538-4357/acb8bc},
archivePrefix = {arXiv},
       eprint = {2302.00445},
 primaryClass = {astro-ph.GA},
       adsurl = {https://ui.adsabs.harvard.edu/abs/2023ApJ...944..228M}
}

@ARTICLE{2023A&A...678A.172P,
       author = {{Pauwels}, T. and {Reggiani}, M. and {Sana}, H. and {Rainot}, A. and {Kratter}, K.},
        title = "{The multiplicity of massive stars in the Scorpius OB1 association through high-contrast imaging}",
      journal = {\aap},
         year = 2023,
        month = oct,
       volume = {678},
          eid = {A172},
        pages = {A172},
          doi = {10.1051/0004-6361/202245324},
archivePrefix = {arXiv},
       eprint = {2307.13500},
 primaryClass = {astro-ph.SR},
       adsurl = {https://ui.adsabs.harvard.edu/abs/2023A&A...678A.172P}
}

@ARTICLE{2004A&A...427..839C,
       author = {{Clark}, J.~S. and {Porter}, J.~M.},
        title = "{Triggered massive star formation in the vicinity of WR 48a}",
      journal = {\aap},
         year = 2004,
        month = dec,
       volume = {427},
        pages = {839-847},
          doi = {10.1051/0004-6361:20040523},
       adsurl = {https://ui.adsabs.harvard.edu/abs/2004A&A...427..839C}
}

@INPROCEEDINGS{2005IAUS..227..389C,
       author = {{Crowther}, Paul A.},
        title = "{Near and mid infrared observations of ultracompact HII regions}",
    booktitle = {Massive Star Birth: A Crossroads of Astrophysics},
         year = 2005,
       editor = {{Cesaroni}, R. and {Felli}, M. and {Churchwell}, E. and {Walmsley}, M.},
       volume = {227},
        month = jan,
        pages = {389-396},
          doi = {10.1017/S1743921305004795},
archivePrefix = {arXiv},
       eprint = {astro-ph/0506324},
 primaryClass = {astro-ph},
       adsurl = {https://ui.adsabs.harvard.edu/abs/2005IAUS..227..389C}
}

@ARTICLE{2023ApJ...953..145V,
       author = {{Verma}, Aayushi and {Sharma}, Saurabh and {Mallick}, Kshitiz K. and {Dewangan}, Lokesh and {Ojha}, Devendra K. and {Yadav}, Ram Kesh and {Pandey}, Rakesh and {Ghosh}, Arpan and {Kaur}, Harmeen and {Panwar}, Neelam and {Chand}, Tarak},
        title = "{Exploring Stellar Cluster and Feedback-driven Star Formation in the Galactic Mid-infrared Bubble [HKS2019] E70}",
      journal = {\apj},
         year = 2023,
        month = aug,
       volume = {953},
       number = {2},
          eid = {145},
        pages = {145},
          doi = {10.3847/1538-4357/acdeef},
archivePrefix = {arXiv},
       eprint = {2306.09031},
 primaryClass = {astro-ph.GA},
       adsurl = {https://ui.adsabs.harvard.edu/abs/2023ApJ...953..145V}
}

@ARTICLE{2010ARA&A..48..431P,
       author = {{Portegies Zwart}, Simon F. and {McMillan}, Stephen L.~W. and {Gieles}, Mark},
        title = "{Young Massive Star Clusters}",
      journal = {\araa},
         year = 2010,
        month = sep,
       volume = {48},
        pages = {431-493},
          doi = {10.1146/annurev-astro-081309-130834},
archivePrefix = {arXiv},
       eprint = {1002.1961},
 primaryClass = {astro-ph.GA},
       adsurl = {https://ui.adsabs.harvard.edu/abs/2010ARA&A..48..431P}
}

@ARTICLE{2020SSRv..216...69A,
       author = {{Adamo}, Angela and {Zeidler}, Peter and {Kruijssen}, J.~M. Diederik and {Chevance}, M{\'e}lanie and {Gieles}, Mark and {Calzetti}, Daniela and {Charbonnel}, Corinne and {Zinnecker}, Hans and {Krause}, Martin G.~H.},
        title = "{Star Clusters Near and Far; Tracing Star Formation Across Cosmic Time}",
      journal = {\ssr},
         year = 2020,
        month = jun,
       volume = {216},
       number = {4},
          eid = {69},
        pages = {69},
          doi = {10.1007/s11214-020-00690-x},
archivePrefix = {arXiv},
       eprint = {2005.06188},
 primaryClass = {astro-ph.GA},
       adsurl = {https://ui.adsabs.harvard.edu/abs/2020SSRv..216...69A}
}

@ARTICLE{2019ARA&A..57..227K,
       author = {{Krumholz}, Mark R. and {McKee}, Christopher F. and {Bland-Hawthorn}, Joss},
        title = "{Star Clusters Across Cosmic Time}",
      journal = {\araa},
         year = 2019,
        month = aug,
       volume = {57},
        pages = {227-303},
          doi = {10.1146/annurev-astro-091918-104430},
archivePrefix = {arXiv},
       eprint = {1812.01615},
 primaryClass = {astro-ph.GA},
       adsurl = {https://ui.adsabs.harvard.edu/abs/2019ARA&A..57..227K}
}

@ARTICLE{2024AJ....168...98V,
       author = {{Verma}, Aayushi and {Sharma}, Saurabh and {Dewangan}, Lokesh K. and {Ojha}, Devendra K. and {Mallick}, Kshitiz and {Yadav}, Ram Kesh and {Kaur}, Harmeen and {Chand}, Tarak and {Mamta} and {Gupta}, Archana},
        title = "{Investigating the Star-forming Sites in the Outer Galactic Arm}",
      journal = {\aj},
         year = 2024,
        month = sep,
       volume = {168},
       number = {3},
          eid = {98},
        pages = {98},
          doi = {10.3847/1538-3881/ad5a8b},
archivePrefix = {arXiv},
       eprint = {2406.10388},
 primaryClass = {astro-ph.GA},
       adsurl = {https://ui.adsabs.harvard.edu/abs/2024AJ....168...98V}
}

@ARTICLE{2024AJ....167..106S,
       author = {{Sharma}, Saurabh and {Verma}, Aayushi and {Mallick}, Kshitiz and {Dewangan}, Lokesh K. and {Kaur}, Harmeen and {Yadav}, Ram Kesh and {Panwar}, Neelam and {Ojha}, Devendra K. and {Chand}, Tarak and {Agarwal}, Mamta},
        title = "{Cluster Formation in a Filamentary Cloud: The Case of the Stellar Cluster NGC 2316}",
      journal = {\aj},
         year = 2024,
        month = mar,
       volume = {167},
       number = {3},
          eid = {106},
        pages = {106},
          doi = {10.3847/1538-3881/ad19cd},
archivePrefix = {arXiv},
       eprint = {2312.17699},
 primaryClass = {astro-ph.GA},
       adsurl = {https://ui.adsabs.harvard.edu/abs/2024AJ....167..106S}
}

@ARTICLE{2009A&A...496..177D,
       author = {{Deharveng}, L. and {Zavagno}, A. and {Schuller}, F. and {Caplan}, J. and {Pomar{\`e}s}, M. and {De Breuck}, C.},
        title = "{Star formation around RCW 120, the perfect bubble}",
      journal = {\aap},
         year = 2009,
        month = mar,
       volume = {496},
       number = {1},
        pages = {177-190},
          doi = {10.1051/0004-6361/200811337},
archivePrefix = {arXiv},
       eprint = {0902.0903},
 primaryClass = {astro-ph.GA},
       adsurl = {https://ui.adsabs.harvard.edu/abs/2009A&A...496..177D}
}

@ARTICLE{2022MNRAS.514.6038Z,
       author = {{Zhou}, Jian-Wen and {Liu}, Tie and {Evans}, Neal J. and {Garay}, Guido and {Goldsmith}, Paul F. and {G{\'o}mez}, Gilberto C. and {V{\'a}zquez-Semadeni}, Enrique and {Liu}, Hong-Li and {Stutz}, Amelia M. and {Wang}, Ke and {Juvela}, Mika and {He}, Jinhua and {Li}, Di and {Bronfman}, Leonardo and {Liu}, Xunchuan and {Xu}, Feng-Wei and {Tej}, Anandmayee and {Dewangan}, L.~K. and {Li}, Shanghuo and {Zhang}, Siju and {Zhang}, Chao and {Ren}, Zhiyuan and {Tatematsu}, Ken'ichi and {Shing Li}, Pak and {Won Lee}, Chang and {Baug}, Tapas and {Qin}, Sheng-Li and {Wu}, Yuefang and {Peng}, Yaping and {Zhang}, Yong and {Liu}, Rong and {Luo}, Qiu-Yi and {Ge}, Jixing and {Saha}, Anindya and {Chakali}, Eswaraiah and {Zhang}, Qizhou and {Kim}, Kee-Tae and {Ristorcelli}, Isabelle and {Shen}, Zhi-Qiang and {Li}, Jin-Zeng},
        title = "{ATOMS: ALMA Three-millimeter Observations of Massive Star-forming regions - XI. From inflow to infall in hub-filament systems}",
      journal = {\mnras},
         year = 2022,
        month = aug,
       volume = {514},
       number = {4},
        pages = {6038-6052},
          doi = {10.1093/mnras/stac1735},
archivePrefix = {arXiv},
       eprint = {2206.08505},
 primaryClass = {astro-ph.GA},
       adsurl = {https://ui.adsabs.harvard.edu/abs/2022MNRAS.514.6038Z}
}

@ARTICLE{2024MNRAS.527.9626P,
       author = {{Pandey}, Rakesh and {Sharma}, Saurabh and {Dewangan}, Lokesh and {Ojha}, D.~K. and {Panwar}, Neelam and {Ghosh}, Arpan and {Sinha}, Tirthendu and {Verma}, Aayushi and {Kaur}, Harmeen},
        title = "{Dissecting the morphology of star forming complex S193}",
      journal = {\mnras},
         year = 2024,
        month = feb,
       volume = {527},
       number = {4},
        pages = {9626-9642},
          doi = {10.1093/mnras/stad2944},
archivePrefix = {arXiv},
       eprint = {2309.12410},
 primaryClass = {astro-ph.GA},
       adsurl = {https://ui.adsabs.harvard.edu/abs/2024MNRAS.527.9626P}
}

@ARTICLE{2023JApA...44...52V,
       author = {{Verma}, Aayushi and {Sharma}, Saurabh and {Dewangan}, Lokesh and {Pandey}, Rakesh and {Baug}, Tapas and {Ojha}, Devendra K. and {Ghosh}, Arpan and {Kaur}, Harmeen},
        title = "{Kronberger 55: A candidate for end-dominated collapse scenario}",
      journal = {Journal of Astrophysics and Astronomy},
         year = 2023,
        month = jun,
       volume = {44},
       number = {1},
          eid = {52},
        pages = {52},
          doi = {10.1007/s12036-023-09932-0},
archivePrefix = {arXiv},
       eprint = {2301.06310},
 primaryClass = {astro-ph.GA},
       adsurl = {https://ui.adsabs.harvard.edu/abs/2023JApA...44...52V}
}

@ARTICLE{2021ApJ...914..115Y,
       author = {{Yasui}, Chikako},
        title = "{Spitzer Mid-infrared Study of Sh 2-208: Evolution of Protoplanetary Disks in Low-metallicity Environments}",
      journal = {\apj},
         year = 2021,
        month = jun,
       volume = {914},
       number = {2},
          eid = {115},
        pages = {115},
          doi = {10.3847/1538-4357/abf8b7},
archivePrefix = {arXiv},
       eprint = {2104.11764},
 primaryClass = {astro-ph.GA},
       adsurl = {https://ui.adsabs.harvard.edu/abs/2021ApJ...914..115Y}
}

@ARTICLE{GaiaDR3_2023,
       author = {{Gaia Collaboration} and {Vallenari}, A. and {Brown}, A.~G.~A. and {Prusti}, T. and {de Bruijne}, J.~H.~J. and {Arenou}, F. and {Babusiaux}, C. and {Biermann}, M. and {Creevey}, O.~L. and {Ducourant}, C. and {Evans}, D.~W. and {Eyer}, L. and {Guerra}, R. and {Hutton}, A. and {Jordi}, C. and {Klioner}, S.~A. and {Lammers}, U.~L. and {Lindegren}, L. and {Luri}, X. and {Mignard}, F. and {Panem}, C. and {Pourbaix}, D. and {Randich}, S. and {Sartoretti}, P. and {Soubiran}, C. and {Tanga}, P. and {Walton}, N.~A. and {Bailer-Jones}, C.~A.~L. and {Bastian}, U. and {Drimmel}, R. and {Jansen}, F. and {Katz}, D. and {Lattanzi}, M.~G. and {van Leeuwen}, F. and {Bakker}, J. and {Cacciari}, C. and {Casta{\~n}eda}, J. and {De Angeli}, F. and {Fabricius}, C. and {Fouesneau}, M. and {Fr{\'e}mat}, Y. and {Galluccio}, L. and {Guerrier}, A. and {Heiter}, U. and {Masana}, E. and {Messineo}, R. and {Mowlavi}, N. and {Nicolas}, C. and {Nienartowicz}, K. and {Pailler}, F. and {Panuzzo}, P. and {Riclet}, F. and {Roux}, W. and {Seabroke}, G.~M. and {Sordo}, R. and {Th{\'e}venin}, F. and {Gracia-Abril}, G. and {Portell}, J. and {Teyssier}, D. and {Altmann}, M. and {Andrae}, R. and {Audard}, M. and {Bellas-Velidis}, I. and {Benson}, K. and {Berthier}, J. and {Blomme}, R. and {Burgess}, P.~W. and {Busonero}, D. and {Busso}, G. and {C{\'a}novas}, H. and {Carry}, B. and {Cellino}, A. and {Cheek}, N. and {Clementini}, G. and {Damerdji}, Y. and {Davidson}, M. and {de Teodoro}, P. and {Nu{\~n}ez Campos}, M. and {Delchambre}, L. and {Dell'Oro}, A. and {Esquej}, P. and {Fern{\'a}ndez-Hern{\'a}ndez}, J. and {Fraile}, E. and {Garabato}, D. and {Garc{\'\i}a-Lario}, P. and {Gosset}, E. and {Haigron}, R. and {Halbwachs}, J.-L. and {Hambly}, N.~C. and {Harrison}, D.~L. and {Hern{\'a}ndez}, J. and {Hestroffer}, D. and {Hodgkin}, S.~T. and {Holl}, B. and {Jan{\ss}en}, K. and {Jevardat de Fombelle}, G. and {Jordan}, S. and {Krone-Martins}, A. and {Lanzafame}, A.~C. and {L{\"o}ffler}, W. and {Marchal}, O. and {Marrese}, P.~M. and {Moitinho}, A. and {Muinonen}, K. and {Osborne}, P. and {Pancino}, E. and {Pauwels}, T. and {Recio-Blanco}, A. and {Reyl{\'e}}, C. and {Riello}, M. and {Rimoldini}, L. and {Roegiers}, T. and {Rybizki}, J. and {Sarro}, L.~M. and {Siopis}, C. and {Smith}, M. and {Sozzetti}, A. and {Utrilla}, E. and {van Leeuwen}, M. and {Abbas}, U. and {{\'A}brah{\'a}m}, P. and {Abreu Aramburu}, A. and {Aerts}, C. and {Aguado}, J.~J. and {Ajaj}, M. and {Aldea-Montero}, F. and {Altavilla}, G. and {{\'A}lvarez}, M.~A. and {Alves}, J. and {Anders}, F. and {Anderson}, R.~I. and {Anglada Varela}, E. and {Antoja}, T. and {Baines}, D. and {Baker}, S.~G. and {Balaguer-N{\'u}{\~n}ez}, L. and {Balbinot}, E. and {Balog}, Z. and {Barache}, C. and {Barbato}, D. and {Barros}, M. and {Barstow}, M.~A. and {Bartolom{\'e}}, S. and {Bassilana}, J.-L. and {Bauchet}, N. and {Becciani}, U. and {Bellazzini}, M. and {Berihuete}, A. and {Bernet}, M. and {Bertone}, S. and {Bianchi}, L. and {Binnenfeld}, A. and {Blanco-Cuaresma}, S. and {Blazere}, A. and {Boch}, T. and {Bombrun}, A. and {Bossini}, D. and {Bouquillon}, S. and {Bragaglia}, A. and {Bramante}, L. and {Breedt}, E. and {Bressan}, A. and {Brouillet}, N. and {Brugaletta}, E. and {Bucciarelli}, B. and {Burlacu}, A. and {Butkevich}, A.~G. and {Buzzi}, R. and {Caffau}, E. and {Cancelliere}, R. and {Cantat-Gaudin}, T. and {Carballo}, R. and {Carlucci}, T. and {Carnerero}, M.~I. and {Carrasco}, J.~M. and {Casamiquela}, L. and {Castellani}, M. and {Castro-Ginard}, A. and {Chaoul}, L. and {Charlot}, P. and {Chemin}, L. and {Chiaramida}, V. and {Chiavassa}, A. and {Chornay}, N. and {Comoretto}, G. and {Contursi}, G. and {Cooper}, W.~J. and {Cornez}, T. and {Cowell}, S. and {Crifo}, F. and {Cropper}, M. and {Crosta}, M. and {Crowley}, C. and {Dafonte}, C. and {Dapergolas}, A. and {David}, M. and {David}, P. and {de Laverny}, P. and {De Luise}, F. and {De March}, R.},
        title = "{Gaia Data Release 3. Summary of the content and survey properties}",
      journal = {\aap},
         year = 2023,
        month = jun,
       volume = {674},
          eid = {A1},
        pages = {A1},
          doi = {10.1051/0004-6361/202243940},
archivePrefix = {arXiv},
       eprint = {2208.00211},
 primaryClass = {astro-ph.GA},
       adsurl = {https://ui.adsabs.harvard.edu/abs/2023A&A...674A...1G}
}

@ARTICLE{2025ApJ...990..155V,
       author = {{Verma}, Aayushi and {Sharma}, Saurabh and {Dewangan}, Lokesh K. and {Chand}, Tarak and {Hoque}, Ariful and {Ojha}, Devendra K. and {Kaur}, Harmeen and {Yadav}, Ram Kesh and {Mamta} and {Chakraborty}, Manojit and {Gupta}, Archana},
        title = "{Unraveling the Feedback-regulated Star Formation Activities around the Expanding Galactic Mid-infrared Bubble [HKS2019] E71}",
      journal = {\apj},
         year = 2025,
        month = sep,
       volume = {990},
       number = {2},
          eid = {155},
        pages = {155},
          doi = {10.3847/1538-4357/adf21d},
archivePrefix = {arXiv},
       eprint = {2507.13232},
 primaryClass = {astro-ph.GA},
       adsurl = {https://ui.adsabs.harvard.edu/abs/2025ApJ...990..155V}
}

@ARTICLE{2018ARA&A..56...41M,
       author = {{Motte}, Fr{\'e}d{\'e}rique and {Bontemps}, Sylvain and {Louvet}, Fabien},
        title = "{High-Mass Star and Massive Cluster Formation in the Milky Way}",
      journal = {\araa},
         year = 2018,
        month = sep,
       volume = {56},
        pages = {41-82},
          doi = {10.1146/annurev-astro-091916-055235},
archivePrefix = {arXiv},
       eprint = {1706.00118},
 primaryClass = {astro-ph.GA},
       adsurl = {https://ui.adsabs.harvard.edu/abs/2018ARA&A..56...41M}
}

@ARTICLE{2020A&A...642A..87K,
       author = {{Kumar}, M.~S.~N. and {Palmeirim}, P. and {Arzoumanian}, D. and {Inutsuka}, S.~I.},
        title = "{Unifying low- and high-mass star formation through density-amplified hubs of filaments. The highest mass stars (>100 M$_{☉}$) form only in hubs}",
      journal = {\aap},
         year = 2020,
        month = oct,
       volume = {642},
          eid = {A87},
        pages = {A87},
          doi = {10.1051/0004-6361/202038232},
archivePrefix = {arXiv},
       eprint = {2008.00295},
 primaryClass = {astro-ph.GA},
       adsurl = {https://ui.adsabs.harvard.edu/abs/2020A&A...642A..87K}
}

@ARTICLE{2023MNRAS.522.3719L,
       author = {{Liu}, Hong-Li and {Tej}, Anandmayee and {Liu}, Tie and {Sanhueza}, Patricio and {Qin}, Sheng-Li and {He}, Jinhua and {Goldsmith}, Paul F. and {Garay}, Guido and {Pan}, Sirong and {Morii}, Kaho and {Li}, Shanghuo and {Stutz}, Amelia and {Tatematsu}, Ken'ichi and {Xu}, Feng-Wei and {Bronfman}, Leonardo and {Saha}, Anindya and {Issac}, Namitha and {Baug}, Tapas and {Toth}, L. Viktor and {Dewangan}, Lokesh and {Wang}, Ke and {Zhou}, Jianwen and {Lee}, Chang Won and {Yang}, Dongting and {Luo}, Anxu and {Shen}, Xianjin and {Zhang}, Yong and {Wu}, Yue-Fang and {Ren}, Zhiyuan and {Liu}, Xun-Chuan and {Soam}, Archana and {Zhang}, Siju and {Luo}, Qiu-Yi},
        title = "{Evidence of high-mass star formation through multiscale mass accretion in hub-filament-system clouds}",
      journal = {\mnras},
         year = 2023,
        month = jul,
       volume = {522},
       number = {3},
        pages = {3719-3734},
          doi = {10.1093/mnras/stad047},
archivePrefix = {arXiv},
       eprint = {2301.03144},
 primaryClass = {astro-ph.GA},
       adsurl = {https://ui.adsabs.harvard.edu/abs/2023MNRAS.522.3719L}
}

@article{10.1093/mnras/stv1521,
    author = {Koch, Eric W. and Rosolowsky, Erik W.},
    title = {Filament identification through mathematical morphology},
    journal = {Monthly Notices of the Royal Astronomical Society},
    volume = {452},
    number = {4},
    pages = {3435-3450},
    year = {2015},
    month = {10},
    issn = {0035-8711},
    doi = {10.1093/mnras/stv1521},
    url = {https://doi.org/10.1093/mnras/stv1521},
    eprint = {https://academic.oup.com/mnras/article-pdf/452/4/3435/18238514/stv1521.pdf},
}

@ARTICLE{2012A&A...542A..10A,
       author = {{Anderson}, L.~D. and {Zavagno}, A. and {Deharveng}, L. and {Abergel}, A. and {Motte}, F. and {Andr{\'e}}, Ph. and {Bernard}, J.-P. and {Bontemps}, S. and {Hennemann}, M. and {Hill}, T. and {Rod{\'o}n}, J.~A. and {Roussel}, H. and {Russeil}, D.},
        title = "{The dust properties of bubble H II regions as seen by Herschel}",
      journal = {\aap},
         year = 2012,
        month = jun,
       volume = {542},
          eid = {A10},
        pages = {A10},
          doi = {10.1051/0004-6361/201117283},
archivePrefix = {arXiv},
       eprint = {1203.5721},
 primaryClass = {astro-ph.GA},
       adsurl = {https://ui.adsabs.harvard.edu/abs/2012A&A...542A..10A}
}

@ARTICLE{2024MNRAS.528.1460R,
       author = {{Rawat}, Vineet and {Samal}, M.~R. and {Eswaraiah}, Chakali and {Wang}, Jia-Wei and {Elia}, Davide and {Panigrahy}, Sandhyarani and {Zavagno}, A. and {Yadav}, R.~K. and {Walker}, D.~L. and {Jose}, J. and {Ojha}, D.~K. and {Zhang}, C.~P. and {Dutta}, S.},
        title = "{Understanding the relative importance of magnetic field, gravity, and turbulence in star formation at the hub of the giant molecular cloud G148.24+00.41}",
      journal = {\mnras},
         year = 2024,
        month = feb,
       volume = {528},
       number = {2},
        pages = {1460-1475},
          doi = {10.1093/mnras/stae053},
archivePrefix = {arXiv},
       eprint = {2401.05310},
 primaryClass = {astro-ph.GA},
       adsurl = {https://ui.adsabs.harvard.edu/abs/2024MNRAS.528.1460R}
}

@ARTICLE{2025ApJ...994..222C,
       author = {{Rishi}, C. and {Panwar}, Neelam and {Haworth}, Thomas J. and {Sun}, Yan and {Sharma}, Saurabh and {Yadav}, R.~K. and {Ojha}, D.~K. and {Singh}, H.~P. and {Jose}, Jessy and {Singh}, Ajay Kumar and {Jose}, Jincen and {Yadav}, Shubham},
        title = "{Investigating the Influence of Radiative Feedback in the Bright-rimmed Cloud 44}",
      journal = {\apj},
         year = 2025,
        month = dec,
       volume = {994},
       number = {2},
          eid = {222},
        pages = {222},
          doi = {10.3847/1538-4357/ae0f03},
archivePrefix = {arXiv},
       eprint = {2510.08229},
 primaryClass = {astro-ph.SR},
       adsurl = {https://ui.adsabs.harvard.edu/abs/2025ApJ...994..222C}
}

@ARTICLE{2022MNRAS.510..658P,
       author = {{Potdar}, Ajay and {Das}, Swagat R. and {Issac}, Namitha and {Tej}, Anandmayee and {Vig}, Sarita and {Chandra}, C.~H. Ishwara},
        title = "{Galactic H II region IRAS 17149 - 3916 - a multiwavelength study}",
      journal = {\mnras},
         year = 2022,
        month = feb,
       volume = {510},
       number = {1},
        pages = {658-673},
          doi = {10.1093/mnras/stab3479},
archivePrefix = {arXiv},
       eprint = {2111.13869},
 primaryClass = {astro-ph.GA},
       adsurl = {https://ui.adsabs.harvard.edu/abs/2022MNRAS.510..658P}
}

@ARTICLE{2025ARA&A..63....1B,
       author = {{Beuther}, H. and {Kuiper}, R. and {Tafalla}, M.},
        title = "{Star Formation from Low to High Mass: A Comparative View}",
      journal = {\araa},
         year = 2025,
        month = aug,
       volume = {63},
       number = {1},
        pages = {1-44},
          doi = {10.1146/annurev-astro-013125-122023},
archivePrefix = {arXiv},
       eprint = {2501.16866},
 primaryClass = {astro-ph.GA},
       adsurl = {https://ui.adsabs.harvard.edu/abs/2025ARA&A..63....1B}
}

@ARTICLE{2024A&A...690A..94P,
       author = {{Polak}, Brooke and {Mac Low}, Mordecai-Mark and {Klessen}, Ralf S. and {Teh}, Jia Wei and {Cournoyer-Cloutier}, Claude and {Andersson}, Eric P. and {Appel}, Sabrina M. and {Tran}, Aaron and {Lewis}, Sean C. and {Wilhelm}, Maite J.~C. and {Portegies Zwart}, Simon and {Glover}, Simon C.~O. and {Rieder}, Steven and {Wang}, Long and {McMillan}, Stephen L.~W.},
        title = "{Massive star cluster formation: I. High star formation efficiency while resolving feedback of individual stars}",
      journal = {\aap},
         year = 2024,
        month = oct,
       volume = {690},
          eid = {A94},
        pages = {A94},
          doi = {10.1051/0004-6361/202348840},
archivePrefix = {arXiv},
       eprint = {2312.06509},
 primaryClass = {astro-ph.GA},
       adsurl = {https://ui.adsabs.harvard.edu/abs/2024A&A...690A..94P}
}

@article{1993A&A...275...67B,
  author  = {Brand, J. and Blitz, L.},
  title   = {{The velocity field of the outer Galaxy}},
  journal = {Astronomy and Astrophysics},
  year    = {1993},
  volume  = {275},
  pages   = {67--90},
  adsurl  = {https://ui.adsabs.harvard.edu/abs/1993A&A...275...67B}
}

@article{2014ApJ...783..130R,
  author  = {Reid, M. J. and Menten, K. M. and Brunthaler, A. and
             Zheng, X. W. and Dame, T. M. and Xu, Y. and Wu, Y. and
             Zhang, B. and Sanna, A. and Sato, M. and Hachisuka, K. and
             Choi, Y. K. and Immer, K. and Moscadelli, L. and
             Rygl, K. L. J. and Bartkiewicz, A.},
  title   = {{Trigonometric Parallaxes of High Mass Star Forming Regions:
              The Structure and Kinematics of the Milky Way}},
  journal = {The Astrophysical Journal},
  year    = {2014},
  volume  = {783},
  number  = {2},
  pages   = {130},
  doi     = {10.1088/0004-637X/783/2/130}
}

@article{2018ApJ...856...52W,
  author  = {Wenger, Trey V. and Balser, Dana S. and
             Anderson, L. D. and Bania, T. M.},
  title   = {{Kinematic Distances: A Monte Carlo Method}},
  journal = {The Astrophysical Journal},
  year    = {2018},
  volume  = {856},
  number  = {1},
  pages   = {52},
  doi     = {10.3847/1538-4357/aaaec8}
}

@article{2009ApJ...700..137R,
  author  = {Reid, M. J. and Menten, K. M. and Zheng, X. W. and
             Brunthaler, A. and Moscadelli, L. and Xu, Y. and
             Zhang, B. and Sato, M. and Honma, M. and Hirota, T. and
             Hachisuka, K. and Choi, Y. K. and Moellenbrock, G. A. and
             Bartkiewicz, A.},
  title   = {Trigonometric Parallaxes of Massive Star-forming Regions.
             VI. Galactic Structure, Fundamental Parameters, and
             Noncircular Motions},
  journal = {The Astrophysical Journal},
  year    = {2009},
  volume  = {700},
  number  = {1},
  pages   = {137--148},
  doi     = {10.1088/0004-637X/700/1/137}
}

@article{2018MNRAS.474.2028R,
  author  = {Ram{\'o}n-Fox, F. G. and Bonnell, I. A.},
  title   = {Streaming Motions and Kinematic Distances to Molecular Clouds},
  journal = {Monthly Notices of the Royal Astronomical Society},
  year    = {2018},
  volume  = {474},
  number  = {2},
  pages   = {2028--2038},
  doi     = {10.1093/mnras/stx2866}
}

@article{2012MNRAS.424..377D,
  author  = {Dale, J. E. and Ercolano, B. and Bonnell, I. A.},
  title   = {Ionizing Feedback from Massive Stars in Massive Clusters.
             II. Disruption of Bound Clusters by Photoionization},
  journal = {Monthly Notices of the Royal Astronomical Society},
  year    = {2012},
  volume  = {424},
  number  = {1},
  pages   = {377--392},
  doi     = {10.1111/j.1365-2966.2012.21205.x}
}
\bibliographystyle{aasjournal}

\appendix

        \section{Membership Probability}\label{mp}

         \begin{figure*}
        \centering
        \includegraphics[width=8cm,height=9cm]{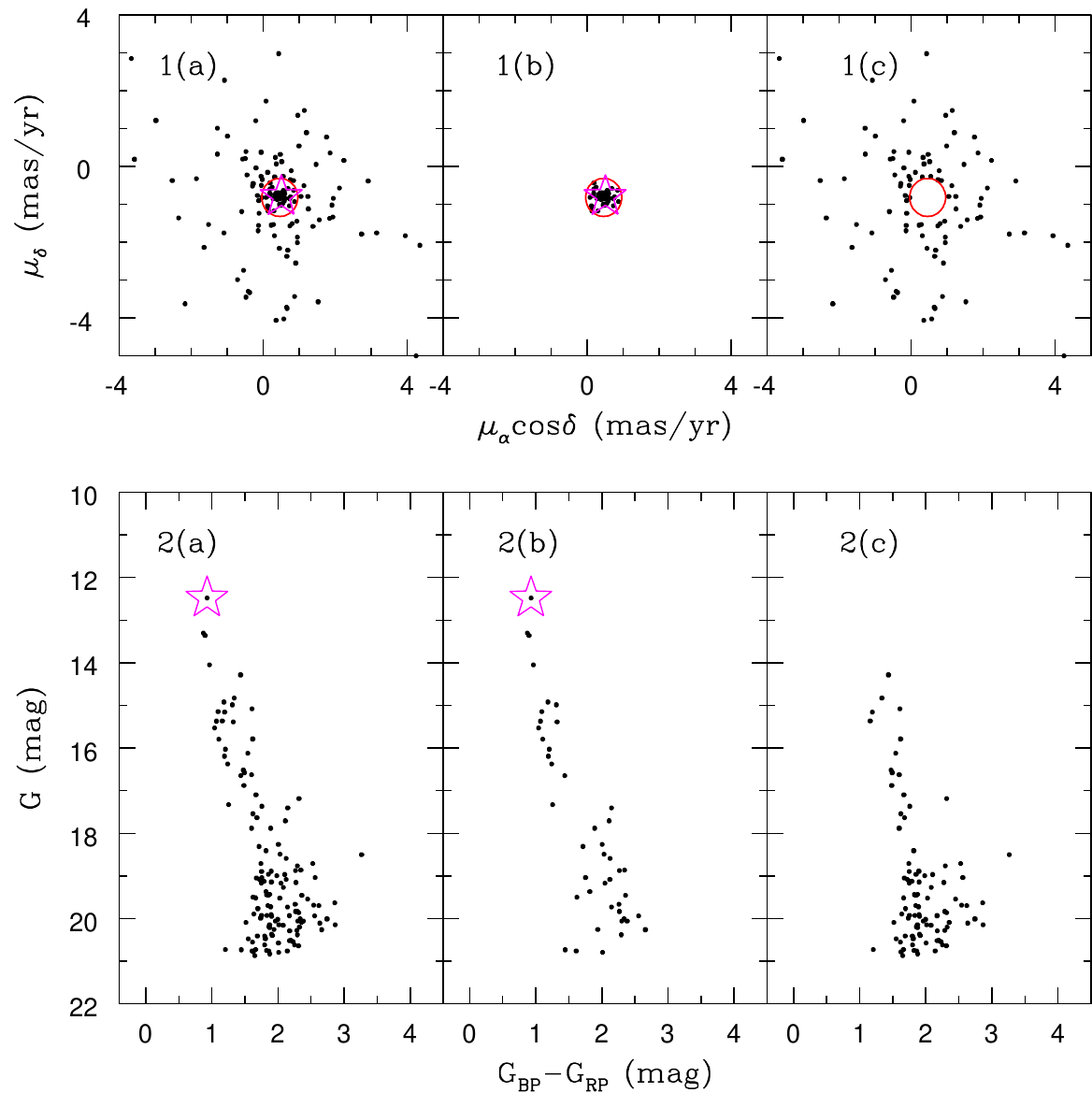}
        \includegraphics[width=8cm,height=9cm]{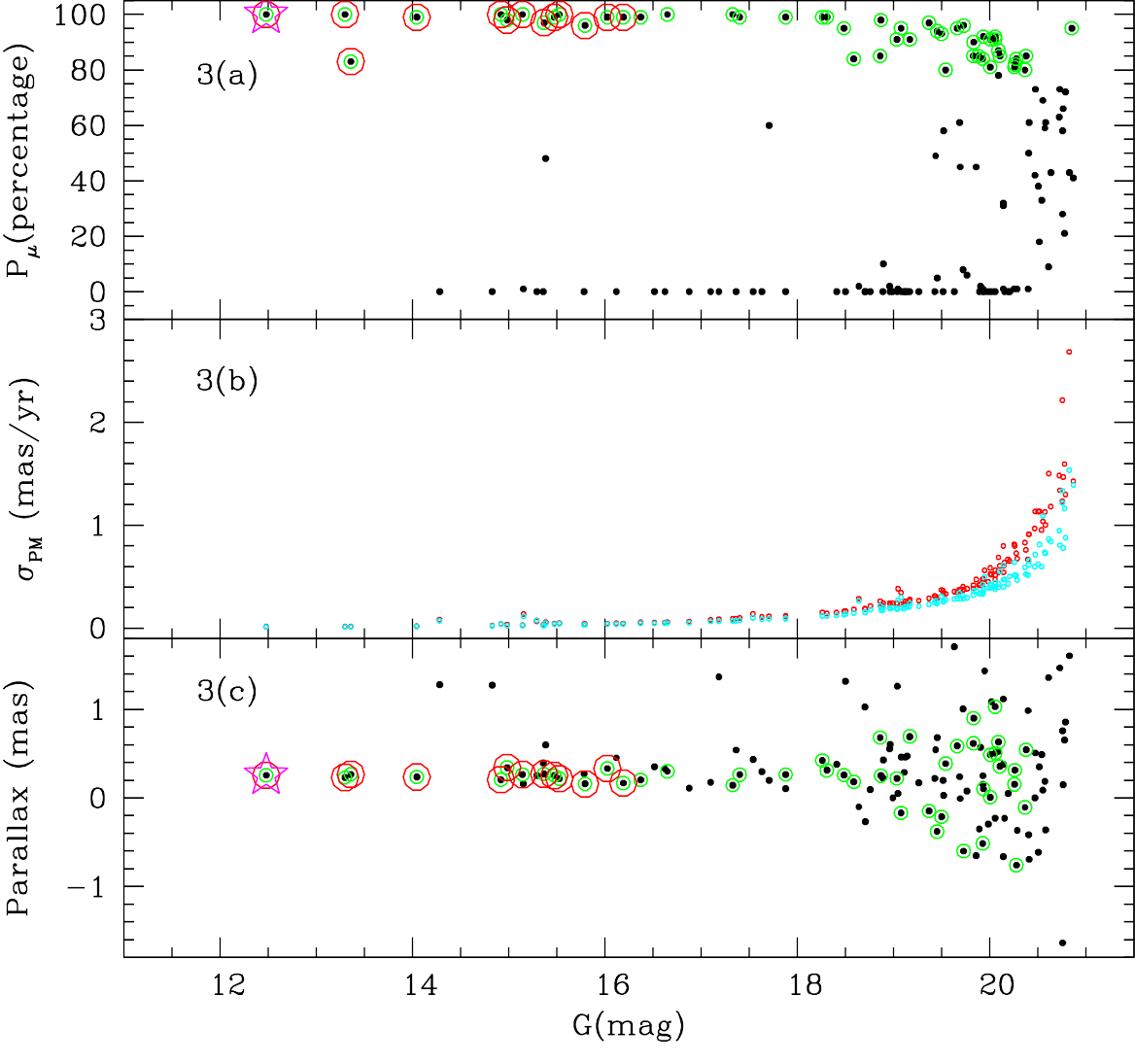}
        \caption{\label{pm} {\it Panel 1 and 2}:  represents PM VPDs and corresponding Gaia DR3,  $G$ vs. ($G_{BP} - G_{RP}$) CMDs. The left sub-panels indicate all the stars within the convex hull of the Wat01 cluster,
        while the middle and right sub-panels show the
        potential cluster members and field stars, respectively (for details see Appendix~\ref{mp}).
        {\it Panel 3}: Membership probability P$\mu$, PM errors $\sigma_{PM}$, and parallax of stars
        as a function of G magnitude for stars within the hull of the Wat01 region. A total of 52 stars with P$\mu$ $>$ 80\%  are identified as probable member stars of the Wat01 cluster and are represented by green circles with black rings. Member stars with parallax errors $<~0.05$ mas are shown as red circles.}
        \end{figure*}

        We adopted the method described in \citet{1998A&AS..133..387B} to determine the membership probability of cluster stars using the proper motion (PM) data of  $Gaia$ DR3\footnote{https://www.cosmos.esa.int/web/gaia/dr3} catalog \citep{2021A&A...649A...1G}. In the literature, this method has been considerably 
        used  for $\omega$ Centauri \citep{2009A&A...493..959B}, NGC 6809 \citep{2012A&A...543A..87S}, 
        NGC 6366 \citep{2015A&A...584A..59S},
        NGC 3201 \citep{2017AJ....153..134S}, 
        Czernik 3 \citep{2020MNRAS.498.2309S}, NGC 6910 \citep{2020ApJ...896...29K}, and Bochum~2 \citep{2023JApA...44...66K}.
        The $Gaia$ PM data located within the convex hull of the Wat01 cluster region are used here to determine the membership probability of member stars.
        In Figure~\ref{pm}, Panel 1 depicts the PM $\mu_\alpha$cos($\delta$) and $\mu_\delta$ distribution of all the stars as vector point diagrams (VPDs), here
        a tight circular clump (within the red circle) observable visually at $\mu_{xc}$ = 0.46
        mas yr$^{-1}$, $\mu_{yc}$ = -0.81 mas yr$^{-1}$  having radius of 0.5 mas yr$^{-1}$.
        The radius is chosen to strike a balance between preserving cluster members with good PMs and including field stars that contribute to the cluster's mean PM. 
        The population of stars in panel 1(b) has nearly identical PMs and a high likelihood of belonging to a cluster, whereas the remaining stars in panel 1(c) have scattered PM values and are most likely part of a field population.
         The panels 2(a), 2(b), and 2(c) display the respective $G$ versus $(G_{BP} -G_{RP})$ CMDs of all the stars inside the hull, probable cluster members, and field stars, respectively. The star symbol represents the location of a massive star within the cluster region. 
        The field stars in the CMD exhibit a dispersed distribution, while the potential cluster members appear to follow a well-defined MS. 
        Given that the Wat01 cluster is located at a distance of 3.53 kpc (current estimate; see Section~\ref{iso}) and the radial velocity dispersion for open cluster \citep{1989AJ.....98..227G} is 1 km~s$^{-1}$, the anticipated dispersion ($\sigma_c$) in PMs is approximately 0.053 mas~yr$^{-1}$. 
        For field stars, we have calculated $\mu_{xf}$ = 0.46 mas yr$^{-1}$, $\mu_{yf}$ = -1.21 mas~yr$^{-1}$, 
        $\sigma_{xf}$ = 1.78 mas~yr$^{-1}$ and $\sigma_{yf}$ = 2.65 mas~yr$^{-1}$, these values are further used to construct frequency distribution  of
        cluster stars $(\phi^\nu_c)$  and field stars $(\phi^\nu_f)$ using the equations given in \citet{2013MNRAS.430.3350Y}.
        The membership probability for the $i^{th}$ star is estimated as:

        \begin{equation}
        P_\mu(i) = {\frac{n_c\times\phi^\nu_c(i)}{n_c\times\phi^\nu_c(i)+n_f\times\phi^\nu_f(i)}}
        \end{equation}
        
	    where $n_c$ (=0.33) and $n_f$(=0.67) are the normalized numbers of stars for the cluster and field region ($n_c$+$n_f = 1$). Panel 3 of Figure \ref{pm} illustrates the parallax values of stars located inside the hull of the Wat01 cluster, respective errors in the PM ($\sigma_{pm}$), and estimated membership probability ($P_{\mu}$) as a function of G magnitude.
         As seen in this plot, a high membership probability 
         (P$\mu$ $\geq$ 80\%) goes all the way down to G $\sim$20 mag. 
         At faint limits, errors in PMs become quite high, and the maximum probability steadily declines at those values. The distinct separation between cluster members and field stars at brighter magnitudes demonstrates the usefulness of this method. 
          Hence, based on their high membership probability (P$\mu$ $\geq$80\%; 
         green circles with black outlines in panel 3 of Figure \ref{pm}), 
         52 stars located within the convex hull were identified as cluster members of the Wat01 region. These members are listed in Table \ref{tab:wat01_mem}.

    \begin{table*}[htbp]
    \centering
    \caption{Sample of the 52 cluster members in Wat01 region identified from their
    Gaia proper-motion data.}
    \label{tab:wat01_mem}

    \setlength{\tabcolsep}{1pt}
    \renewcommand{\arraystretch}{1.15}
    \scriptsize

    \resizebox{\textwidth}{!}{%
    \begin{tabular}{lccccccccccccc}
    \toprule
    \textbf{ID} & \textbf{RA} & \textbf{Dec} & \textbf{$V\pm\sigma$} & \textbf{$B\pm\sigma$} & \textbf{$I\pm\sigma$} & \textbf{$R\pm\sigma$} & \textbf{$U\pm\sigma$} & \textbf{$\varpi\pm\sigma$} & \textbf{$\mu$} &
    \textbf{$\mu_{\alpha}\cos\delta\pm\sigma$} & \textbf{$\mu_{\delta}\pm\sigma$} & \textbf{$G\pm\sigma$} & \textbf{$P_{\mu}$} \\
     & \textbf{(deg; J2000)} & \textbf{(deg; J2000)} & \textbf{(mag)} & \textbf{(mag)} & \textbf{(mag)} & \textbf{(mag)} & \textbf{(mag)} & \textbf{(mas)} &
    \textbf{(mas yr$^{-1}$)} & \textbf{(mas yr$^{-1}$)} & \textbf{(mas yr$^{-1}$)} & \textbf{(mag)} & \textbf{(\%)} \\
    \midrule
    1 & 64.665231 & 52.862928 & $20.015 \pm 0.084$ & $22.558 \pm 0.277$ & $17.688 \pm 0.014$ & $18.732 \pm 0.045$ & $99.999 \pm 9.9999$ & $0.2509 \pm 0.210$ & 0.929 & $0.591 \pm 0.210$ & $-0.717 \pm 0.176$ & $18.869 \pm 0.004$ & 98 \\
    2 & 64.662345 & 52.857778 & $20.074 \pm 0.027$ & $21.625 \pm 0.321$ & $18.275 \pm 0.015$ & $19.140 \pm 0.037$ & $99.999 \pm 9.9999$ & $-0.1476 \pm 0.279$ & 0.887 & $0.270 \pm 0.290$ & $-0.845 \pm 0.231$ & $19.368 \pm 0.004$ & 97 \\
    3 & 64.672074 & 52.859156 & $20.573 \pm 0.075$ & $99.999 \pm 9.9999$ & $18.634 \pm 0.022$ & $19.523 \pm 0.045$ & $99.999 \pm 9.9999$ & $-0.6008 \pm 0.357$ & 0.974 & $0.477 \pm 0.373$ & $-0.850 \pm 0.290$ & $19.729 \pm 0.005$ & 96 \\
    4 & 64.658009 & 52.857747 & $16.024 \pm 0.004$ & $16.803 \pm 0.006$ & $15.131 \pm 0.007$ & $15.596 \pm 0.007$ & $16.955 \pm 0.013$ & $0.6298 \pm 0.497$ & 1.308 & $1.045 \pm 0.545$ & $-0.786 \pm 0.399$ & $20.090 \pm 0.011$ & 87 \\
   5 & 64.656742 & 52.857906 & $16.024 \pm 0.004$ & $16.803 \pm 0.006$ & $15.131 \pm 0.007$ & $15.596 \pm 0.007$ & $16.955 \pm 0.013$ & $0.1578 \pm 0.036$ & 0.801 & $0.315 \pm 0.040$ & $-0.737 \pm 0.032$ & $15.793 \pm 0.003$ & 96 \\
    \bottomrule
    \end{tabular}%
    }

\vspace{1ex}
\begin{minipage}{0.98\textwidth}
\footnotesize
\textit{Note.} This table is available in its entirety in machine-readable form. Here, $\varpi$ is the Gaia parallax, $\mu$ is the total proper motion, and $P_{\mu}$ is the proper-motion membership probability.
\end{minipage}
\end{table*}

        \section{Identification of young stellar objects}\label{yso}

        The identification and classification of YSOs are crucial for probing the physical processes governing star formation. YSOs are identified based on their excess IR emission. We used NIR data from UKIDSS/2MASS  and MIR data from \textit{Spitzer}/WISE to identify YSOs. We used the GLIMPSE360 catalog of the \textit{Spitzer} Science Center, employing the classification scheme of \citet{2009ApJS..184...18G}. The [K$-$[3.6]]0 versus [[3.6]$-$[4.5]]0 TCD 
        gives 9 Class I and 39 Class II YSOs in the Wat01-S208 region. We also used UKIDSS data along with 2MASS data by choosing brighter sources (with J magnitude $\leq$13) from
         2MASS and the fainter sources (with J magnitude $>$13) from UKIDSS. In this way, we created an NIR catalog.
        We used the classification scheme of \citet{2004ApJ...608..797O} on these data. The (J$-$H) versus (H$-$K) TCD gives 7 Class I and 62 Class II YSOs in the region. Furthermore, we used the MIR data of the ALLWISE catalog of WISE and followed the scheme of \citet{2014ApJ...791..131K}. According to this scheme, a selection criterion is applied to all four WISE bands to first get good-quality WISE data; then the extragalactic contaminants, such as AGNs, star-forming galaxies, and transition disks, are separated
        out. We identified 13 Class II using ([3.4] - [4.6]) versus ([4.6] - [12]) TCD.  We cross-matched the YSOs identified by the three classification schemes using a matching radius of 
        \(3^{\prime\prime}\), obtaining 105 unique YSOs in our region: 8 Class I and 97 Class II sources (Table \ref{tab:yso-sample}), excluding 40 of those found in \citet{2021ApJ...914..115Y}.

    \providecommand{\nodata}{\ensuremath{\cdots}}

\begin{table}[ht!]
\centering
\scriptsize
\setlength{\tabcolsep}{1pt}
\caption{A sample of identified 105 YSOs in the Wat01-S208 star-forming complex.}
\label{tab:yso-sample}
\resizebox{\textwidth}{!}{%
\begin{tabular}{@{}l c c *{11}{c} c@{}}
\toprule
\textbf{ID} & \textbf{RA} & \textbf{Dec} & \textbf{$J\pm\sigma$} & \textbf{$H\pm\sigma$} &
\textbf{$K\pm\sigma$} & \textbf{$[3.6]\pm\sigma$} & \textbf{$[4.5]\pm\sigma$} &
\textbf{$[5.8]\pm\sigma$} & \textbf{$[8.0]\pm\sigma$} & \textbf{$[3.4]\pm\sigma$} &
\textbf{$[4.6]\pm\sigma$} & \textbf{$[12]\pm\sigma$} & \textbf{$[22]\pm\sigma$} & \textbf{Class} \\
 & \textbf{(deg; J2000)} & \textbf{(deg; J2000)} & \textbf{(mag)} & \textbf{(mag)} & \textbf{(mag)} & \textbf{(mag)} & \textbf{(mag)} & \textbf{(mag)} & \textbf{(mag)} & \textbf{(mag)} & \textbf{(mag)} & \textbf{(mag)} &\textbf{(mag)} & \\
\midrule
1 & 64.849453 & 52.960757 & \nodata & \nodata & \nodata &
\(13.831\pm0.014\) & \(12.543\pm0.009\) &
\(11.458\pm0.006\) & \(10.818\pm0.006\) &
\nodata & \nodata & \nodata & \nodata & II \\
2 & 65.022881 & 53.674088 &
\(16.383\pm0.011\) & \(15.250\pm0.009\) & \(14.528\pm0.008\) &
\nodata & \nodata & \nodata & \nodata & \nodata &
\nodata & \nodata & \nodata & II \\
3 & 64.828909 & 53.014177 &
\(16.018\pm0.009\) & \(14.957\pm0.006\) & \nodata &
\(13.259\pm0.010\) & \(12.711\pm0.012\) &
\(12.312\pm0.422\) & \(12.052\pm0.018\) &
\(13.592\pm0.033\) & \(12.845\pm0.030\) &
\(9.739\pm0.120\) & \(7.123\pm0.099\) & II \\
5 & 64.728829 & 52.844092 &
\(15.537\pm0.006\) & \(14.908\pm0.006\) & \(14.557\pm0.009\) &
\(14.048\pm0.043\) & \(13.848\pm0.036\) &
\nodata & \nodata &
\(13.709\pm0.028\) & \(13.584\pm0.036\) &
\(12.325\pm0.014\) & \(7.811\pm0.189\) & II \\
9 & 64.894609 & 52.965336 & \nodata & \nodata & \nodata &
\(11.799\pm0.052\) & \(10.974\pm0.004\) &
\(10.234\pm0.003\) & \(9.610\pm0.002\) &
\(12.151\pm0.026\) & \(10.762\pm0.022\) &
\(7.491\pm0.026\) & \(3.919\pm0.024\) & I \\
\bottomrule
\end{tabular}%
}
\par\vspace{1ex}
\begin{minipage}{\textwidth}
\footnotesize\textit{Note.} Full table is available in machine-readable form. \nodata\ denotes missing data.
\end{minipage}
\end{table}

    \begin{table*}[ht!]
    \centering
    \caption{\label{fp} The top three rows list the total pressure exerted by massive stars `W1' (O9V) and `W2' (B1.5V) at various projected distances ($D_{s}$) along the rim of the bubble/ring-like structure and peaks of the globules (see panel (b) of Figure~\ref{akari}). The bottom two rows list the total pressure exerted by massive star ‘S1’ at the projected distance of the bubble and embedded cluster (see panel (c) of Figure~\ref{all}).}
    \begin{tabular}{@{}cccccc@{}}
    \toprule
   
    \textbf{Projected $D_s$} & \textbf{$D_s$} & \textbf{P$_{HII}$} & \textbf{P$_{rad}$} & \textbf{P$_{wind}$} & \textbf{P$_{total}$} \\
    
     &  \textbf{(pc)} & \textbf{(dynes cm$^{-2}$)} & \textbf{(dynes cm$^{-2}$)} & \textbf{(dynes cm$^{-2}$)}  &  \textbf{(dynes cm$^{-2}$)} \\
    \midrule
    
	d1 (GB1/Bubble) & 7.8  & $1.33\times10^{-11}$ & $1.32\times10^{-12}$  & $1.26\times10^{-14}$ & $1.46\times10^{-11}$  \\
    
    d2 (GB2) & 5.8  & $2.07\times10^{-11}$  &  $2.39\times10^{-12}$ & $2.28\times10^{-14}$ & $2.31\times10^{-11}$   \\
   
    d3 (GB3) &  5.4 & $2.31\times10^{-11}$  & $2.76\times10^{-12}$  &  $2.63\times10^{-14}$    &  $2.59\times10^{-11}$      \\
    
     \midrule
     
      d1 (Ycl) & 1.23 & $3.18\times10^{-11}$ & $6.74\times10^{-12}$ & $4.26\times10^{-13}$ & $3.90\times10^{-11}$ \\%

    d2 (Bubble) & 0.58 & $9.83\times10^{-11}$ & $3.03\times10^{-11}$ & $1.92\times10^{-12}$ & $1.47\times10^{-10}$  \\
   
     \bottomrule
    \end{tabular}
    \end{table*}

    \section{Feedback from the Massive Stars}\label{pre}

    To quantify the feedback impact of massive stars associated with Wat01 and S208 on their surrounding medium, we have calculated the  feedback pressure through different pressure components expressed as \citet{{2012ApJ...758L..28B}}:
       
        \renewcommand{\theenumi}{\Roman{enumi}}%
        \begin{enumerate}
        \item Pressure imposed by H\,{\sc ii} region-
        
        \begin{equation}
       P_{\mathrm{HII}} = \mu\, m_{\mathrm{H}}\, c_{\mathrm{s}}^2 \left( \sqrt{ \frac{3 N_{\mathrm{uv}}}{4 \pi\, \alpha_{B}\, D_{\mathrm{s}}^3} } \right)
        \end{equation}
        
        \item The Radiation Pressure
        
        \begin{equation}
        P_{\mathrm{rad}} = L_{\mathrm{bol}}/ 4\pi c D_{\mathrm{s}}^2; \\ 
        \end{equation}
         
         and 
        \item The stellar wind Ram Pressure
        
        \begin{equation}
        P_{\mathrm{wind}} = \dot{M}_{\mathrm{w}} V_{\mathrm{w}} / 4 \pi D_{\mathrm{s}}^2; \\
        \end{equation}
        \end{enumerate}

        Therefore, we calculated the total feedback pressure from the massive stars `W1' (O9.5V) and `W2' (B1.5V), associated with Wat01, along the bubble/ring for various projected distances (D$_\mathrm{s}$ $\approx$ d1 $\sim10$ pc, d2 $\sim 5.9$ pc, and d3 $\sim5.4$ PC 
        ; see panel (b) of Figure~\ref{akari} ). It is calculated by considering Lyman continuum photons (N$_{\mathrm{uv}}$), mass-loss rate  ($\dot{M}_{\mathrm{w}}$), wind velocity (V$_{\mathrm{w}}$), and bolometric luminosity (L$_{\mathrm{bol}}$) as  1.2 $\times$ 10$^{48}$ photons s$^{-1}$ \citep[][]{1973AJ.....78..929P}, $\approx$ 1.58 $\times$ 10$^{-9}$ M$_{\odot}$ yr$^{-1}$ \citep[][]{2009A&A...498..837M},  $\approx$ 1500 km\,s$^{-1}$ \citep[][]{2009A&A...498..837M} and 66070 L$_{\odot}$ \citep[][]{1973AJ.....78..929P}, respectively for O9.5V (``W1'') type star; and $1.1\times 10^{46}$ photons s$^{-1}$ \citep{2004A&A...427..839C}, $\approx$ 1.02$\times$10$^{-8}$ M$_{\odot}$ yr$^{-1}$ \citep{2023A&A...678A.172P}, $\approx$ 1200 km\,s$^{-1}$ \citep{2023A&A...678A.172P} and 9549.93  L$_{\odot}$ \citep{2005IAUS..227..389C}, respectively, for B1.5V (`W2') type star \citep[for more details on pressure estimation, please refer][]{2020ApJ...896...29K, 2023ApJ...953..145V, 2025ApJ...990..155V}. We also calculated the total pressure from massive star `S1' (B 1.5V) located in the S208 region to the bubble boundary and embedded cluster Ycl at projected distances d1 and d2 (see panel c of Figure~\ref{all}).  
        The total pressure calculated for different distances is listed in Table~\ref{fp}.

\end{document}